\documentclass[sn-mathphys]{sn-jnl_edited}

\jyear{2022}%

\theoremstyle{thmstyleone}%
\theoremstyle{thmstyletwo}%

\theoremstyle{thmstylethree}%

\newcommand{\covid}{\textsc{covid}-19}
\newcommand{\sars}{\textsc{sars}-\textsc{c}o\textsc{v}-2}

\usepackage{bm}

\usepackage{siunitx} 

\usepackage{pdflscape}

\graphicspath{{fig/}}

\begin{document}

\title{A Stochastic Mobility-Driven Spatially Explicit SEIQRD \covid{} Model with VOCs, Seasonality, and Vaccines}


\author*[1,2]{\fnm{Tijs W.}
\sur{Alleman}}\email{tijs.alleman@ugent.be}
\equalcont{These authors contributed equally to this work.}

\author[1]{\fnm{Michiel} \sur{Rollier}}
\equalcont{These authors contributed equally to this work.}

\author[1,2]{\fnm{Jenna} \sur{Vergeynst}}

\author[1]{\fnm{Jan M.}
\sur{Baetens}}

\affil[1]{\orgdiv{BIOSPACE}, \orgname{Department of Data Analysis and Mathematical Modelling, Ghent University}, \orgaddress{\street{Coupure Links 653}, \city{Ghent}, \postcode{9000}, \country{Belgium}}}

\affil[2]{\orgdiv{BIOMATH}, \orgname{Department of Data Analysis and Mathematical Modelling, Ghent University}, \orgaddress{\street{Coupure Links 653}, \city{Ghent}, \postcode{9000}, \country{Belgium}}}


\maketitle

\begin{small}



\noindent\textbf{Abstract}
In this work, we extend our previously developed compartmental SEIQRD model for \sars{} in Belgium. We introduce \sars{} variants of concern, vaccines, and seasonality in our model, as their addition has proven necessary for modelling \sars{} transmission dynamics during the 2020-2021 \covid{} pandemic in Belgium. The model is geographically stratified into eleven spatial patches (provinces), and a telecommunication dataset provided by Belgium's biggest operator is used to incorporate interprovincial mobility. We calibrate the model using the daily number of hospitalisations in each province and serological data. We find the model adequately describes these data, but the addition of interprovincial mobility was not necessary to obtain an accurate description of the 2020-2021 \sars{} pandemic in Belgium. We further demonstrate how our model can be used to help policymakers decide on the optimal timing of the release of social restrictions.We find that adding spatial heterogeneity by geographically stratifying the model results in more uncertain model projections as compared to an equivalent nation-level model, which has both communicative advantages and disadvantages. We finally discuss the impact of imposing local mobility or social contact restrictions to contain an epidemic in a given province and find that lowering social contact is a more effective strategy than lowering mobility.\\


\noindent\textbf{Word count} 6640 (main text), excluding captions of figures, tables and titles.\\


\end{small}

\section{Introduction}\label{sec:introduction}

\begin{figure}[b]
    \centering
    \includegraphics[width=0.95\textwidth]{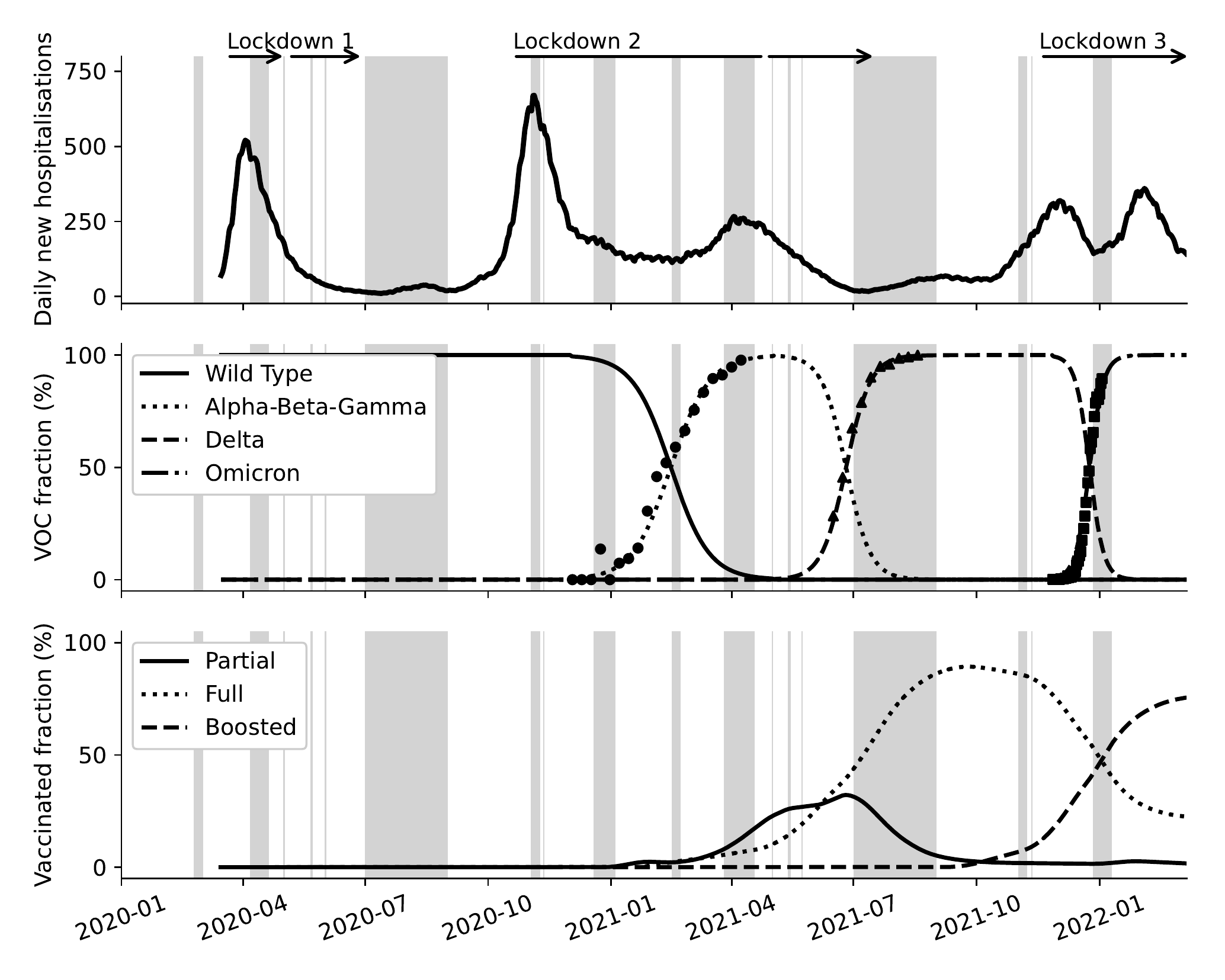}
    \caption{(top) Seven-day moving average of daily new \covid{} hospitalisations in Belgium during 2020 and 2021. The horizontal arrows denote the periods with severe social restrictions (\textit{lockdown}), along with their subsequent release. (middle) Fraction of laboratory confirmed cases caused by a given variant of concern (VOC). (bottom) Fraction of the adult population with partial, full or boosted vaccine immunity. From 2021 onward, dynamic transmission models had to be extended to accommodate the effects of VOCs and vaccines. A grey background is used to indicate a holiday period.}
    \label{fig:overview-hosp-VOC-vacc}
\end{figure}

Coronavirus Disease 2019 (\covid{}) is a respiratory disease caused and spread by Severe Acute Respiratory Syndrome Coronavirus 2 (\sars{}). The virus most likely originated in Wuhan, China in December 2019 \citep{Zhu2020} and has since spread globally. The pandemic continues up to the moment of writing and is characterised by sequential waves of COVID-19 cases and hospitalisations, warranting a series of preventive governmental policies (top panel, Figure \ref{fig:overview-hosp-VOC-vacc}). An more detailed overview of key events that influenced \sars{} prevalence in Belgium during the 2020-2021 \covid{} pandemic can be found in Figure \ref{fig:timeline_2020-2021}.\\

\noindent To better understand the spread of \sars{} in Belgium and inform policymakers during the 2020-2021 \covid{} pandemic, a deterministic, nation-level compartmental model stratified in nine age groups was developed \citep{Alleman2021}. However, by the end of 2020, more transmissible variants of concern (VOCs) emerged. In addition, a large-scale vaccination campaign against \sars{} was conducted. Thus, to accommodate policymakers requests for projections, the existing models had to be extended to accommodate the quickly expanding knowledge on \sars{}. In addition to VOCs and vaccines, we obtained detailed information on mobility patterns in Belgium during the 2020-2021 \covid{} pandemic, which we use to construct a spatially-explicit variant of our previously established model. Further, the transmissibility of \sars{} might be influenced by seasonal factors and the model can benefit from the use of a stochastic solution algorithm. This work thus focusses on the extension of the existing dynamic transmission model to incorporate these aspects, as they have proven useful during the 2020-2021 \covid{} pandemic.\\

\noindent The importance of spatial heterogeneity and mobility in Belgium is demonstrated first by the fact that viral spread was not uniform, but rather linked to the location of the initial clusters \citep{rollier2023}. Similar conclusions were drawn for France, Italy, and Spain as well \citep{Iacus2020a}. Second, a weak correlation was found between the mobility on the one hand, and the morphology and timing of local \covid{}-related time series on the other hand \citep{rollier2023}. Third, a national model cannot take into account local differences in immunity, both natural and due to vaccination, possibly leading to local herd immunity \citep{Barker2021, Aschwanden2021}. Fourth, the national model assumes population mixing is spatially homogeneous. This simplifying assumption may result in overestimating future hospitalisation incidence, which may compromise credibility when modelling for policymakers. These four reasons suggest that a Belgian epidemiological model may benefit from the introduction of spatial heterogeneity and mobility, as was successfully done for e.g.~Spain \citep{Arenas2020}, Brazil \citep{Costa2020} and France \citep{Roques2020}. Further, the inclusion of mobility and spatial heterogeneity into the metapopulation models allows scientists to advise policy-makers on the effect of localised measures, by predicting on a local level which areas face imminent danger, as well as which areas play a pivotal role in controlling the spread of the virus \citep{alleman_reportv1p1, alleman_reportv1p2}.\\

\noindent The importance of including VOCs and vaccines in the model is motivated by their influence on \sars{} dynamics. Subsequent VOCs are associated with different transmissibilities and hospitalisation propensities \cite{Grint2021, Bager2021, VENETI2022}, and speculation on increase in severity is an important factor in policy advice \cite{RESTORE7}. Vaccination has the explicit goal of reducing viral transmission and/or disease severity and has been shown to do this in both clinical trials \cite{doi:10.1056/NEJMoa2034577} and society-scale follow-up studies \cite{Tartof2021}. However, the protection offered by vaccination is imperfect (``leaky'') and was shown to decrease over time, from hereon referred to as \textit{waning} \cite{Braeye2022a}. Furthermore, the protection against severe \covid{} is more long-lasting than protection against \sars{} transmission \citep{Tartof2021}. In addition, vaccine efficacies differ between VOCs \cite{Braeye2022a}. The direct or indirect effect of seasonal changes on the \sars{} transmission rate is not supported by the same overwhelming amount of data due to the limited time since the start of the pandemic. However, seasonality plays a crucial role in many viral diseases \cite{martinez2018}, and has been required in recent \covid{} modeling efforts \cite{Liu2021season}.\\

\noindent As explained in more detail in the next section, the Belgian population of 11 million individuals is subdivided into 13 disease compartments, each stratified in 11 provinces, 10 age groups, and 4 vaccination states. The large number of compartments results in relatively small populations where stochastic effects may become noticeable \citep{daley_gani_1999}. Our previously developed nation-level \sars{} model \citep{Alleman2021} was simulated deterministically because it was stratified in only nine age groups and thus stochastic effects were expected to average out. To simulate the model stochastically, an approximation to the exact Stochastic Simulation Method \citep{Gillespie1977}, the so-called Tau-leaping method \citep{Gillespie2001}, is used. Instead of simulating each transition between the compartments explicitly, which, for large populations, results in a very small timestep, the Tau-leaping method fixes the length of the timestep at $\tau$ and approximates the exact number of transitions in the interval $[t,t+\tau]$ by a binomial process \citep{Higham2021}.\\

\noindent In this work, we first describe the model's equations, its extensions, and parameters. We then calibrate the model and demonstrate its adequacy in describing past \covid{}-related time series. Second, we examine the effect of disabling the interprovincial mobility in the model on the description of the past \covid{}-related time series. Third, we demonstrate how the model can be used to set up hypothetical future scenarios to inform policymakers about the effects of releasing social policies. Fourth, we compare the difference when the aforementioned scenarios for policymakers are simulated with the national versus the spatially explicit model. Finally, in a purely hypothetical setup, we study if \sars{} outbreaks can be contained within a province, or if a province can be shielded from a \sars{} outbreak in another province, by imposing local mobility restrictions or social restrictions.

\section{Methods}\label{sec:model}

\subsection{Disease compartments and stratifications}
\label{subsec:age-stratified-model}

\begin{figure}[h]
\centering
\includegraphics[width=\linewidth]{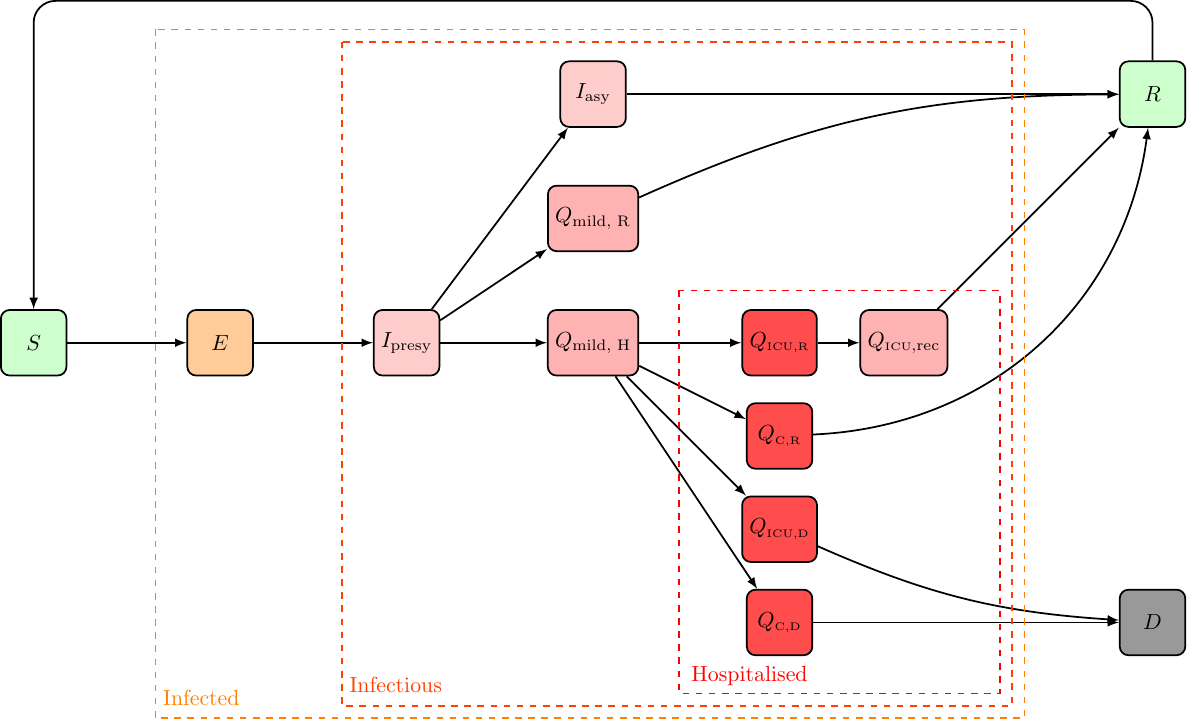}
\caption{Flowchart of the SEIQRD model. Here, $S$ stands for susceptible, and $E$ for exposed but not yet infectious. Infected individuals in the $I$ compartments are those that are considered to actively drive the pandemic because they are either presymptomatic ($I_\text{presy}$), or asymptomatic ($I_\text{asy}$). Individuals in the $Q$ compartments are assumed not to be infectious, supposedly due to heightened symptom awareness. Some have mild symptoms that don't become worse ($Q_\text{mild, R}$), while others will have mild symptoms that do, up to the point of seeking medical attention at a hospital ($Q_\text{mild, H}$). In the hospital, some are hospitalised in a regular cohort ward ($Q_\text{C,R}$ and $Q_\text{C,D}$), while others are hospitalised in the ICU ($Q_\text{ICU,R}$ and $Q_\text{ICU,D}$). After a stay in an ICU, patients remain in a recovery stay in cohort ($Q_\text{ICU,rec}$). After infection, individuals are either deceased ($D$) or recovered ($R$). Recovered individuals may again become susceptible. Every disease state is stratified into 10 age classes, 11 provinces and 4 vaccination states.}
\label{fig:flowchart_SEIQRD}
\end{figure}

A compartmental disease model assumes that a population is well mixed and distributed over a number of compartments, each of them  corresponding to a stage in the disease development. The model presented here has similar disease compartments as the one we developed previously at nation level for Belgium \citep{Alleman2021}. A flowchart depicting the various compartments in our \covid{} model is shown in Fig. \ref{fig:flowchart_SEIQRD}. The infectious compartment (I) in the original SEIRD formulation \citep{Kermack1927} is subdivided into nine compartments based on expert knowledge on \sars{}. In this way, the model accounts for pre-symptomatic and asymptomatic transmission of \sars{} \citep{Ganyani2020,Wei2020,Gudbjartsson2020}, and for different \covid{} severities, ranging from mild disease to hospitalisation. Our model distinguishes between regular hospital wards (cohort) and intensive care units (ICUs) and further accounts for a recovery stay in cohort after an ICU stay resulting in a total of 13 disease compartments. Using data from \num{22 136} \covid{} patients admitted to Belgian hospitals, we previously obtained the probability of needing intensive care, the mortalities in both hospital wards, and the residence time distributions in both hospital wards \cite{Alleman2021}. Waning of antibodies (seroreversion) is included, enabling re-susceptibility to \sars{} after a prior infection.

\noindent Every disease compartment is first stratified into ten age groups: 0-12, 12-18, 18-25, 25-35, 35-45, 45-55, 55-65, 65-75, 75-85 and 85-120 years of age, to account for the fact that social contact and disease severity differs substantially between individuals of different ages \cite{Alleman2021}. The model is then stratified further in eleven spatial patches representing the 10 Belgian provinces and the Brussels capital region, and  four possible vaccination stages: unvaccinated, partially vaccinated, fully vaccinated or boosted. Thus, in total, the model presented here contains $13 \times 10 \times 11 \times 4 = 5720$ states (or metapopulations).

\subsection{Governing equations and stochastic simulation}\label{section:governing_equations}

\noindent\textbf{Governing equations} The dynamics shown in Fig.~\ref{fig:flowchart_SEIQRD} are simulated by iteratively working through two phases: first vaccination and then disease progression. First, at each subsequent time step $t + \tau$, individuals are transferred between the vaccination metapopulations based on vaccine incidence data. Individuals in the susceptible ($S$) and recovered ($R$) disease states are considered eligible for vaccination in the model, and their respective fractions on day $t$ are computed as follows,
\begin{align*}
f^{g}_{S,v,i}(t) &= \frac{S^{g}_{v,i}(t)}{S^{g}_{v,i}(t) + R^{g}_{v,i}(t)}, \text{ \ and \ } \\
f^{g}_{R,v,i}(t) &= 1 - f^{g}_{S,v,i}(t),
\end{align*}
where $g$ refers to the spatial stratification, $i$ to the age stratification, and $v$ to the vaccination stage. We assume that vaccines administered between times $t$ and $t + \tau$ are distributed among the susceptible ($S$) and recovered ($R$) disease states according to their respective fractions. The number of individuals in province $g$ and age group $i$ entering the new vaccination stage $v'$ from vaccination stage $v$ between day $t$ and $t+\tau$, is used to compute the vaccine-modified populations of susceptible ($S$) and recovered ($R$) individuals,
\begin{align*}
\widetilde{S}^{g}_{i,v'}(t) &= {S}^{g}_{i,v'}(t) + f^{g}_{S,v,i}(t)[\phi^g_{i,v'}(t+\tau) - \phi^g_{i,v'}(t)], \\
\widetilde{S}^{g}_{i,v}(t) &= {S}^{g}_{i,v}(t) - f^{g}_{S,v,i}(t)[\phi^g_{i,v'}(t+\tau) - \phi^g_{i,v'}(t)], \text{ \ and \ } \\
\widetilde{R}^{g}_{i,v'}(t) &= {R}^{g}_{i,v'}(t) + f^{g}_{R,v,i}(t) [\phi^g_{i,v'}(t+\tau) - \phi^g_{i,v'}(t)], \\
\widetilde{R}^{g}_{i,v}(t) &= {R}^{g}_{i,v}(t) - f^{g}_{R,v,i}(t) [\phi^g_{i,v'}(t+\tau) - \phi^g_{i,v'}(t)].
\end{align*}
Here $\phi^g_{v',i}(t)$ is the cumulative number of vaccines of type $v'$ administered in province $g$ and age class $i$ up to times $t$. Second, we model the progress of the \covid{} disease proper, using the Tau-leaping method \citep{Gillespie2001}. This algorithm advances the system discretely in time by drawing the number of transitions between the model's states on the interval $[t, t+\tau]$ from a binomial distribution. To compute the probabilities of the transitions, the disease dynamics presented in Fig.~\ref{fig:flowchart_SEIQRD} are translated into the rates associated with the 17 possible transitions of the system (per unit time),
\begin{align*}
            \mathcal{R}(\widetilde{S} \rightarrow E)^{g}_{i,v} &= \sum\limits_{h=1}^G \Bigg( P^{gh} (t) \sum\limits_{w} \sum\limits_{j=1}^{N} \bar{\beta}^{gh}_{ij,vw}(t) \\
            & \qquad \times \bar{N}_{ij}^{gh} (t) 
            \dfrac{(I_\text{presy})_{j,w,\text{eff}}^h (t) + (I_\text{asy})_{j,w,\text{eff}}^h (t)}{T_{j,w,\text{eff}}^g (t)} \Bigg), \\
            \mathcal{R}(E \rightarrow I_{\text{presy}})^{g}_{i,v} &= \sigma^{\text{-}1} (t),\\
            \mathcal{R}(I_{\text{presy}} \rightarrow I_{\text{asy}})^{g}_{i,v} &= a_i \omega ^{\text{-}1}, \\
            \mathcal{R}(I_{\text{presy}} \rightarrow Q_{\text{mild,R}})^{g}_{i,v} &= (1-a_i) (1-h_i(t)) \omega ^{\text{-}1},\\
            \mathcal{R}(I_{\text{presy}} \rightarrow Q_{\text{mild,H}})^{g}_{i,v} &= (1-a_i) h_i(t) \omega ^{\text{-}1},\\
            \mathcal{R}(I_{\text{asy}} \rightarrow R)^{g}_{i,v} &= d_a^{\text{-}1},\\
            \mathcal{R}(Q_{\text{mild,R}} \rightarrow R)^{g}_{i,v} &= d_m^{\text{-}1},\\
            \mathcal{R}(Q_{\text{mild,H}} \rightarrow Q_{\text{C,R}})^{g}_{i,v} &= c_i (1-m_{\text{C,i}}) d_{\text{hospital}}^{\text{-}1}, \\
            \mathcal{R}(Q_{\text{mild,H}} \rightarrow Q_{\text{C,D}})^{g}_{i,v} &= c_i m_{\text{C,i}} d_{\text{hospital}}^{\text{-}1}, \\
            \mathcal{R}(Q_{\text{mild,H}} \rightarrow Q_{\text{ICU,R}})^{g}_{i,v} &= (1-c_i) (1-m_{\text{ICU,R}}) d_{\text{hospital}}^{\text{-}1}, \\
            \mathcal{R}(Q_{\text{mild,H}} \rightarrow Q_{\text{ICU,D}})^{g}_{i,v} &= (1-c_i) m_{\text{ICU,R}} d_{\text{hospital}}^{\text{-}1}, \\
            \mathcal{R}(Q_{\text{C,R}} \rightarrow R)^{g}_{i,v} &= d_{C,R,i}^{\text{-}1},\\
            \mathcal{R}(Q_{\text{C,D}} \rightarrow D)^{g}_{i,v} &= d_{C,D,i}^{\text{-}1},\\
            \mathcal{R}(Q_{\text{ICU,R}} \rightarrow Q_{\text{ICU,rec}} )^{g}_{i,v} &= d_{\text{ICU},R,i}^{\text{-}1},\\
            \mathcal{R}(Q_{\text{ICU,rec}} \rightarrow R)^{g}_{i,v} &= d_{\text{ICU,rec}}^{\text{-}1},\\
            \mathcal{R}(Q_{\text{ICU,D}} \rightarrow D)^{g}_{i,v} &= d_{\text{ICU},D,i}^{\text{-}1}, \\
            \mathcal{R}(\widetilde{R} \rightarrow S)^{g}_{i,v} &= \zeta, \\
    \label{eq:all_ODEs_spatial}
\end{align*}
where subscript $g$ is used to index the $G=11$ provinces, subscript $i$ to index the $N=10$ age groups and subscript $v$ to index the four vaccination states. The meaning and value of all model parameters are listed in Table \ref{tab:SEIQRD_params}. The introduction of interprovincial mobility, dynamic social policies, variants of concern, seasonal changes in the transmissibility of \sars{}, and vaccination make the interprovincial mobility matrix  $\bm{P}(t)$, the social contact matrix $\bm{\bar{N}}(t)$, the fraction of mildly symptomatic patients requiring hospitalisation $\bm{h}(t)$, the average duration of the latent period $\sigma(t)$, and, the mobility-weighted (\textit{effective}) populations of presymptomatic infectious $\bm{I}_\text{presy,eff}(t)$, asymptomatic infectious $\bm{I}_\text{asy,eff}(t)$, and the total population $\bm{T}_\text{eff}(t)$, time-dependent. An overview of the additional parameters associated with these extensions is listed in Table \ref{tab:SEIQRD_timedep_params}. The addition of interprovincial mobility is discussed in more detail in Section \ref{subsec:spatial-extension} while the addition of VOCs, seasonality, and vaccines are discussed in Section \ref{subsec:voc_and_vac}.\\

\noindent{\textbf{Stochastic simulation} Assuming the aforementioned transition rates from a generic state $X$ to a state $Y$ in province $g$,  age class $i$ and vaccination state $v$, denoted $\mathcal{R}(X \rightarrow Y)^g_{i,v}$, are constant over the interval $[t, t+\tau]$, the probability of a transition from a generic state $X$ to $Y$ happening in the interval $[t, t+\tau]$ is exponentially distributed, mathematically,
$$
\mathcal{P}(X \rightarrow Y)^g_{i,v} = 1 - e^{- \tau \mathcal{R}(X \rightarrow Y)^g_{i,v}}.
$$
The corresponding number of transitions $X \rightarrow Y$ in province $g$, age class $i$ and vaccination state $v$ between time $t$ and $t+\tau$ can then be drawn from a binomial distribution,
$$
\mathcal{N}(X \rightarrow Y)^{g}_{i,v} = \text{Binom}(\mathcal{P}(X \rightarrow Y)^{g}_{i,v}, A_{i,v}^g)\,.
$$
Finally, the number of individuals in each of the compartments at time $t+\tau$ is then updated as follows,
{
\begin{align*}
S^{g}_{i,v}(t + \tau) &= \widetilde{S}^{g}_{i,v}(t) - \mathcal{N}(\widetilde{S} \rightarrow E)^{g}_{i,v} + \mathcal{N}(\widetilde{R} \rightarrow S)^{g}_{i,v},\\
E^{g}_{i,v}(t + \tau) &= \mathcal{N}(\widetilde{S} \rightarrow E)^{g}_{i,v} - \mathcal{N}(E \rightarrow I_{\text{presy}})^{g}_{i,v},\\
\vdots \\
R^{g}_{i,v}(t + \tau) &= \widetilde{R}^{g}_{i,v}(t) + \mathcal{N}(I_{\text{asy}} \rightarrow R)^{g}_{i,v} + \mathcal{N}(Q_{\text{mild,R}} \rightarrow R)^{g}_{i,v} \\
& \qquad + \mathcal{N}(Q_{\text{C,R}} \rightarrow R)^{g}_{i,v} + \mathcal{N}(Q_{\text{ICU,rec}} \rightarrow R)^{g}_{i,v} - \mathcal{N}(\widetilde{R} \rightarrow S)^{g}_{i,v}.
\end{align*}
}

\noindent The leap value was determined by balancing the accuracy of the obtained results with the need for computational resources. As a reference for accuracy, model trajectories obtained using the (exact) stochastic simulation algorithm (SSA) \citep{Gillespie1977} were used. A leap value of half a day ($\tau = 0.5~d$) was chosen. Our models are simulated and calibrated using our in-house code pySODM \citep{alleman2023}, a Python 3 software for the simulation and calibration of dynamical systems with n-dimensional, labeled states. The code is freely available on GitHub and fully documented.

\begin{landscape}
\thispagestyle{empty}
\begin{table}
    \centering
    \caption{\small{Parameters used for calculating the dynamics between the various SEIQRD compartments shown in Fig. \ref{fig:flowchart_SEIQRD}. For parameters with time dependencies, the baseline value, prior to the introduction of time dependency, is given.}}
    \begin{tabular}{p{1.5cm}p{9.5cm}lp{3.2cm}}
        \textbf{Symbol} & \textbf{Parameter} & \textbf{Migration} & \textbf{Value} \\ \toprule
        $\bm{a}$ & Fraction of infected individuals remaining asymptomatic & $I_\text{presy} \rightarrow I_\text{asy}$ & Table \ref{tab:ageDistributionAsymptomatic}, \citep{Poletti2021} \\
        $\bm{h}(t)$ & Fraction of mildly symptomatic individuals requiring hospitalisation. Time-dependent due to VOCs and vaccination. & $I_\text{presy} \rightarrow Q_\text{mild, H} \text{ or } Q_\text{mild, R}$ & Table \ref{tab:ageDistributionAsymptomatic}, \textit{inferred}\\
        $\bm{c}$ & Fraction of hospitalisations admitted in regular cohort hospital ward & $Q_\text{mild, H} \rightarrow Q_\text{hosp}$ or $Q_\text{ICU}$ & Table \ref{tab:results_hospital_age}, \citep{Alleman2021}\\
        $\bm{m}_C$ & Mortality of patients in a cohort hospital ward & $Q_\text{hosp} \rightarrow D$ & Table \ref{tab:results_hospital_age}, \citep{Alleman2021} \\
        $\bm{m}_\text{ICU}$ & Mortality of patients in an IC unit & $Q_\text{ICU} \rightarrow D$ & Table \ref{tab:results_hospital_age}, \citep{Alleman2021} \\ \midrule
        $\bm{d}_{C,R}$ & Length-of-stay in hospital cohort ward (outcome: recovered) & $Q_\text{hosp} \rightarrow R$ & Table \ref{tab:results_hospital_days}, \citep{Alleman2021} \\
        $\bm{d}_{C,D}$ & Length-of-stay in hospital cohort ward (outcome: deceased) & $Q_\text{hosp} \rightarrow D$ & Table \ref{tab:results_hospital_days}, \citep{Alleman2021} \\
        $\bm{d}_{\text{ICU},R}$ & Length-of-stay in an IC unit (outcome: recovered) & $Q_\text{ICU} \rightarrow Q_{\text{ICU, rec}}$ & Table \ref{tab:results_hospital_days}, \citep{Alleman2021} \\
        $\bm{d}_{\text{ICU},D}$ & Length-of-stay in an IC unit (outcome: deceased) & $Q_\text{ICU} \rightarrow D$ & Table \ref{tab:results_hospital_days}, \citep{Alleman2021} \\
        $\bm{d}_{\text{ICU},\text{rec}}$ & Average recovery stay in a cohort ward after ICU & $Q_{\text{ICU,rec}} \rightarrow R$ & Table \ref{tab:results_hospital_days}, \citep{Alleman2021} \\
        \midrule
        $\sigma(t)$ & Average duration of latent period. Time-dependent due to VOCs. & $E \rightarrow I_{\text{presy}}$ & Table \ref{tab:VOC-dependent-variables} \\
        $d_a$ & Average duration of asymptomatic infection & $I_\text{asy} \rightarrow R$ & 5.0 d, \textit{assumed}\\
        $d_m$ & Average duration of mild infection before recovery & $Q_\text{mild, R} \rightarrow R$ & 5.0 d, \textit{assumed} \\ $d_\text{hosp}$ & Average duration between symptom onset and hospitalisation & $Q_\text{mild, H} \rightarrow Q_\text{C} \text{ or } Q_\text{ICU}$ & 6.4 d, \citep{Alleman2021} \\
        $\omega$ & Average duration of presymptomatic infectious period & $I_\text{presy} \rightarrow I_\text{asy} \text{ or } Q_\text{mild, R}, Q_\text{mild, H}$ & 0.7 d, \citep{Wei2020, He2020} \\
        $\zeta$ & Average seroreversion rate & $R \rightarrow S$ & $\ln (2)/365\ \text{d}^{\text{-}1}$, \textit{assumed} \\ \midrule
        $\bar{N}_{ij}^{g}(t)$ & Social contact matrix. Number of social contacts made by age group $i$ with individuals of age group $j$ in province $g$ on day $t$. Time-dependent due to dynamic rescaling with GCMR indicators. & $S \rightarrow E$ & Retrieved using SOCRATES by Willem et al. \citep{Willem2020a}. Section \ref{subsec:spatial-extension}. \\ \midrule
        $\bar{\beta}^{gh}_{ij,vw}(t)$ & Transmission coefficient. Probability of infection when an infectious individual in age group $i$, province $g$ and vaccination stage $v$ contacts an individual of age group $j$, province $h$ and vaccination stage $w$ on day $t$. Time-dependent due to seasonality, VOC prevalence, and vaccination. & $S \rightarrow E$ & $\beta = 0.027$ (corresponds to $R_0 = 3.3$, \citep{Alimohamadi2020}). Section \ref{subsec:spatial-extension}. \\ \midrule
        $P^{gh}(t)$ & Interprovincial mobility matrix. Estimated fraction of all the time available to all individuals in
province $g$, spent in province $h$, in the day corresponding to time $t$. & $S \rightarrow E$ & Confidential telecommunication data, \textit{Proximus}. Section \ref{subsec:spatial-extension}.\\
        \bottomrule
    \end{tabular}
    \label{tab:SEIQRD_params}
\end{table}
\end{landscape}

\begin{landscape}
\thispagestyle{empty}
\begin{table}
    \centering
    \caption{\small{Additional parameters of model extensions making model parameters $\bar{\bm{\beta}}(t)$, $\bar{\bm{N}}(t)$, $\bm{h}(t)$ time-dependent.}}
    \begin{tabular}{p{1.5cm}p{11.5cm}lp{4.5cm}}
        \textbf{Symbol} & \textbf{Parameter} & \textbf{Alters} & \textbf{Value} \\ \toprule
        \multicolumn{4}{l}{Social contact, Section \ref{subsec:spatial-extension}}\\ \midrule
        $\Omega$ & Relative effectivity of social contacts in workplaces, schools and during leisure activities for the transmission of \sars{} as compared to contacts at home. & $\bm{\bar{N}}(t)$ & 0.40, \textit{inferred}\\
        $\Psi$ & Intervention parameter. Additional reduction of social contacts under lockdown on top of reductions indicated by the GCMRs. Arguably due to mentality changes induced by government measures. & $\bm{\bar{N}}(t)$ & 0.65, \textit{inferred} \\ 
        $P^{gh}(t)$ & Interprovincial mobility matrix. Estimated fraction of all the time available to all individuals in province $g$, spent in province $h$, in the day corresponding to time $t$. & $\bm{\bar{N}}(t)$ & Confidential telecommunication data, \textit{Proximus}.\\
        \midrule
        \multicolumn{4}{l}{Variants of concern, Section \ref{subsec:voc_and_vac}.} \\ \midrule
        $\bm{K}_{\text{inf}}$ & Infectivity gain of VOCs (Alpha-Beta-Gamma, Delta) as compared to wild type & $\bar{\bm{\beta}}(t)$ & $[40~\%,100\%]$, \textit{inferred} \\
        $\bm{K}_{\text{hosp}}$ & Hospitalisation propensity gain due to VOCs as compared to wild type & $\bm{h}(t)$ & Table \ref{tab:VOC-dependent-variables}\\ \midrule
        \multicolumn{4}{l}{Seasonality, Section \ref{subsec:voc_and_vac}.} \\ \midrule
        $A$ & Amplitude of seasonality in the viral transmissibility of \sars{}  & $\bar{\bm{\beta}}(t)$ & 18~\%, \textit{inferred}  \\ \midrule
        \multicolumn{4}{l}{Vaccination, Section \ref{subsec:voc_and_vac}.} \\ \midrule
        $\bm{E}_{\text{v, susc}}(t)$ & Efficacy of vaccine dose $v$ in lowering susceptibility to \sars{}. Elements for every age group $i$ and province $g$. Computed using vaccine incidence data and reported vaccine efficacies to obtain the average efficacy subject to waning. & $\bar{\bm{\beta}}(t)$ & Dynamically scaled, see section \ref{app_section:vaccination}\\
        $\bm{E}_{\text{v, inf}}(t)$ & Efficacy of vaccine dose $v$ in lowering the infectiousness when infected with \sars{}. Elements for every age group $i$ and province $g$. Computed using vaccine incidence data and reported vaccine efficacies to obtain the average efficacy subject to waning. & $\bar{\bm{\beta}}(t)$ & Dynamically scaled, see section \ref{app_section:vaccination}\\
        $\bm{E}_{\text{v, hosp}}(t)$ & Efficacy of vaccine dose $v$ in lowering the hospitalisation propensity. Elements for every age group $i$ and province $g$. Computed using vaccine incidence data and reported vaccine efficacies to obtain the average efficacy subject to waning. & $\bm{h}(t)$ & Dynamically scaled, see section \ref{app_section:vaccination} \\
        \bottomrule
    \end{tabular}
    \label{tab:SEIQRD_timedep_params}
\end{table}
\end{landscape}

\subsection{Spatially explicit model extension}
\label{subsec:spatial-extension}

\noindent\textbf{Interprovincial mobility} Belgium is stratified in a collection of 11 geographical units: 10 provinces and the arrondissement Brussels-capital (NUTS2 level, Fig. \ref{fig:beta_classes_prov}, Table \ref{tab:class-NIS-name}). We will refer to the latter as the `11th province' for convenience. Key to the quantification of interprovincial connectivity is the telecommunication dataset provided by Belgium's largest telecom operator Proximus (Appendix \ref{app:proximus-mobility-data}). The use of this type of data as a proxy for mobility has been shown to be legitimate \citep{Palchykov2014}, and has been done in the particular context of \covid{} in other analyses and modelling efforts \citep{agren2021, santamaria2020, kishore2020}. The geographical spread of individuals between $G$ provinces is quantified in a $G\times G$ time-dependent mobility matrix $\bm{P}(t)$ with elements $P^{gh}(t)$. Element $P^{gh}(t)$ represents the estimated fraction of all the time available to all individuals in province $g$, spent in province $h$, in the day corresponding to time $t$. Fig. \ref{fig:diagram-spatial-model} schematically depicts the mobility between three provinces. As an example, two time series of $P^{gh}(t)$ for two different $(g,h)$ pairs are shown in Fig. \ref{fig:staytime_percentage_timeseries}. \\

\begin{figure}[h!]
    \centering
    \includegraphics[width=0.4\linewidth]{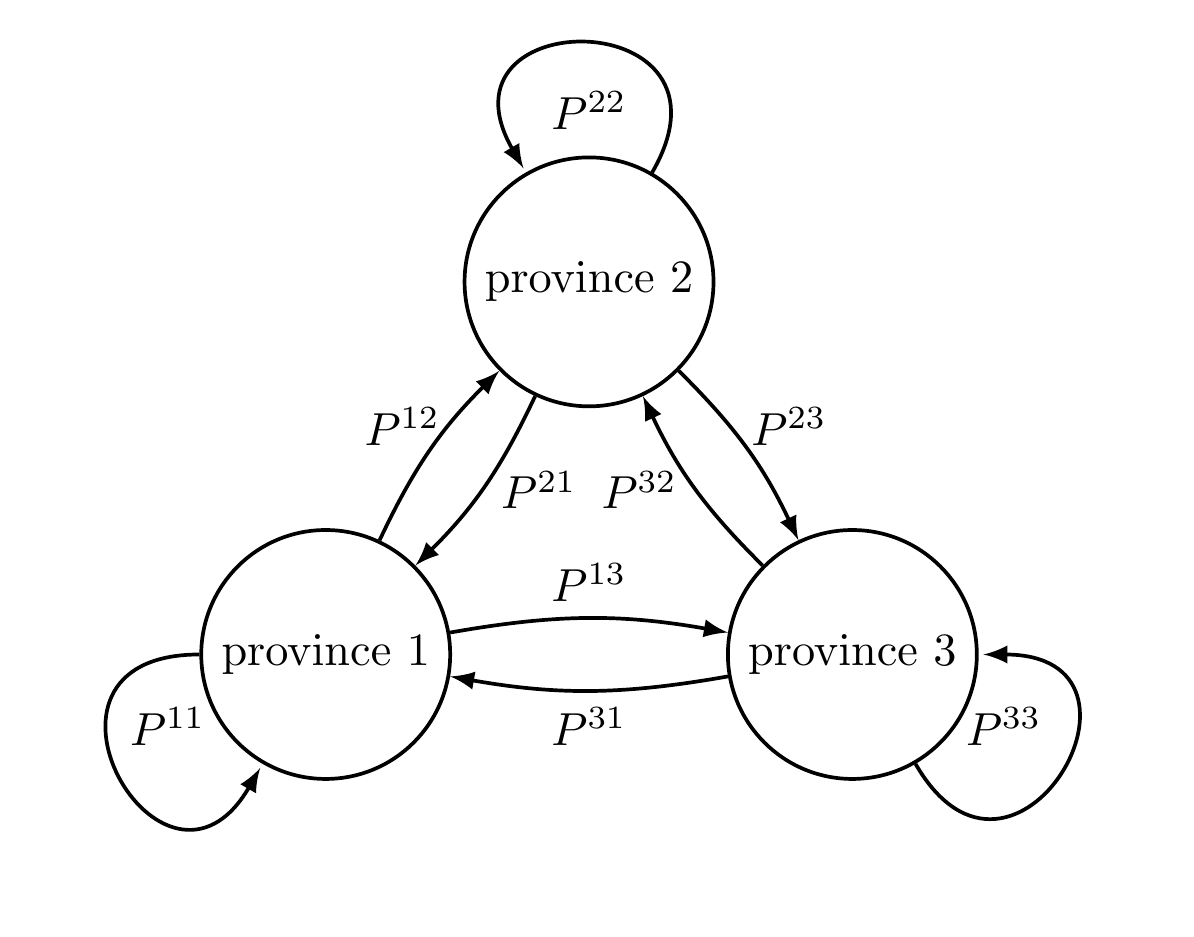}
    \caption{\small{Schematic representation of the interpretation of the inter-provincial mobility matrix $\bm{P}$ for only three provinces. In the model, we consider 11 Belgian provinces and the mobility matrix elements are updated daily (Fig. \ref{fig:staytime_percentage_timeseries}).}}
    \label{fig:diagram-spatial-model}
\end{figure}

\begin{figure}[h]
    \centering
    \includegraphics[width=1.02\linewidth]{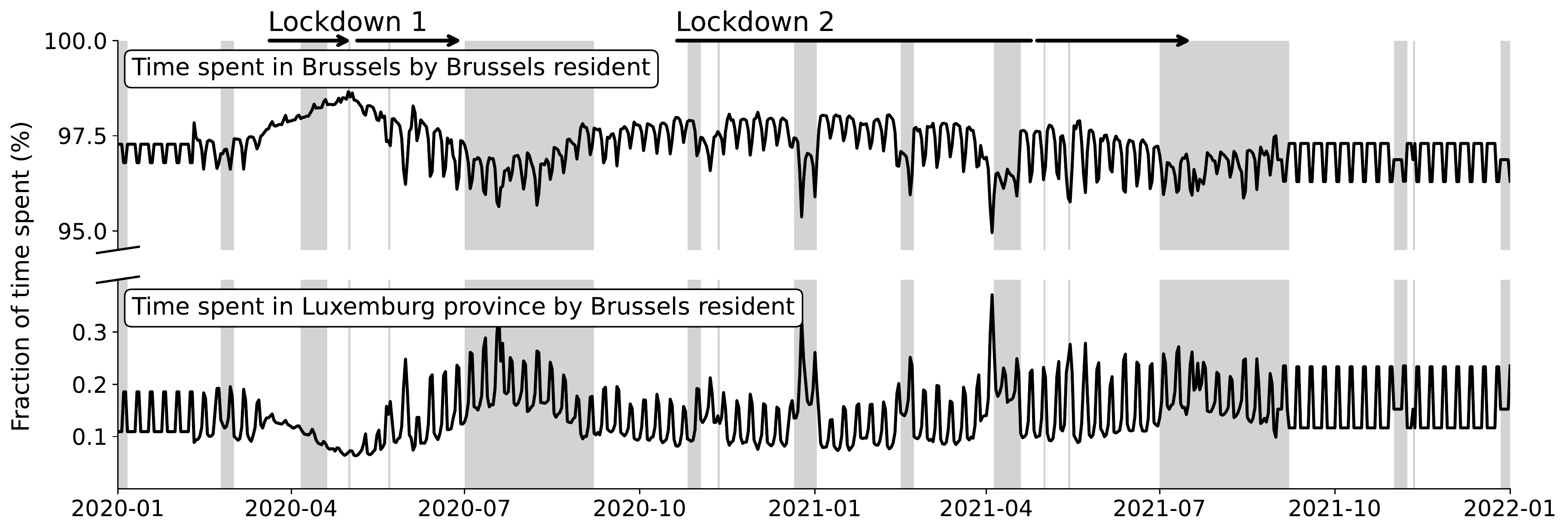}
    \caption{Two of the $11^2$ time series $P^{gh}(t)$, here representing the daily percentage of time that all residents of Brussels spent in their home province (top), or in Luxembourg province (bottom). The hatched regions indicate periods where an estimate was used because no data was available. A grey background is used to indicate a holiday period.}
    \label{fig:staytime_percentage_timeseries}
\end{figure}

\noindent The mobility matrix $\bm{P}(t)$ is used to determine the number of susceptible people from province $g$ that visit province $h$ at time $t$, and the \textit{effective} population sizes per model compartment in the visited province $h$ \cite{Arenas2020}. Mathematically, the effective population size is computed as follows:
\begin{equation}\label{eq:effective_population}
    X^g_{i,v,\text{eff}}(t) = \sum_{h=1}^G P^{hg}(t)X^h_{i,v}(t),
\end{equation}
where $X$ is a generic disease state. Daily effective populations are calculated for states $\bm{I}_\text{presy}$ and $\bm{I}_\text{asy}$, and for the total population $\bm{T}$.\\

\noindent\textbf{Provincial Social contact model} Social behavior is an important driver of \sars{} spread and varies throughout the pandemic \citep{Coletti2020, Gimma2022, google_mobility, Alleman2021}. The behavioral changes are likely based on an individual's risk perception \citep{Coletti2020}, (social) media consumption \citep{Liu2021social}, and vaccination state \citep{Bauch2013}. In our model, social behavior during the pandemic is translated into a number of social contacts. These contacts depend on the geographical location in two ways. First, prepandemic contact matrices are scaled with activity indicators at the Belgian provincial level to construct a pandemic contact matrix $\widetilde{\bm{N}}^g(t)$ for every province $g$ at time $t$. The linear combination of prepandemic interaction matrices used to model pandemic social contact is,
\begin{multline}
    \widetilde{\bm{N}}^g(t) = \alpha^g(t) \bm{N}^\text{home} + \beta^g(t) \bm{N}^\text{schools} + \gamma^g(t)\bm{N}^\text{work}  \\
    + \delta^g(t)\bm{N}^\text{transport} + \epsilon^g(t)\bm{N}^\text{leisure} + \zeta(t)^g\bm{N}^\text{other},
\end{multline}
where,
\begin{equation}
    \left\{
        \begin{array}{rl}
            \alpha^g(t) &= \Psi^g(t),\\
            \beta^g(t) &= \Psi^g(t)\ \Omega\ G^{g,\ \text{schools}}(t),\\
            \gamma^g(t) &= \Psi^g(t)\ \Omega\ G^{g,\ \text{work}}(t),\\
            \delta^g(t) &= \Psi^g(t)\ \Omega\ G^{g,\ \text{transport}}(t),\\
            \epsilon^g(t) &= \Psi^g(t)\ \Omega\ G^{g,\ \text{leisure}}(t),\\
            \zeta(t)^g &= \Psi^g(t)\ \Omega\ G^{g,\ \text{other}}(t).
        \end{array}
    \right.
    \label{eq:coefficients_matrices}
\end{equation}\\

\noindent where $\widetilde{\bm{N}}^g(t)$ is the total pandemic social contact matrix in province $g$ at time $t$. $\bm{N}^{\text{home}}$, $\bm{N}^{\text{schools}}$, $\bm{N}^{\text{work}}$, $\bm{N}^{\text{transport}}$, $\bm{N}^{\text{leisure}}$ and $\bm{N}^{\text{other}}$ are the prepandemic social contact matrices retrieved from Willem et al. \citep{Willem2020a}. Our model distinguishes between weekdays and weekends. $G^{g,\ \text{schools}}$, $G^{g,\ \text{work}}$, $G^{g,\ \text{transport}}$, $G^{g,\ \text{leisure}}$, and $G^{g,\ \text{other}}$ represent the mobility reductions during the \covid{} pandemic in province $g$ retrieved from the Google Community Mobility Reports (GCMRs) \citep{google_mobility}. $\Omega$ is the relative effectivity of social contacts in workplaces, schools, and during leisure activities to the spread of \sars{} as compared to social contacts at home ($\Omega^{\text{home}} = 1$). This parameter is introduced because household secondary attack rates are higher than secondary attack rates in other settings \citep{Thompson2021} for \sars{}. Finally, $\Psi^g(t)$ is the phenomenological \textit{intervention} parameter in province $g$ at time $t$. It is introduced to reduce the number of social contacts when lockdown measures are taken beyond what the GCMRs suggest, arguebly due to government induced \textit{mentality} changes. It is gradually introduced during a two-week period using a ramp function when lockdown measures are taken (2020-03-15 and 2020-10-19) and kept in place throughout lockdowns. Once the lockdown measures are released, it is gradually released from the social contact model over a two-month period using a ramp function. Its value, which will be the same in every province $g$, will be obtained during the calibration of the model. An in-depth, bottom-up discussion of the social contact is presented in Appendix \ref{app:social_contact}.\\

\noindent Second, the number of social contacts differs depending on whether or not an individual is in their home province. All individuals are assumed to have their home contacts in their \textit{home} province $g$, whereas all other contacts happen in the \textit{visited} province $h$. Mathematically, the number of contacts of individual residing in province $g$ and visiting province $h$ on day $t$ is,
\begin{equation}
    {\bm{\bar{N}}}^{gh}(t) = \delta^{gh}\bm{N}^{\text{home}} + (1-\delta^{gh}) \bigg[ \bm{\widetilde{N}}^h(t) - \bm{N}^{\text{home}} \bigg],
    \label{eq:time-dep_social-contact}
\end{equation}
where $\delta^{gh}$ is the Kronecker delta, $\bm{N}^{\text{home}}$ is the prepandemic matrix of home contacts, and $\widetilde{\bm{N}}^h(t)$ is the total pandemic contact matrix in the visited province $h$ at time $t$ introduced before.\\

\begin{figure}
    \centering
    \includegraphics[width=\linewidth]{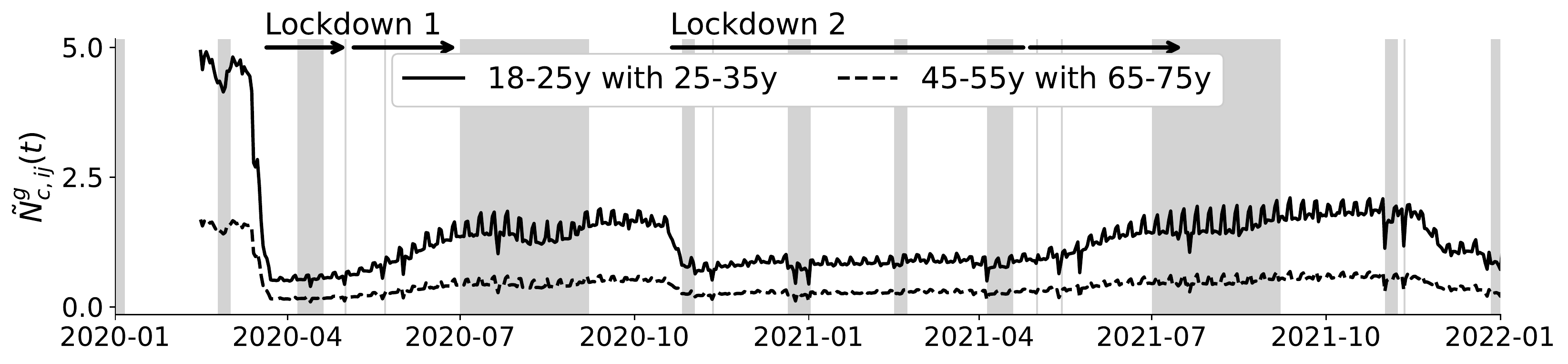}
    \caption{Example of time series $\widetilde{N}_{ij}^g(t)$ for Brussels, which represent the local effective social contact between two age classes $i$ and $j$. The solid curve shows effective contact between 18-25 year-olds and 25-35 year-olds. The dashed curve shows the same information, but 45-55 year-olds contacting 65-75 year-olds, clearly following a similar overall trend but involving fewer contacts. A grey background is used to indicate a holiday period.}
    \label{fig:resulting_Nc_21000_18-25-with-25-35_45-55-with-65-75}
\end{figure}

\noindent\textbf{Population density dependence} Following the example of other work \citep{Arenas2020}, we have previously used different transmission coefficients $\beta$ for Belgium's regions in our model (Table \ref{tab:class-NIS-name}, Figure \ref{fig:beta_classes_prov}). After calibrating these transmission coefficients to hospitalisation data, Brussels seemed to accommodate an increased viral transmission while its neighbouring provinces Vlaams-Brabant and Brabant Wallon seemed to accomodate a much lower viral transmission. This observation is likely due to the way hospitalisation data were collected: Patients were recorded in the province they were hospitalised, not their province of residence. Thus, a patient residing in province $g$ but hospitalised in province $h$ is counted as a data point in province $h$. As metropolitan Brussels is surrounded by the much more sparsely populated Vlaams-Brabant and Brabant Wallon provinces (Fig. \ref{fig:beta_classes_prov}), it is likely that individuals from these provinces have been hospitalised in Brussels. To avoid incorporating imperfections in the data collection process into the model's structure, we calibrate the model to the sum of hospitalisations in Brussels, Vlaams-Brabant and Brabant Wallon and assume the transmission rate does not depend on population density. We choose $\beta = 0.027$, corresponding to a basic reproduction number of 3.3 \citep{Alimohamadi2020}.

\subsection{Dynamical rescaling of model parameters to include VOCs, seasonality, and vaccines}
\label{subsec:voc_and_vac}

\noindent\textbf{Variants of concern} Beyond the wild-type \sars{} variant, we consider four VOCs identified by the World Health Organisation \citep{10.3389/fimmu.2022.825256}: Alpha, Beta, Gamma, and Delta. Due to their similar properties in our model \citep{Braeye2021}, we aggregate the first three VOCs and denote it as the Alpha-Beta-Gamma variant. To model the emergence of these variants, national prevalence data were used \cite{Wenseleers2021} (see Fig. \ref{fig:VOC_prevalence}, top). At every time $t$, a weighted average infectiousness of \sars{} variants was computed using the variant fractions, which effectively turns the (geographically stratified) transmission coefficient into a time-dependent function, i.e.
\begin{equation}
    \beta(t) = \beta \sum_n\alpha_n(t)K_{\text{inf},n}
    \label{eq:beta-from-VOC}
\end{equation}
Here, $\beta$ is the previously introduced probability of infection upon contact with an infectious individual. $\alpha_n(t)$ represents the fraction of variant $n$ present in Belgium at time $t$, and $K_{\text{inf},n}$ is the increase in infectivity compared to the wild type, whose distributions are determined during the calibration procedure. The variants were assumed to alter the serial interval and disease severity as well, which translates to dynamically changing the length of the average latent time ($\sigma$) and the hospitalisation propensity ($\bm{h}$) in a similar fashion (see Fig. \ref{fig:VOC_prevalence}). These parameters have been derived from literature \citep{Grint2021, Bager2021, VENETI2022, Hart2022} and are listed in Table \ref{tab:VOC-dependent-variables}.\\

\noindent\textbf{Seasonality} Changes in climate have been recognised to play a role in the spread of many viral diseases amongst humans, notably influenza \citep{martinez2018}. Seasonality is included in our model by scaling the transmission coefficient of \sars{} with a cosine function \citep{Liu2021season}. Its period is one year, and its amplitude is denoted by $A$, i.e.
\begin{equation}
    \bar{\beta}(t) = \beta(t)\left[ 1 + A \cos\left(2\pi \frac{t}{365 \text{ days}}\right) \right].
    \label{eq:seasonality}
\end{equation}
Here $t$ is expressed in days since January 1st, at which time we assume the $\bar{\beta}(t)$ values are maximal. Its simplicity reflects the current lack of understanding of seasonality's actual effect on \sars{}, mainly due to lack of long-term data. The amplitude $A$ is determined during the calibration procedure.\\

\noindent\textbf{Vaccination} Vaccine incidence data are publicly available for all Belgian provinces and per age class \cite{Sciensano2022}. The fraction of individuals in the population with a partial, full, or boosted vaccination is shown in Figs. \ref{fig:vaccination_timeseries_NIS} and \ref{fig:vaccination_timeseries_age} of the supplementary materials. As shown in Subsection \ref{section:governing_equations}, cumulative vaccination data $\bm{\phi}_v(t)$ are used to transfer people between vaccination stages. These data can be further exploited to estimate the effective (average) protection individuals in vaccination stage $v$ experience at time $t$, as we elucidate below.\\

\noindent In every vaccination state, the vaccine is assumed to offer protection through three mechanisms: 1) Vaccines lower the susceptibility to \sars{}, 2) Vaccines lower the infectiousness of an individual infected with \sars{}, 3) Vaccines lower the hospital admission propensity of \covid{}. To incorporate the effect of \textit{waning} vaccine immunity, the vaccine efficacies within each vaccination metapopulation are rescaled dynamically based on vaccination history. This approach is preferred over adding more vaccination states to limit the computational burden, optimally use the available vaccine incidence data, and allow the three vaccine efficacies to wane at different rates, which decrease to zero, which is not possible using a metapopulation approach.\\

\noindent We compute a population average dynamic vaccine efficacies $\bm{E}_{v,n,\text{susc}}(t), \bm{E}_{v,n,\text{inf}} (t)$ and $\bm{E}_{v,n,\text{hosp}}(t)$ for every vaccine stage $v$, VOC $n$ (and age group $i$ and province $g$) assuming exponential waning of the vaccine. For that purpose, we rely on 1) the past vaccine incidence, available per province $g$, age group $i$ and vaccination stage $v$ \citep{Sciensano2022}, 2) the vaccine efficacies for every protective mechanism, vaccination stage $v$ and every VOC $n$, both 25 and 175 days after vaccination (see Table \ref{tab:vaccine_properties}) \citep{Braeye2022a}, and 3) the assumption that waning is governed by a decreasing exponential,
\begin{equation}
    \widetilde{E}_{v,n,\text{susc}}(t) = E_{v,n,0,\text{susc}} \exp\left(-t/l\right),
\end{equation}
where,
\begin{equation}
    l = \frac{150\text{ d}}{\ln\left(\dfrac{E_{v,n,0,\text{susc}}}{E_{v,n,w,\text{susc}}}\right)} > 0,
\end{equation}
and similarly for $\widetilde{E}_{\text{full},n,\text{inf}}(t)$ and $\widetilde{E}_{\text{full},n,\text{hosp}}$(t) (see Fig. \ref{fig:effect_of_waning_delayed}). These waning functions can be combined with the (previously administered) vaccines to compute the weighted average vaccine efficacy in province $g$ and age class $i$ at time $t$ as follows,
\begin{equation}
    \bm{E}_{v,n,\text{susc}}(t) = \frac{1}{\bm{\phi}_v(t)} \int\limits_{-\infty}^t\dot{\bm{\phi}}_v(t')\widetilde{E}_{v,n,\text{susc}}(t-t')dt',
    \label{eq:effective_rescaling_param}
\end{equation}
where $\dot{\bm{\phi}}_v(t)$ is the time derivative of the \textit{cumulative} number of individuals that reached vaccination stage $v$ at time $t$ for the various age and province populations. Hence, when a large number of individuals are vaccinated to vaccination stage $v$ at once, the average vaccine efficacy in that metapopulation will temporarily increase. The efficacies $\bm{E}_{v,n,\text{susc}}$ and $\bm{E}_{v,n,\text{inf}}$ are then used to scale the transmission coefficient,
\begin{equation}
    \bar{\beta}^{gh}_{ij,vw} = \bar{\beta}(t) \sum_n \alpha_n(t)(1-E_{v,n,\text{susc},i}^{g})(1-E_{w,n,\text{inf},j}^{h}),
\end{equation}
where $i$, $g$ and $v$ denote the age group, province and vaccination stage of the individual `making" contact and $j$, $h$, $w$ denote the age group, province and vaccination stage of the individual ``receiving" contact. Similarily, the values $\bm{E}_{v,\text{hosp}}(t)$ scale the hospitalisation propensities ($\bm{h}$). The timeseries of $\bm{E}_{v,\text{susc}}$, $\bm{E}_{v,\text{inf}}$ and $\bm{E}_{v,\text{hosp}}$ are shown in Figure \ref{fig:vaccine_rescaling_effect}. An overview of all parameters introduced as part of a time-dependency of the parameters $\bar{\bm{N}}(t)$, $\bm{\beta}(t)$ and $\bm{h}(t)$ is provided in Table \ref{tab:SEIQRD_timedep_params}. For the sake of simplicity it is assumed that the number of social contacts made by an individual is assumed independent of the vaccination stage.

\subsection{Model calibration}

\noindent\textbf{Calibrated parameters} Five model parameters are considered to be a priori unknown and must be calibrated using the available data: 1) The relative effectivities of social contacts outside the household as compared to social contacts within the household ($\Omega$), 2) The additional reduction of the contacts under lockdown measures, which could not be explained by activity reductions observed in the GCMR, named ``intervention" parameter ($\Psi$), 3-4) The infectivity gains of the relevant VOCs ($K_{\text{inf},\alpha \beta \gamma}$, $K_{\text{inf},\delta}$) and 5) The seasonal amplitude of the infectiousness ($A$). The simulated daily number of hospitalisations is matched to the eleven time series of daily new hospitalisations in each province, starting on March 15th, 2020 and ending on October 1st, 2021. Further, assuming that on average half of the recovered individuals are again susceptible after one year (associated with seroreversion rate parameter $\zeta$), the simulated numbers of recovered individuals are matched to seven serological measurements from Herzog et al., 2020 \cite{Herzog2020} and 21 serological measurements from Sciensano \cite{Sciensano2020}, spanning the period from March 30th, 2020 until January 8th, 2021.\\

\noindent\textbf{Initial condition} To find an appropriate initial condition on March 15th, 2020, the model was initialised with one presymptomatic infectious individual on February 5th, 2020, corresponding to the day the first case was detected in Belgium. The model was simulated without seasonality or mobility, with the contact effectivity set to one, and with a reproduction number of $R_0=3.3$. The number of initial infected that resulted in the best fit to the provincial time series between March 15th, 2020 and March 22nd, 2020 was determined. The model states on March 15th, 2020 were then used as the initial condition for all simulations depicted in this work.\\

\noindent\textbf{Statistical model} We found that a quadratic relationship best describes the relationship between the mean and variance of the daily hospitalisations time series data, indicating that a negative binomial model is best fit to describe the relationship between the model outcome and observed data \cite{Chan2021} (see Appendix \ref{app:appropriate_statistical_model}). We therefore iteratively optimise the following log-likelihood function,
\begin{multline}
    \log \mathcal{L}(\bm{\widetilde{x}} \vert \bm{x}) = -\sum_{g=1}^G\sum_{t=1}^n \left( \log\left[\frac{\Gamma(x^g_t + 1/\alpha^{g})}{\Gamma(x^g_t + 1) \Gamma(1/\alpha^{g})}\right] + \right.\\ \left.\frac{1}{\alpha^{g}}\log\left[ \frac{1/\alpha^{g}}{1/\alpha^{g} + \widetilde{x}^g_t} \right] + x^g_t\log\left[ \frac{\widetilde{x}^g_t}{1/\alpha^{g} + \widetilde{x}^g_t} \right]\right).
    \label{eq:calibration_loglikelihood}
\end{multline}
Here the outer sum is over all $G=11$ provinces. The inner sum is over all $n$ observed data points at times $t$. $\Gamma$ is the gamma function. $\bm{\widetilde{x}}$ represents the simulated time series of daily hospitalisations (obtained after summing over all age groups and vaccination states), and $\bm{x}$ the equivalent observed time series. By fitting the negative binomial model to the mean-variance relation of the provincial time series, the overdispersion parameter $\alpha^g$, which quantifies the presumed error on the data per province $g$, was computed and subsequently used in the optimisation procedure (see Table \ref{tab:overdispersions}). Maximising the result of Eq. \eqref{eq:calibration_loglikelihood} is computationally demanding and has local minima. A good technique to initially broadly identify the region where the global maximum is situated is Particle Swarm Optimisation (PSO) \cite{kennedy1995}. Subsequently, once a region of interest has been identified, we use the maximum-likelihood estimates as initial values for the ensemble sampler for Markov Chain Monte Carlo (MCMC) \cite{goodman2010}. For all parameters, uniform prior distributions were used. More details are found in Appendix \ref{app:calibration}, and section \ref{sec:results_model-calibration} contains calibration results.

\subsection{Scenario analyses for policymakers}

\subsubsection{Combined impact of VOCs and vaccines}

Next, we illustrate how our model can be used to simulate counterfactual scenarios on the combined impacts of the emergence of new variants, an ongoing nation-wide vaccination campaign and social relaxations. Such simulations can be used to provide policymakers with insights on the optimal timing of the release of social restrictions. We therefore calibrate our model up to March 1st, 2021, a point in time interesting because the Alpha-Beta-Gamma VOCs had just become dominant, the Belgian vaccination campaign was picking up speed, and there was a high pressure to relax social restrictions. Under the emergence of the Alpha-Beta-Gamma VOCs and vaccination campaign, we thus define four future scenarios in which social restrictions are released.\\

\noindent The first scenario (S0) assumes that social restrictions are gradually eased over a two-month period, starting on March 1st, 2021. The subsequent scenarios S1 through S3 then postpone the release date of the social restrictions by one month. Thus, measures are released on April 1st, May 1st and June 1st for scenarios S1 through S3. In all scenarios, we make three assumptions: 1) We assume the Delta variant does not emerge. 2) We assume a 40~\% increase in transmissibility of the Alpha-Beta-Gamma VOCs (Table \ref{tab:calibration_parameters}) with no uncertainty. On March 1st, 2021, it was not yet possible to calibrate the transmissibility increase of the VOC and thus determine the uncertainty on the estimate. 3) We use the observed number of administered vaccines. The number of vaccine doses that would be administered was of course not known on March 1st, 2021, but in policy advice given at that time, realistic projections for the future-administered doses were used. The scenarios are additionally simulated using our equivalent nation-level model to asses the effect of adding spatial heterogeneity on the outcome of these scenarios. The (total) number of contacts associated with each scenario is shown in Fig. \ref{fig:four_projected_nc_1mar2021}).
\begin{figure}[h!]
    \centering
    \includegraphics[width=\linewidth]{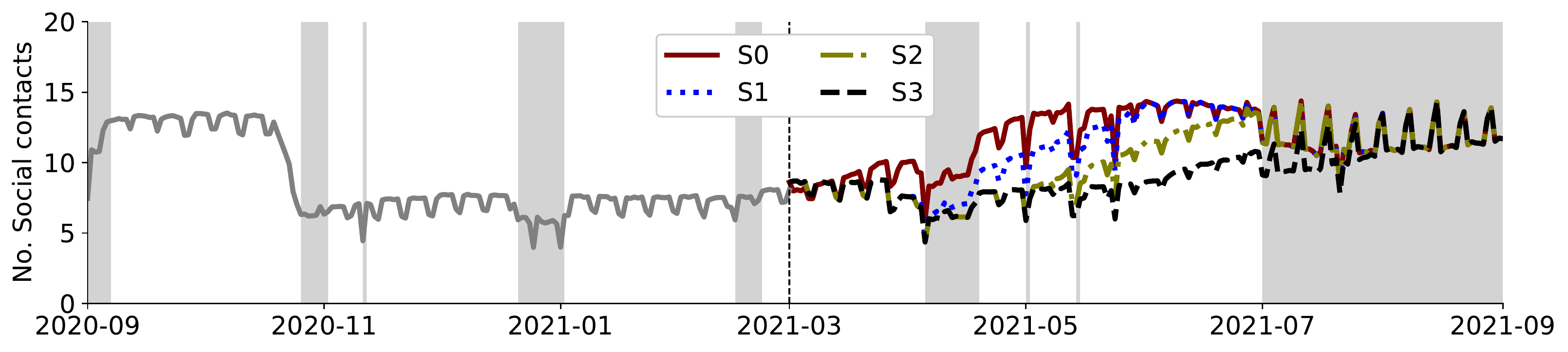}
    \caption{Age-weighted average of the observed social contact matrix elements $N_{ij}(t)$ (grey), plotted alongside the social contact associated with scenarios S0 (release of social restrictions on March 1st, 2021), S1 (April 1st, 2021), S2 (May 1st, 2021), and S3 (June 1st, 2021). A grey background is used to indicate a holiday period.}
    \label{fig:four_projected_nc_1mar2021}
\end{figure}

\subsubsection{Impact of local restrictions}
A particular strength of our model is its ability to explore the impact of mobility restrictions on the spread of \sars{} in Belgium. We inspect if altering the mobility between two provinces can be used to contain an epidemic within a province, or alternatively, if altering the mobility between two provinces can shield one province from an epidemic in the other. Next, we assess in a similar manner the impact of limiting social contacts in one of the provinces. All simulations are started on January 1st 2020, upon which we inspect the resulting hospitalisations until September 1st, 2020. In all scenarios, seasonality, VOCs and vaccines are omitted. Default social contact and mobility are equal to their prepandemic values. Due to their large demographic differences and their relatively weak connectivity, we inspect results for Brussels and Luxembourg.\\

\noindent\textbf{Regulating local mobility} We define a parameter $\bm{p}$ whose elements $p^g \in [0, 1]$ linearly control the mobility to and from province $g$, compared to some static, prepandemic baseline mobility,
\begin{equation}
    \widetilde{P}^{gh} =\bar{P}^{gh} p^g p^h + \delta^{gh}\sum_{f=1}^G \bar{P}^{gf}(1-p^f p^h),
\end{equation}
where $\delta^{gh}$ is the Kronecker delta, and $\bar{\bm{P}}$ is the prepandemic baseline mobility,
\begin{equation}
    \bar{\bm{P}} = \text{avg}\left\{ \bm{P}(t) \right\} \qquad \text{for } t < \text{March 15th 2020}.
\end{equation}
We assume social contact behaviour remains the same and is independent of whether a province is an individual's home province or a visited province. We change the connectivity between Brussels and Luxembourg to \textit{shield} Brussels from an outbreak in Luxembourg (Mob. S.), or we attempt to \textit{contain} an outbreak in Brussels (Mob. C.). The simulations are started with 10 exposed individuals, distributed over the model's 10 age groups according to the demographics of the spatial patch. In Mob. S., the index patients are in Luxembourg, while in Mob. C., the index patients are in Brussels. We run the simulation 100 times for each of the 30 $p^g$ values logarithmically spaced between 1 and $10^{-3}$. Because of the stochastic nature of the model, in the analysis, the mean of the 100 repeated simulations is used.\\

\noindent\textbf{Regulating local social contact} We define a parameter $\bm{n}$ whose elements $n^g \in [0, 1]$ determine the local average social contact in province $g$ compared to the total prepandemic number of social contacts $\bm{N}$,
\begin{equation}
    \bar{N}_{ij}^{g} = n^g(N_{ij} - N_{ij}^\text{home}) + N_{ij}^\text{home},
\end{equation}
where $\bm{N}^{\text{home}}$ is the prepandemic matrix of home contacts. Note that this quantity is independent of the province of origin $h$: thus, individuals follow the social rules of the province they \textit{visit}. By altering the social contacts in Brussels, we perform a similar analysis for shielding and containing an initial outbreak in Brussels or Luxembourg. We call these scenarios Soc. S. and Soc. C., respectively. We again run 30 (times 100) simulations, one for every $n^g$ value, now equidistantly spaced between 1 and 0, and all other parameters are fixed (including mobility $\bm{p}=1$). 

\section{Results and discussion}\label{sec:results}

\subsection{Model calibration}
\label{sec:results_model-calibration}

All calibrated parameter values and their 95~\% quantiles are listed in Table \ref{tab:calibration_parameters}. Further, a corner plot showing the two-dimensional posterior distributions of the 9 calibrated parameters is given in Fig. \ref{fig:full-calibration-corner-plot}. Generally, the calibrated parameter's values are tangible and their posterior distributions clearly resolved and free of excessive correlation. Noteworthy are the amplitude of the seasonal transmissibility, $A = 18~\%$ (95~\% CI: 15~\% - 21~\%), implying \sars{} may be 44\% more transmissible during winter compared to summer time. The effectivity of social contacts outside the home environment in spreading \sars{} is lower than for social contacts at home, $\Omega = 0.40$ (95~\% CI: 0.34 - 0.46), which is qualitatively consistent with Thompson et. al \citep{Thompson2021}. \\ 

\begin{table}[!h]
    \centering
    \caption{Values and 95~\% quantiles of the calibrated model parameters.}
    \begin{tabular}{p{1.5cm}>{\raggedright\arraybackslash}p{5.9cm}lp{2cm}}
        \toprule
        \textbf{Param.} & \textbf{Interpretation} & \textbf{Value} & \textbf{$\mathbf{[q_{0.025}, q_{0.975}]}$}\\ \midrule
        $\Omega$ & Relative effectivity of social contacts outside home environment & $0.40$   & $[0.34,0.46]$  \\
        $\Psi$ & Intervention parameter & $0.65$   & $[0.60,0.70]$\\
        $K_{\text{inf},\alpha\beta\gamma}$ & Increased infectivity of the Alpha-Beta-Gamma VOCs compared to the wild type for non-vaccinated individuals. & $1.40$    & $[1.29,1.50]$ \\
        $K_{\text{inf},\delta}$ & Increased infectivity of the Delta VOC compared to the wild type for non-vaccinated individuals. & $2.00$    & $[1.77,2.18]$ \\
        $A$ &  Amplitude of seasonality in the viral transmissibility of \sars{}. & $0.18$   & $[0.15, 0.21]$ \\
      \bottomrule
    \end{tabular}
    \label{tab:calibration_parameters}
\end{table}

\noindent We show the nationally and regionally aggregated simulations between March 15th and October 14th 2021 in Fig. \ref{fig:national-and-regional-complete-model-fit} while the eleven provincial time series are given in Figs. \ref{fig:provincial-complete-model-fit-0} and \ref{fig:provincial-complete-model-fit-1}. In general, over the calibrated period (before the dashed line in Fig. \ref{fig:national-and-regional-complete-model-fit}), both the regional and national aggregates fit the observed number of daily hospitalisations well (Fig. \ref{fig:national-and-regional-complete-model-fit}).  Beyond the model's calibration period (after the dashed line in Fig. \ref{fig:national-and-regional-complete-model-fit}) during the peak of October-December 2021 caused by the Delta variant, the forecasted number of new hospitalisations is adequate at the national level. However, upon examination of the regional breakdown, the daily hospitalisations during the Delta wave are slightly underestimated in Flanders, slightly overestimated in Wallonia, and severely overestimated Brussels. These results are not explained by the regional vaccination degrees prior to the arrival of the Delta variant, Flanders (91.4\% of 18+ by October 1st 2021) has a much higher vaccination coverage than Wallonia (79.8\%) and Brussels (66.5\%) (Fig. \ref{fig:vaccination_timeseries_NIS}). We offer two possible explanations: First, the vaccines might have waned faster or their efficacy against the Delta VOC might have been lower than reported. This would result in the mitigation of the effects of regional differences in vaccine coverage. Second, prior to the Delta wave, all social restrictions were lifted in Flanders while some measures were still in place in Wallonia and Brussels. This difference in the degree of social contact was not adequately captured in the GCMRs, likely because there may still have been mentality changes in the population. A contact study using surveys at the regional level may have given more insights into the observed differences in Delta peak height, although it should be noted such studies may be biased towards greater adherence to restrictions \citep{Kennedy2022}.\\

\begin{figure}[h!]
    \centering
    \includegraphics[width=\linewidth]{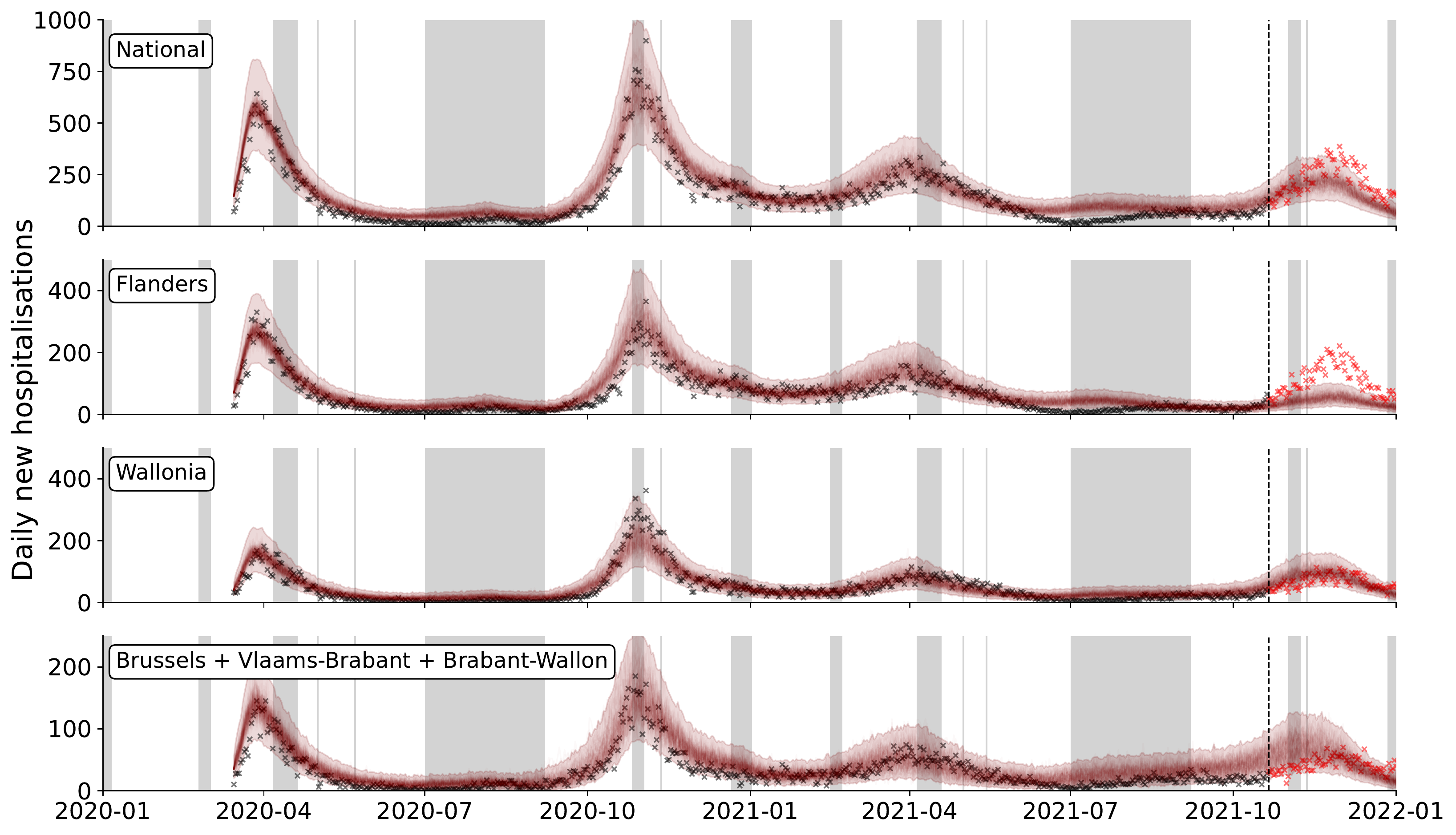}
    \caption{100 model realisations of the daily new hospitalisations between March 15th 2020 and January 1st 2022 (solid lines) with a negative binomial 95\% confidence region (transparent band). Black crosses signify raw data from Sciensano \cite{Sciensano2020} were used in the calibration procedure while red crosses signify data were not used during the calibration procedure. From top to bottom: Nationally aggregated daily number of hospitalisations, daily hospitalisations aggregated over all Flemish provinces (except Vlaams-Brabant), daily hospitalisations aggregated over Walloon provinces (except Brabant Wallon), daily hospitalisations in Brussels, Vlaams-Brabant and Brabant Wallon (see Table \ref{tab:class-NIS-name} and Fig. \ref{fig:beta_classes_prov}). A grey background is used to indicate a holiday period.}
    \label{fig:national-and-regional-complete-model-fit}
\end{figure}

\noindent In Fig. \ref{fig:RMSE-mobility-boxplot} we show the abscence of differences in the time series of the negative binomial log-likelihood score between the observed and simulated daily new hospitalisations of the spatially explicit model when interpovincial mobility is included or excluded. As no differences in goodness-of-fit were found, the interprovincial mobility does not seem to have a significant impact on the model outcome during the calibration period. The impact of interprovincial mobility is likely mitigated because \sars{} has been present in every Belgian province over the calibrated range. We conclude that incorporating interprovincial mobility in Belgium was not necessary to obtain an accurate description of the 2020-2021 \sars{} pandemic in Belgium. This finding is consistent with our previous work by Rollier et al. \citep{rollier2023}, who concluded that the overall correlation between the shape and timing of epidemic peaks and the mobility between them was in general weak for \sars{} in Belgium. The correlation was strongest during epidemic onset but quickly declined once the epidemic became widespread. Noteworthy in Fig. \ref{fig:RMSE-mobility-boxplot} is the goodness-of-fit of the model deteriorates at very low \sars{} prevalences. A similar finding was found for the equivalent national model \ref{fig:RMSE-fit-boxplot}. This is likely because the assumption of homogeneous mixing in every metapopulation is no longer fulfilled.\\

\begin{figure}[h!]
    \centering
    \includegraphics[width=\linewidth]{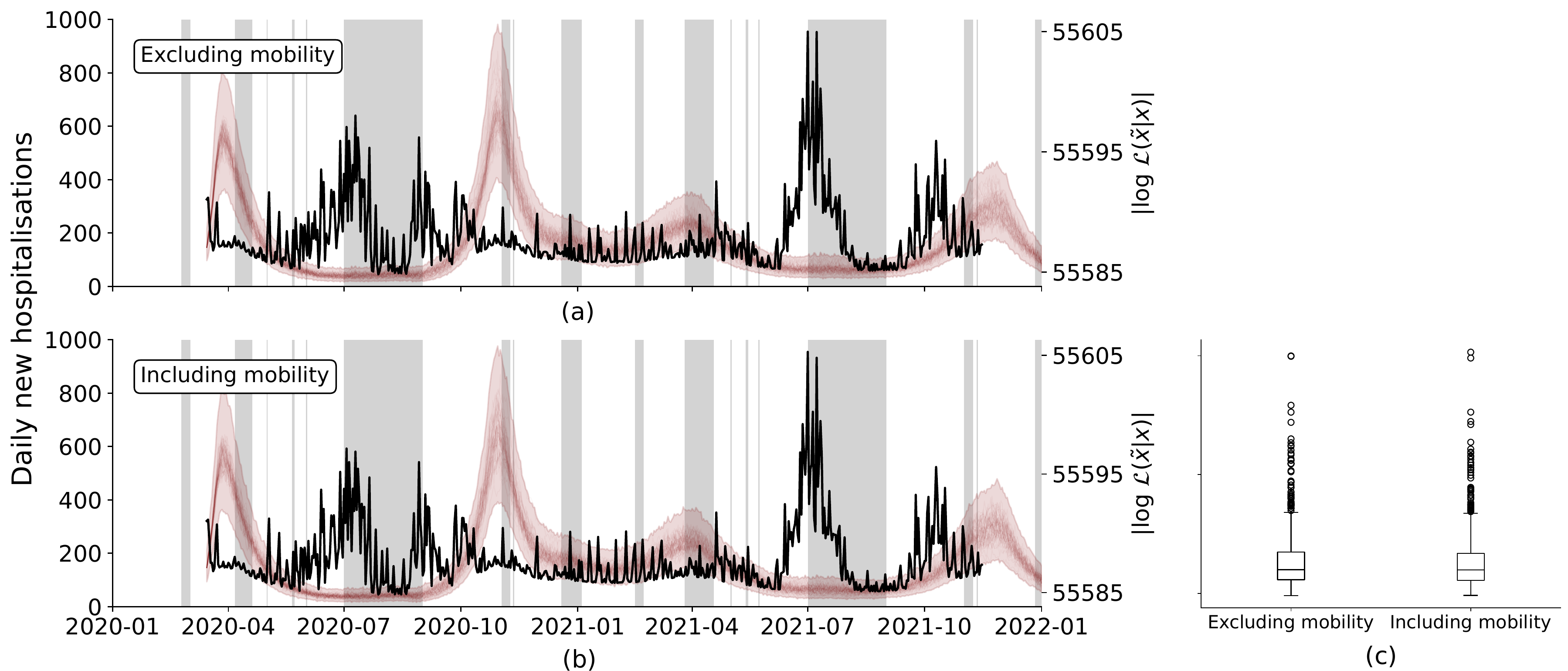}
    \caption{100 realisations of the daily new hospitalisations between March 15th 2020 and January 1st 2022 using the spatially explicit model (a) excluding interprovincial mobility and (b) including interprovincial mobility. (black solid line; right-hand axis) The accompanying negative binomial log-likelihood score of the model predictions. (c) Boxplot of the log-likelihood values at every time $t$. A grey background is used to indicate a holiday period.}
    \label{fig:RMSE-mobility-boxplot}
\end{figure}

\subsection{Scenario analyses for policymakers}
\subsubsection{Combined impact of VOCs and vaccines}
\label{sec:scenarios-for-policymakers}

We start by focusing on the provincial-level model first (top panel, Fig. \ref{fig:four_scenarios}). If the measures are released on March 1st (S0), the provincial model predicts a broad wave of \covid{} hospitalisations, with 550 daily hospitalisations by the first week of June. Releasing the measures results in a total of 8000 \covid{} patients in Belgian hospitals, corresponding to approximately 1400 patients in Belgian ICs (top panel, Figure \ref{fig:four_scenarios_H_tot}). The nominal IC capacity of Belgium is approximately 1000 IC beds, and the extended number of IC beds is approximately equal to 2000 beds. Surpassing the nominal capacity of 1000 IC beds is generally undesirable due to high stress on the Belgian healthcare system. Surpassing the extended capacity of 2000 IC beds corresponds to a scenario where not enough IC beds are available and clinicians need to start prioritising patients, which is undesired. Thus, relaxing measures on March 1st (S0) would most likely severely strain the Belgian healthcare system and in the worst case cause its collapse. Relaxing measures on April 1st (S1) results in a peak of 300 hospitalisations per day by July 1st, 2021. Although the hospitalisation peak of S1 seems much less threatening than the hospitalisation peak of S0, the prolonged strain on the healthcare system is still undesirable. Relaxations starting on May 1st, 2021 (S2) and June 1st, 2021 (S3) contain the epidemic, likely due to the combined effect of vaccination and favourable seasonal changes during summer. Our projections strongly recommend against the relaxation of social relaxations on March 1st (S0), and on April 1st (S1) even if the measures are gradually relaxed over a two-month period. We thus conclude that relaxing social restrictions without straining the Belgian healthcare system is only possible starting May 1st (S2). \\

\begin{figure}[h!]
    \centering
    \includegraphics[width=\linewidth]{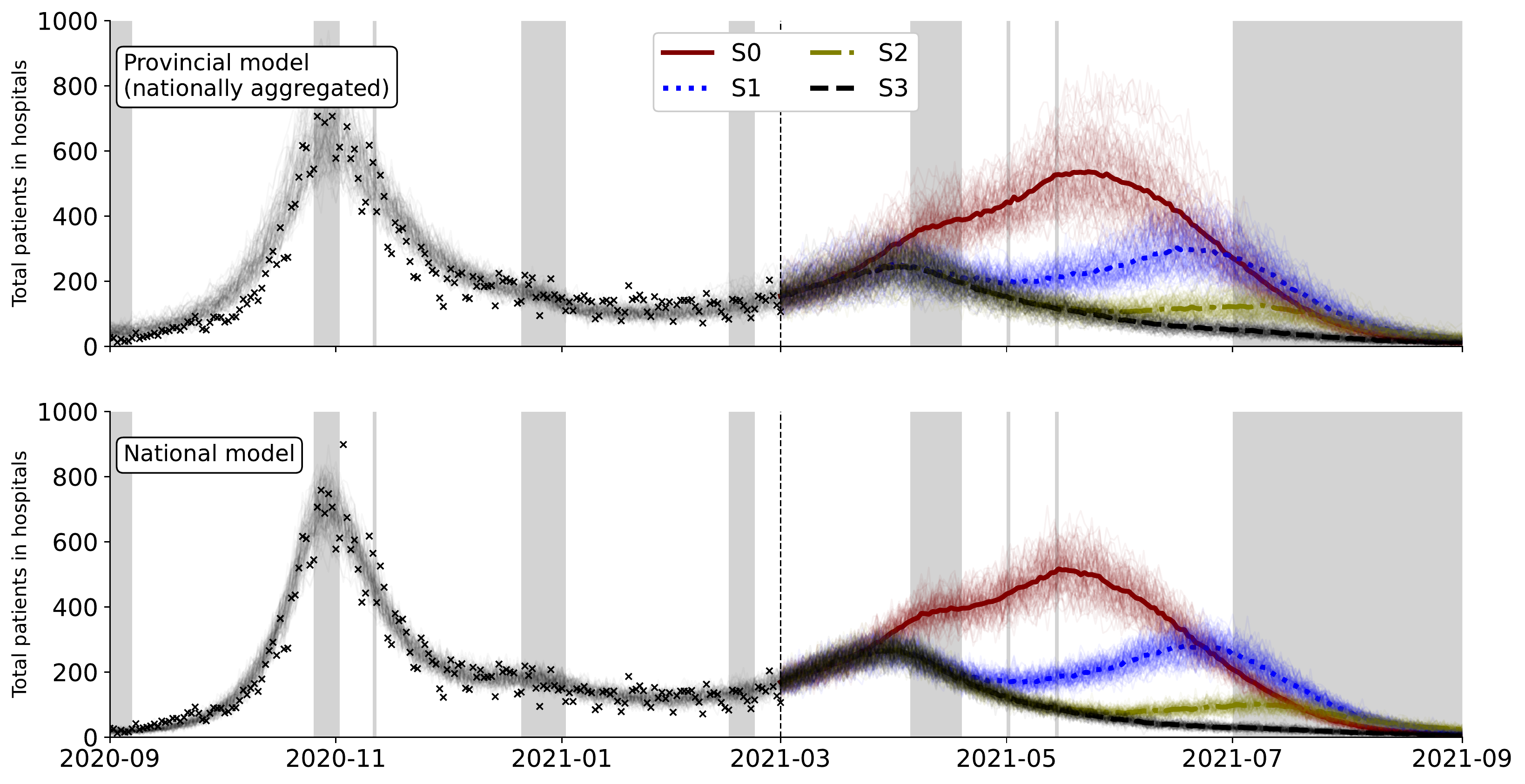}
    \caption{Combined impact of the Alpha-Beta-Gamma VOCs, the ongoing nation-wide vaccination campaign and social relaxations on the number of daily hospitalisations in Belgium. Social relaxations start on March 1st, 2021 (S0), and are postponed by one month in subsequent scenarios. The result of 100 stochastic simulations and their mean are shown. \textit{Top}: Simulated using the provincial-level model. \textit{Bottom}: Simulated using the (equivalent) nation-level model. A grey background is used to indicate a holiday period.}
    \label{fig:four_scenarios}
\end{figure}

\begin{figure}[h!]
    \centering
    \includegraphics[width=\linewidth]{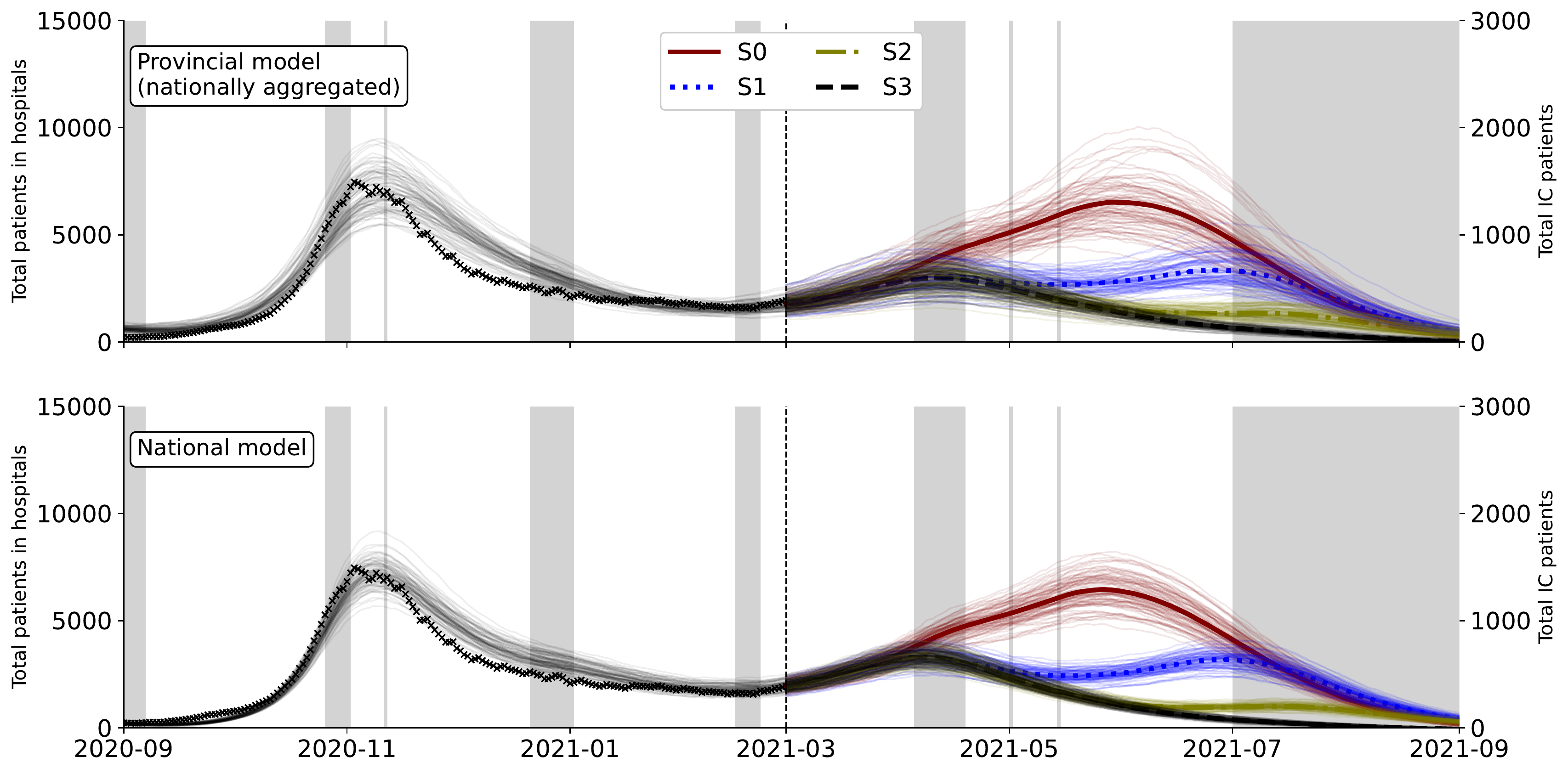}
    \caption{Combined impact of the Alpha-Beta-Gamma VOCs, the ongoing nation-wide vaccination campaign and social relaxations on the total number of \covid{} patients in Belgian hospitals. Social relaxations start on March 1st, 2021 (S0), and are postponed by one month in subsequent scenarios. The model is not fitted to the total number of patients in Belgian hospitals, rather, the hospital dynamics of the model are informed with estimates on the length-of-stay obtained from analyzing a dataset containing 22 136 \covid{} patients \citep{Alleman2021}. Belgium has a nominal capacity of 1000 IC beds and an extended maximum capacity of 2000 IC beds. The result of 100 stochastic simulations and their mean are shown. \textit{Top}: Simulated using the provincial-level model. \textit{Bottom}: Simulated using the (equivalent) nation-level model. A grey background is used to indicate a holiday period.}
    \label{fig:four_scenarios_H_tot}
\end{figure}

\noindent Both qualitatively and quantitatively speaking, the results obtained with the nation-level model (bottom panel, Fig. \ref{fig:four_scenarios}) are in agreement with those of the provincial-level model. Adding spatial heterogeneity (provinces) to the model does not seem to significantly affect the projections. However, the key difference in both projections lay in the amount of uncertainty associated with the projections. The nation-level model has less uncertainty associated with its projections than the provincial-level model. The addition of spatial heterogeneity in a stochastic compartmental model thus seems to increase uncertainty of the outcomes. This is most likely because the number of compartments, and thus the number of potential (stochastic) transitions are larger in the provincial model (5720 vs. 520 compartments). The nation-level model's simplicity may imply it is overconfident in its projections for policymakers, although ranking models in practice is difficult.\\

\noindent In our experience, policymakers and press media tend to disproportionately focus on the quantitative model outcomes in policy advice rather than the qualitative outcomes. The added uncertainty in the provincial-level model is both advantageous and disadvantageous from a communication point-of-view.  The introduction of uncertainty shifts the focus away from minor deviations between the observations and projections, as these deviations will appear less significant. Conversely, the increased uncertainty may result in policymakers focusing on their preferred outcome included in the projection's uncertainty. Or they may choose to disregard the projection altogether due to its perceived lack of reliability, even though the mean predictions of both models are closely aligned.} Our results demonstrate the need for cautious interpretation of projections derived from a single model. Particularly during an urgent crisis, there is little time to rank the suitability of models for providing policy advice. To address the inherent methodological uncertainty associated with mathematical models, it is advisable to combine multiple projections, obtained using different paradigms and developed by different groups, into an ensemble to enhance the robustness of the predictions \cite{RESTORE7, RESTORE8}.\\

\subsubsection{Impact of local restrictions}
\label{sec:results_local-scenarios}

\begin{figure}
  \centering
  \begin{tabular}{@{}c@{}}
    \includegraphics[width=\linewidth]{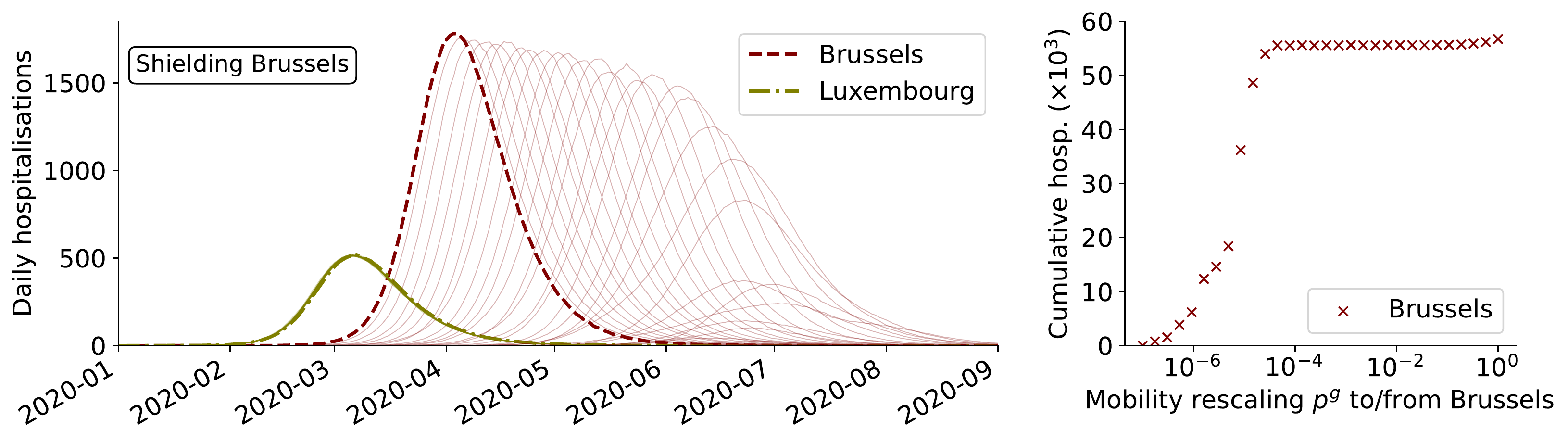} \\
    \small{(a) Scenario Mob. S.}
  \end{tabular}
  \begin{tabular}{@{}c@{}}
    \includegraphics[width=\linewidth]{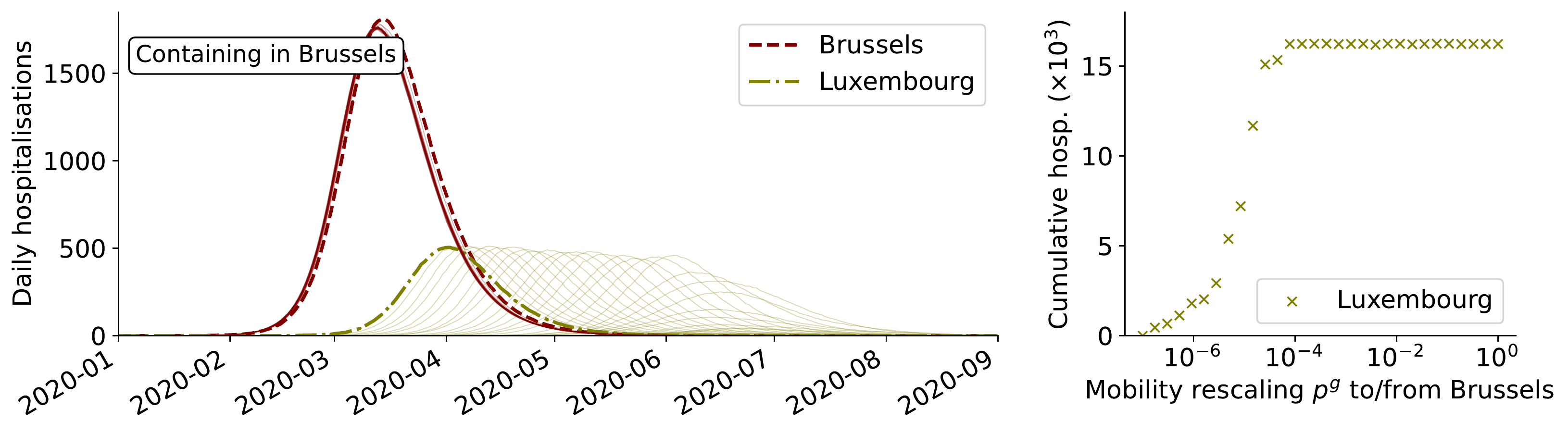} \\
    \small{(b) Scenario Mob. C.}
  \end{tabular}
\caption{Effect of a logarithmic decrease in mobility between Brussels and Luxembourg on the simulated daily number of hospitalisations. Either when (top) shielding Brussels from an outside epidemic, simulated as 10 index patients introduced in Luxembourg or (bottom) containing an epidemic within Brussels, simulated as 10 index patients introduced in Brussels. The red dashed line indicates the daily number of hospitalisations in Brussels when the mobility is unaltered ($p^g = 1$). The thinner, solid red lines indicate how the daily number of hospitalisations in Brussels change when the mobility between Brussels and Luxembourg is decreased ($p^g < 1$). The same applies to the daily number of hospitalisations in Luxembourg.}
\label{fig:mobility-reduction-to-21000}
\end{figure}

\noindent\textbf{Regulating local mobility} In scenario Mob. S. (Fig. \ref{fig:mobility-reduction-to-21000}(a)), where we attempt to shield Brussels from an epidemic originating in Luxembourg, for all practical isolation cases, the effect of lowering the mobility is to \textit{delay} the onset of the epidemic in Brussels without lowering the cumulative number of hospitalisations. Only when extreme isolation values of $p^g < 10^{\text{-}4}$ are reached, which correspond to single commuters between these provinces, does the expected cumulative number of hospitalisations decline. In scenario Mob. C. (Fig. \ref{fig:mobility-reduction-to-21000}(b)), where we attempt to prevent an epidemic originating in Brussels to spread to Luxembourg, a similar result is obtained. During the first, very strict national lockdown, the value for $p^g$ was approximately 0.5, (see Fig. \ref{fig:staytime_percentage_timeseries}). Consequently, we may conclude that \textit{only} reducing mobility can delay the spread of a \sars{} epidemic but can ultimately not contain it. The qualitative relations are of course not unique to Brussels and Luxembourg, but apply to all pairs of provinces.\\

\begin{figure}
  \centering
  \begin{tabular}{@{}c@{}}
    \includegraphics[width=\linewidth]{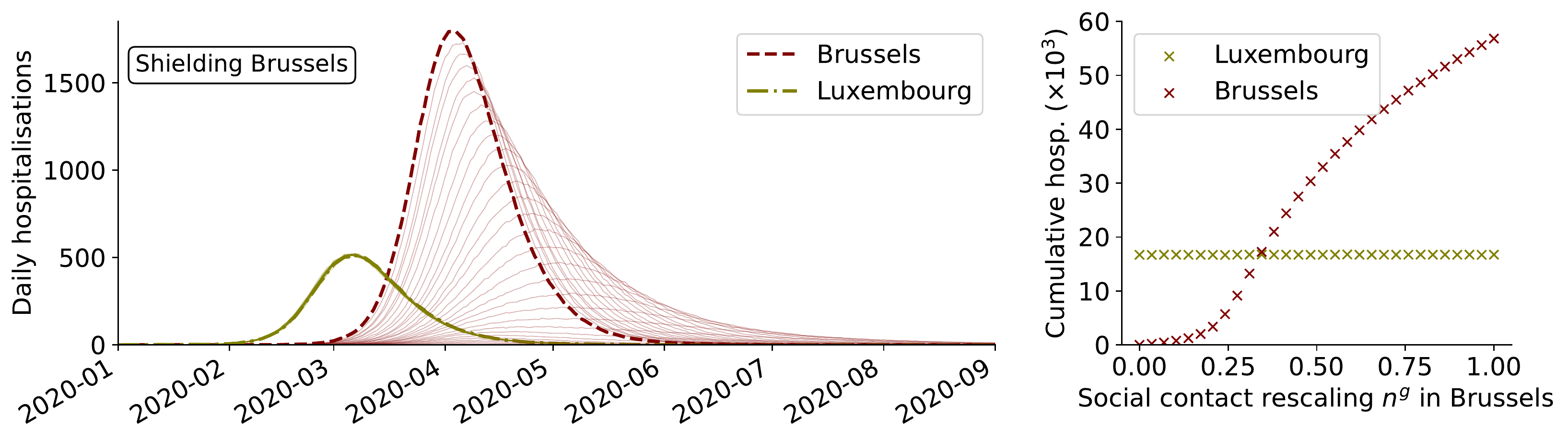} \\
    \small{(a) Scenario Soc. S.}
  \end{tabular}
  \begin{tabular}{@{}c@{}}
    \includegraphics[width=\linewidth]{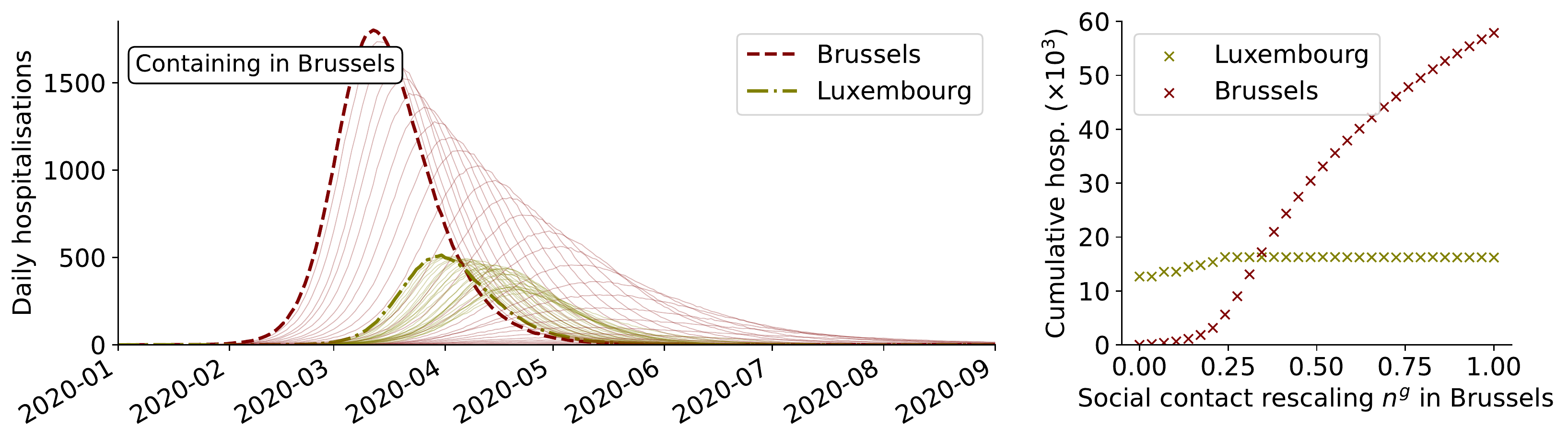} \\
    \small{(b) Scenario Soc. C.}
  \end{tabular}
\caption{Effect of linear decrease of social contact in Brussels on the simulated hospitalisation time series. Either to (top) shield Brussels from an outside epidemic, simulated as 10 index patients introduced in every age class in Luxembourg or to (bottom) contain an epidemic within Brussels, simulated as 10 index patients introduced in every age class in Brussels. The red dashed line indicates the daily number of hospitalisations in Brussels when the social contact is unaltered ($n^g = 1$). The thinner, solid red lines indicate how the daily number of hospitalisations in Brussels change when the social contact in Brussels is decreased ($n^g < 1$). The same applies to the daily number of hospitalisations in Luxembourg.}
\label{fig:contact-reduction-in-21000}
\end{figure}

\noindent\textbf{Regulating local social contact} In scenario Soc. S. (Fig. \ref{fig:contact-reduction-in-21000}(a)) we attempt to shield Brussels from an epidemic originating in Luxembourg by lowering social contact in Brussels. Reducing social contact in Brussels does not influence the epidemic in Luxembourg, while strongly delaying \textit{and} reducing the epidemic in Brussels. Especially for $0.25 < n^{g} < 0.50$, a strong decrease in the expected cumulative number of hospitalisations occurs in Brussels. In the containment scenario Soc. C. (Fig. \ref{fig:contact-reduction-in-21000}(b)) we attempt to prevent an epidemic originating in Brussels from spreading to Luxembourg by lowering social contacts in Brussels. A similar trend as in scenario Soc. S. is seen for Brussels. However, we observe an additional effect of \textit{delaying} and, for large social contact reductions ($n^{g}<0.25$) in Brussels, lowering the expected cumulative number of hospitalisations in Luxembourg. The epidemic in Luxembourg is never fully contained by lowering social contact in Brussels and this is a consequence of the way the scenario is setup. If $n^{\text{Brussels}} = 0$ but $p^{\text{Brussels}} = 1$, Brussels residents are still commuting to other provinces where they have their work contacts, hence the epidemic can, in a fraction of the 100 simulations, still spread across Belgium. From the presented scenarios we conclude that lowering the number of social contacts is a more effective policy measure than lowering mobility to contain a \sars{} epidemic in a given province.

\pagebreak
\section{Conclusions}\label{sec:conclusion}

Starting from our previously developed national model \cite{Alleman2021}, we developed a spatially explicit model variant which we then simulated stochastically. Both models were, over the past two years, extended to account for the emergence of VOCs, seasonality, and vaccines. These were critical model additions that were desired and required for the description, forecasting, and understanding of the \covid{} pandemic in Belgium. \\

\noindent We have demonstrated the spatially explicit model's ability to describe the relevant \covid{} time series in Belgium. We then demonstrated its capabilities to evaluate counterfactual scenarios on the combined  effects of non-pharmaceutical policy interventions, VOCs and vaccination, which can -- and have been -- applied to support the decision-making process. In addition, we used the model to study the effects of locally reducing mobility and of locally reducing social contact to shield or contain an epidemic within a given province. We concluded that lowering social contact is a more effective means of containing a \sars{} epidemic in a given province than lowering mobility.\\

\noindent Additionally, we assessed if adding the impact of adding spatial heterogeneity and interprovincial mobility is worth the added complexity. This question is of importance when deciding wether to use a spatially-explicit model; the answer is nuanced as the choice for a spatially-explicit method depends on the intended purpose. First, we have assessed the differences in goodness-of-fit over the calibrated range with and without the inclusion of interprovincial mobility and found that incorporating interprovincial mobility in Belgium was \textit{not} necessary to obtain an accurate description of the 2020-2021 \sars{} pandemic in Belgium. Second, when modelling a scenario for policymakers, adding spatial heterogeneity into the model did not significantly change the mean projected daily \covid{} hospitalisations but increased the uncertainty on the model projections. Added uncertainty entails both communicative advantages and disadvantages. Third, the spatially-explicit model enables the communication of more fine-grained results to policymakers, for which we experienced a demand during the \covid{} pandemic. Fourth, the spatially explicit model with finely-grained mobility data allows to assess the impact of interprovincial mobility in the early stages of an epidemic, which is not possible with the national model. We found this impact to be limited for \sars{} in Belgium, but this may not necessarily the case in larger, more poorly connected countries or for other infectious diseases; it is possible and interesting to mobilise the presented model for such purposes, which is arguably its most valuable scientific contribution.

\backmatter

\clearpage

\bmhead{Supplementary information}

This paper contains additional information on the geography of Belgium (Appendix \ref{app:sciensano}), the data used in this work (Appendices \ref{app:sciensano}, \ref{app:proximus-mobility-data} and \ref{app:social_contact}), the implementation of the VOCs, seasonality, and vaccines (Appendix \ref{app:VOC_vacc}), an overview of the model parameters and assumptions (Appendix \ref{app:model-equations-and-model-parameters}) and details on the model calibration (Appendix \ref{app:calibration}).

\bmhead{CRediT author statement}

\textbf{Tijs W. Alleman}: Conceptualisation, Software, Methodology, Investigation, Data curation, Writing – original draft. Writing – Review \& Editing. \textbf{Michiel Rollier}: Conceptualisation, Methodology, Investigation, Data curation, Writing – original draft.  Both of the above authors have closely collaborated on the manuscripts contents and should both be regarded as the primary authors of the text. \textbf{Jenna Vergeynst}: Conceptualisation, Investigation, Project administration. \textbf{Jan M. Baetens}: Supervision, Conceptualisation, Funding acquisition, Project administration.

\bmhead{Acknowledgements}

We would like to thank Proximus for providing the telecommunication data free of charge. We would like to thank Lander De Visscher for his methodological help on the \textit{Markov-Chain Monte-Carlo} technique used in this work. This work was financially supported by \textit{Crelan}, the \textit{Ghent University Special Research Fund}, the \textit{Research Foundation Flanders} (FWO), Belgium, project numbers G0G2920 and 3G0G9820 and by \textit{VZW 100 km Dodentocht Kadee}, through the organisation of the 2020 100 km COVID-Challenge.

\bmhead{Conflict of interest} None declared.

\bmhead{Ethics approval} All used data conform to GDPR standards. 

\bmhead{Consent to participate} Not applicable

\bmhead{Consent for publication} All authors consent to the publication in Applied Mathematical Modeling, preceded by pre-print publication in an open-access archive.

\bmhead{Availability of data} The COVID-19 hospitalisation data \citep{Sciensano2020}, seroprevalence data \citep{Sciensano2020, Herzog2020}, Google Community Mobility data \citep{google_mobility} and the prepandemic contact matrices for Belgium \citep{Willem2020a} are publicly available. The telecommunication data to describe the mobility between the Belgian provinces is not publicly available.

\bmhead{Code availability} The source code of the model is freely available on GitHub: \url{https://github.com/UGentBiomath/COVID19-Model}. The model was implemented using our in-house code for simulating $n$-dimensional dynamical systems in Python 3 named \textit{pySODM} \citep{alleman2023}, which is freely available on GitHub: \url{https://github.com/twallema/pySODM}, published on \textit{pyPI}: \url{https://pypi.org/project/pySODM/}, and features an extensive documentation website: \url{https://twallema.github.io/pySODM}. 


\newpage
\begin{appendices}

\section{Data}

\subsection{COVID-19 time series data}
\label{app:sciensano}

\begin{figure}[h!]
    \centering
    \includegraphics[width=\linewidth]{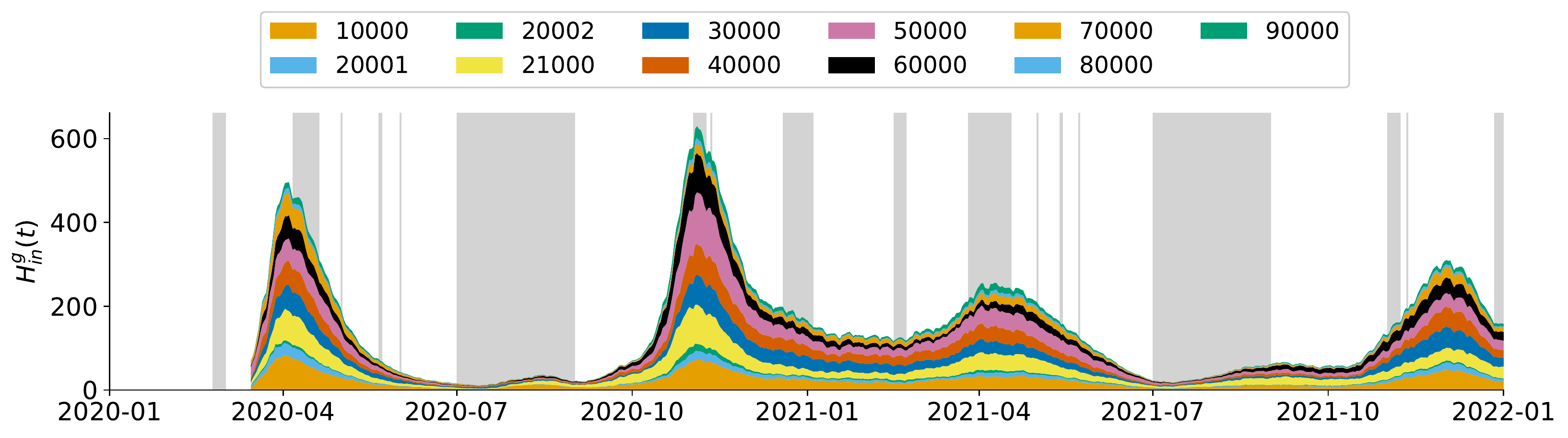}
    \caption{Stacked area plot of all seven-day moving-averaged time series for daily new hospitalisations per province (denoted with NIS code, see Table \ref{tab:class-NIS-name}) \citep{Sciensano2020}. Daily data is available from March 15th 2020 onward. A grey background is used to indicate a holiday period.}
    \label{fig:all-H_in-series_prov}
\end{figure}

\begin{figure}[b]
    \centering
    \includegraphics[width=\textwidth]{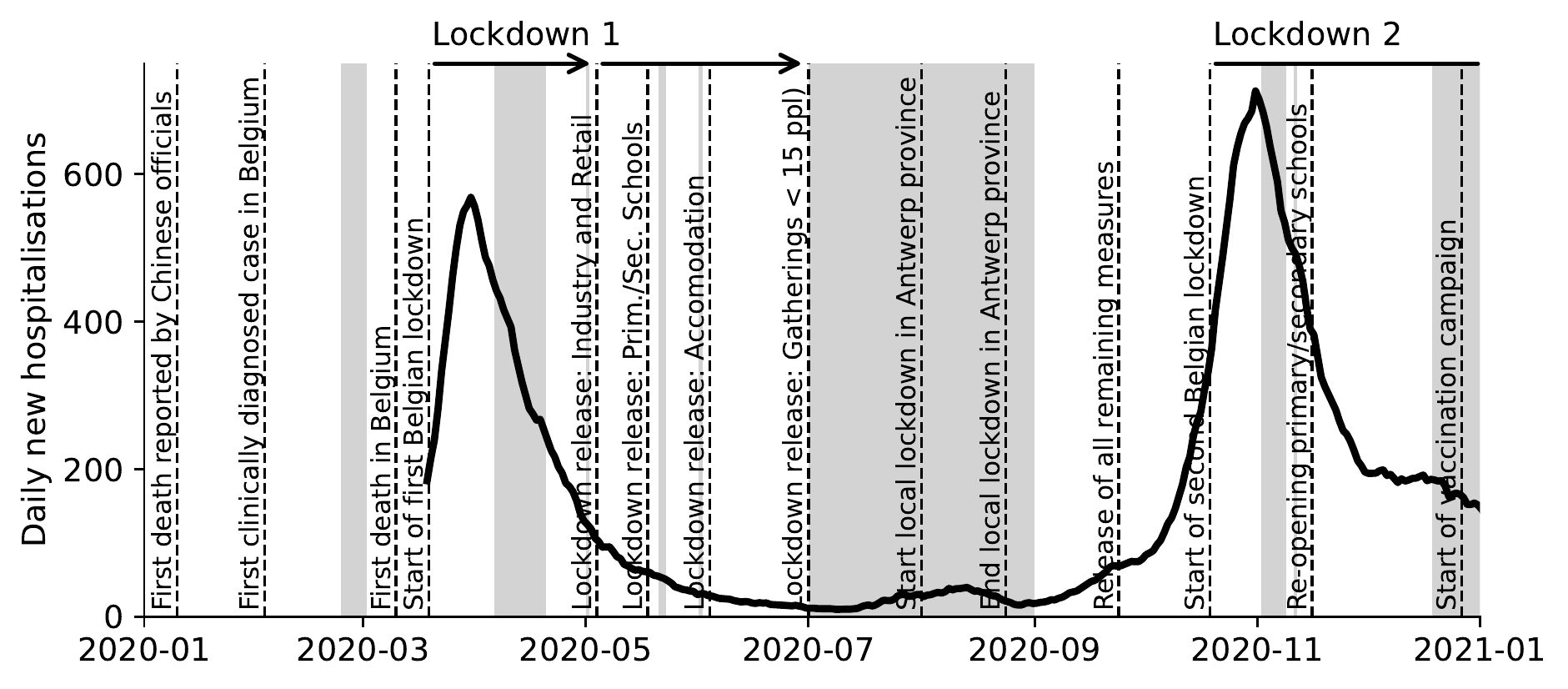}\\
    \includegraphics[width=\textwidth]{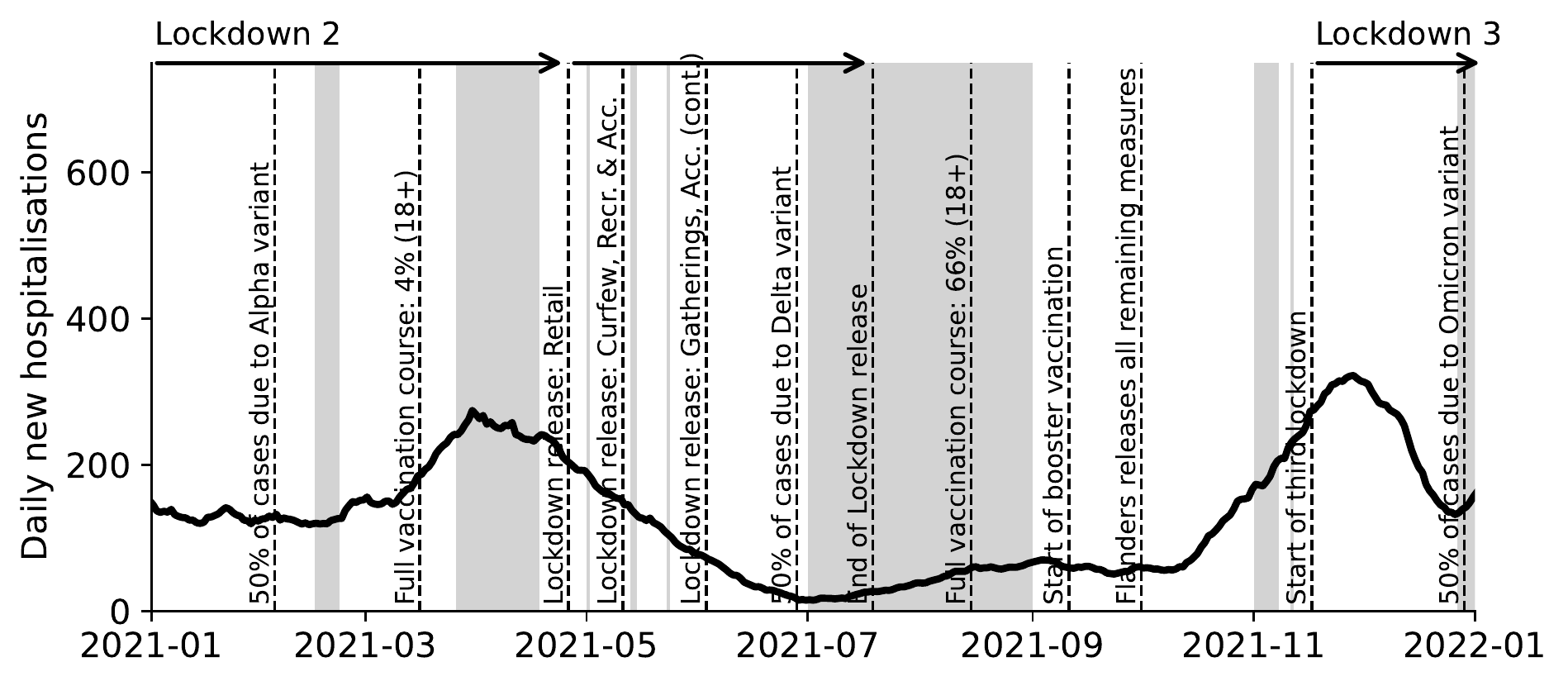}
    \caption{Seven-day moving average of daily new \covid{} hospitalisations in Belgium during 2020 and 2021 (solid black line). Vertical dashed lines are used to indicate events or policy changes with a possible impact on the number of daily new \covid{} hospitalisations. Grey background colour is used to indicates school vacations. The horizontal arrows denote the periods with severe social restrictions, along with their subsequent release.}
    \label{fig:timeline_2020-2021}
\end{figure}

The model parameters $\Omega$, $\Psi$, $K_{\text{inf},\alpha\beta\gamma}$, $K_{\text{inf},\delta}$ and $A$ are calibrated using the 11 provincial time series for daily new hospitalisations (Fig. \ref{fig:all-H_in-series_prov}). The motivation to use these data are fourfold. First, as long as the total hospital capacity is not surpassed, which has not happened in Belgium, the number of hospitalisations is a more objective measure than the daily number of newly detected cases. After all, the latter is highly dependent on the available test capacity. Second, pressure on hospitals is the most relevant measure when informing policy decisions. From a public health perspective, one primarily wants to avoid excess pressure on hospitals, which results in postponement of non-\covid{} care and eventually the collapse of the health care system. Third, these time series are preferred over data for ICU admissions or deaths, because due to the low number of counts, these data are very noisy, especially at the provincial level. Fourth, the daily number of hospitalisations does not depend on hospital dynamics, such as residence times and distributions between wards. Fig. \ref{fig:timeline_2020-2021} provides a timeline of periods with severe preventive policies (\textit{lockdown}) and their subsequent easing, along with an oversights of key events that influenced \sars{} prevalence in Belgium during the 2020-2021 \covid{} pandemic.\\

\begin{figure}[h!]
    \centering
    \includegraphics[width=1.02\linewidth]{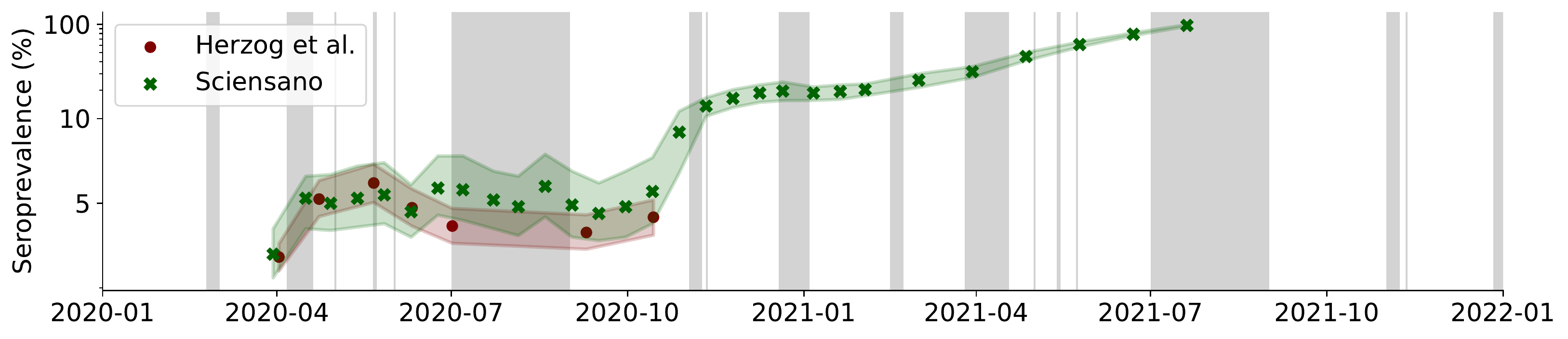}
    \caption{Timeline with seroprevalence data from randomly sampling individuals visiting the general practitioner (Herzog et al. \citep{Herzog2020}, maroon), or Red Cross blood donors (Sciensano \citep{Sciensano2020}, green). The data is space- and age-aggregated and expressed as a percentage of the total population. The band around the data shows the 95\% uncertainty interval. Note the asymmetrical log scale on the y axis. A grey background is used to indicate a holiday period.}
    \label{fig:seroprevalence-data_timeline}
\end{figure}

\noindent The model calibration secondarily relies on seroprevalence data, indicating the rate at which antibodies wane and thus the rate at which humoral immunity is lost (Fig. \ref{fig:seroprevalence-data_timeline}). The seroprevalence time series is the estimated percentage of the population with \sars{} antibodies in the blood, reflecting how many individuals have recovered from \covid{}. Demonstrating the model's ability to match the seroprevalence in the Belgian population is an important gauge for overall model fidelity. In this way it is possible to demonstrate that the model captures the total number of asymptomatic infections. We assume that new VOCs and vaccines do not alter the seroreversion rate over the calibration period.\\

\noindent \textbf{Sciensano hospitalisation data} Sciensano, the national public health institute of Belgium \citep{Sciensano2020}, gathers and processes \covid{}-related hospitalisation time series at the provincial level from all 104 Belgian hospitals. This data set is updated daily, is exhaustive since March 15th 2020, and is anonymous (aggregated over all ages). It contains the number of newly-admitted lab-confirmed \covid{} hospital patients in the last 24 hours, not referred from another hospital. This number excludes the patients that were admitted to the hospital for other reasons but tested positive for \covid{} in a screening context. Seven-day moving-average time series for daily new hospitalisations are shown per province in Fig. \ref{fig:all-H_in-series_prov}. Provinces are denoted according to their NIS code (Table \ref{tab:class-NIS-name}).\\

\noindent The used hospitalisation time series are exhaustive and of high quality, but two limitations should be noted. First, there is a \textit{weekend effect} in the raw time series. This is mainly due to fewer hospitals reporting data over the weekend and does not reflect viral dynamics; the effect is hence not captured by the model. Second, patients are recorded in the province they are hospitalised, not their province of residence. Thus, a patient residing in province $g$ but hospitalised in province $h$ is counted as a data point in province $h$. Since there is no way to circumvent this problem without considerable privacy issues, we must assume that at the level of provinces this effect is negligible.\\

\noindent \textbf{Seroprevalence data} We consider two independent nationally aggregated time series containing information on the extrapolated number of Belgians that have a significant amount of anti-\sars{} antibodies in residual serum samples (i.e. seroprevalence) -- See Fig. \ref{fig:seroprevalence-data_timeline}. The first time series was gathered by Herzog et al. \citep{Herzog2020} between March 30th and October 17th 2020, and contains 7 data points from $3500$ samples per collection period, spread over both sexes, all ages and all provinces (see Table 1 in \citep{Herzog2020}). Residual serum samples in this study originated from ambulatory patients (including people living in nursing homes) visiting their doctor (mainly general practitioners) for any reason including primary care, routine check-up or follow-up of pathology. The second time series was gathered by Sciensano \citep{Sciensano2020} between March 30th 2020 and July 20th 2021, and contains 29 data points from $1000$ samples per collection period, again homogeneously spread throughout Belgium. The blood samples originate from Red Cross blood donors. Combining both data sets is therefore interesting, as it contains both individuals in need of medical attention and healthy individuals capable of donating blood. The larger time period over which the latter study is conducted, implies that the data start to show the prevalence of anti-\sars{} antibodies resulting from vaccination. This, combined with the acquisition of natural immunity, causes the percentage of `immune' individuals to approach 100\% by the summer of 2021 (see Fig. \ref{fig:seroprevalence-data_timeline}).\\

\subsection{Mobility time series data}
\label{app:proximus-mobility-data}

\textbf{Origin and nature of the data} Proximus is Belgium's largest telecommunication company with a market share of 30-40\% in terms of active SIM cards \citep{FOD_economie_proximus_market-share}. Based on the connection between a user's SIM card and the closest transmission tower, the approximate position of a SIM card is known at all times at which the device is operational. The amount of time that this device spends connected to a particular transmission tower is registered, on the condition that it has \textit{reconnected} to a transmission tower and stays connected to this tower for over 15 minutes. Reconnecting occurs either by switching on a disabled device, or by travelling around -- either within or outside a particular postal code. For any given Belgian province, the number of tracked SIM cards represents 25-50\% of the province's population. The extrapolation factor is calculated on a daily basis, based on the number of devices used by individuals living in a particular postal code, and the total registered population there.\\ 

\noindent No data is available for times indicated by the hatched periods in Fig. \ref{fig:staytime_percentage_timeseries}, so we estimate $P^{gh}(t)$ values at these times based on particular periods in the available data. For business days (resp. weekends) before February 10th 2020, we take the average $P^{gh}(t)$ values over all business days (resp. weekends) between February 10th and March 1st 2020. For business days (resp. weekends) after August 31st 2021, we take the average over all business days (resp. weekends) between July 1st and August 31st 2021 (the summer holiday).

\begin{table}[h!]
    \centering
    \caption{All 10 provinces and Brussel-Capital Region (the ``11th province'' for convenience). We denote the population density classification, the systematic name (NIS code), and which region it is in (Flanders, Brussels-Capital, Wallonia). We also denote their registered population and the number of hospitals that report the daily number of new \covid{} patients.}
    \begin{tabular}{p{1.6cm}p{0.8cm}p{2.3cm}rrr}
        \toprule
        \textbf{Type} & \textbf{NIS} & \textbf{Name} & \textbf{Region} & \textbf{Population} & \textbf{\# hospitals} \\ \midrule
        Metropolitan & 21000 & Brussels & B & \num{1218255} & 15 \\ \midrule
        Urban & 10000 & Antwerpen & F & \num{1869730} & 14 \\ 
         & 20001 & Vlaams-Brabant & F & \num{1155843} & 6 \\
         & 40000 & Oost-Vlaanderen & F & \num{1525255} & 14 \\ \midrule
         Rural & 20002 & Brabant Wallon & W & \num{406019} & 2 \\
          & 30000 & West-Vlaanderen & F & \num{1200945} & 11 \\
          & 50000 & Hainaut & W & \num{1346840} & 14 \\
          & 60000 & Li\`ege & W & \num{1109800} & 12 \\
          & 70000 & Limburg & F & \num{877370} & 7 \\
          & 80000 & Luxembourg & W & \num{286752} & 3 \\
          & 90000 & Namur & W & \num{495832} & 6 \\ \bottomrule 
    \end{tabular}
    \label{tab:class-NIS-name}
\end{table}

\begin{figure}[h]
    \centering
    \includegraphics[width=0.99\linewidth]{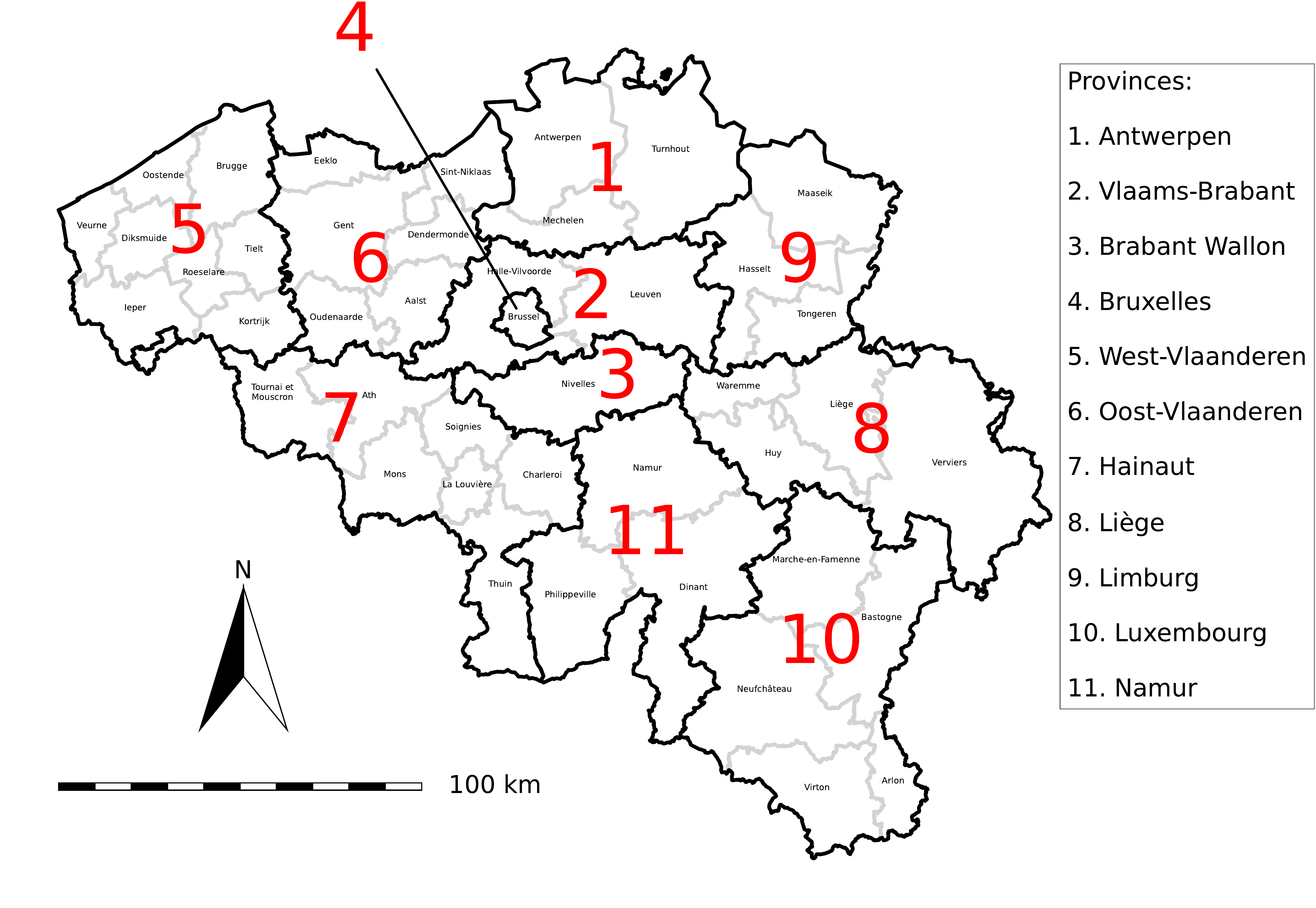}
    \caption{Map of Belgium containing: The 10 Belgian provinces and Brussels bounded by solid black lines and labeled using red numbers (NUTS2). The 43 Belgian arrondissments bounded by grey lines and labeled in small script on the map (NUTS3).}
    \label{fig:beta_classes_prov}
\end{figure}

\clearpage

\section{Social contact model}
\label{app:social_contact}

\noindent\textbf{Scaling prepandemic contacts to model pandemic contacts} The pandemic social behaviour of the Belgian population must be translated into a linear combination of prepandemic interaction matrices, mathematically denoted as,
\begin{multline}
    \widetilde{\bm{N}}^g(t) = \alpha^g(t) \bm{N}^\text{home} + \beta^g(t) \bm{N}^\text{schools} + \gamma^g(t)\bm{N}^\text{work}  \\
    + \delta^g(t)\bm{N}^\text{transport} + \epsilon^g(t)\bm{N}^\text{leisure} + \zeta(t)^g\bm{N}^\text{other},
    \label{matrices}
\end{multline}
where $\widetilde{\bm{N}}^g(t)$ is the pandemic social contact matrix in province $g$ at time $t$ and $\bm{N}^l$ is the prepandemic interaction matrix in location $l$. These prepandemic matrices are available for six locations $l$: at home, in schools, in workplaces, during leisure activities, on public transport and during \textit{other} activities \citep{Willem2020a}. We have to find sensible time-dependent coefficients $\alpha^g(t), \beta^g(t), \dots, \zeta^g(t)$ so that the linear combination of prepandemic interaction matrices in Eq.~\eqref{matrices} is a good representation of macroscopic social behaviour throughout the pandemic. This supplementary read elaborates on the reasoning behind our choice of coefficients in Eq.~\eqref{matrices}.\\

\noindent Ideally, pandemic contact matrices, gathered by performing surveys, are used as these are more likely to adequately represent mixing behaviour under lockdown measures. However, our models have been built upon prepandemic knowledge of social behaviour to make a prediction on pandemic social behaviour for two reasons. First, data on pandemic mixing were not available at the start of the pandemic, and setting up surveys requires a substantial investment of time and resources. Thus, the development of a readily available alternative may still be useful. Second, leveraging the general interest in \covid{} to gather social contact data risks creating unrepresentative samples due to differential interest in the topic (selection bias) \citep{Kennedy2022}. Unrepresentative sampling may skew data in the direction of more adherence to government measures, as it can be expected individuals adhering to government restrictions against \covid{} are more likely to fill in a survey on their behavior. The effect of selection bias likely becomes larger as measures are prolonged, and public debate becomes more polarised. The use of aggregated mobility indicators does not suffer from such bias.\\

\noindent\textbf{Google Community Mobility Reports} Social contact is rescaled daily based on data publicly provided in the GCMR. These data are available (virtually) every day since February 15th, 2020, and are expressed as fractions of ``activity'' compared to the median value from the 5‑week period between January 3rd and February 6th, 2020. This activity is quantified as an anonymous aggregated GPS-informed visitation frequency to six activity types $l'$: retail \& recreation, grocery, parks, transport, work, and residential, which differ slightly from the locations/activities used in the prepandemic contact matrices. We call these unprocessed time series the GCMR indicators, or mathematically, $\bm{\mathcal{G}}(t)$ with elements $\mathcal{G}^{g,l'}(t)$ for every province $g$ and every activity type $l'$. The time series $\bm{G}^{l}(t)$, with elements $G^{g,l}(t)$, used to rescale the prepandemic contact matrices $\bm{N}^l$, are derived from $\bm{\mathcal{G}}(t)$ as follows,
\begin{equation}
    \left\{
        \begin{array}{rl}
            \bm{G}^\text{home}(t) &= 1,\\
            \bm{G}^\text{school}(t) &= \bm{H}(t),\\
            \bm{G}^\text{work}(t) &= \bm{\mathcal{G}}^\text{work}(t),\\
            \bm{G}^\text{transport}(t) &= \bm{\mathcal{G}}^\text{transport}(t),\\
            \bm{G}^\text{leisure}(t) &= \bm{\mathcal{G}}^\text{retail~\&~recreation}(t),\\
            \bm{G}^\text{other}(t) &= \bm{\mathcal{G}}^\text{grocery}(t).
        \end{array}
    \right.
    \label{eq:gcm-to-alpha}
\end{equation}

\begin{figure}[h!]
    \centering
    \includegraphics[width=\linewidth]{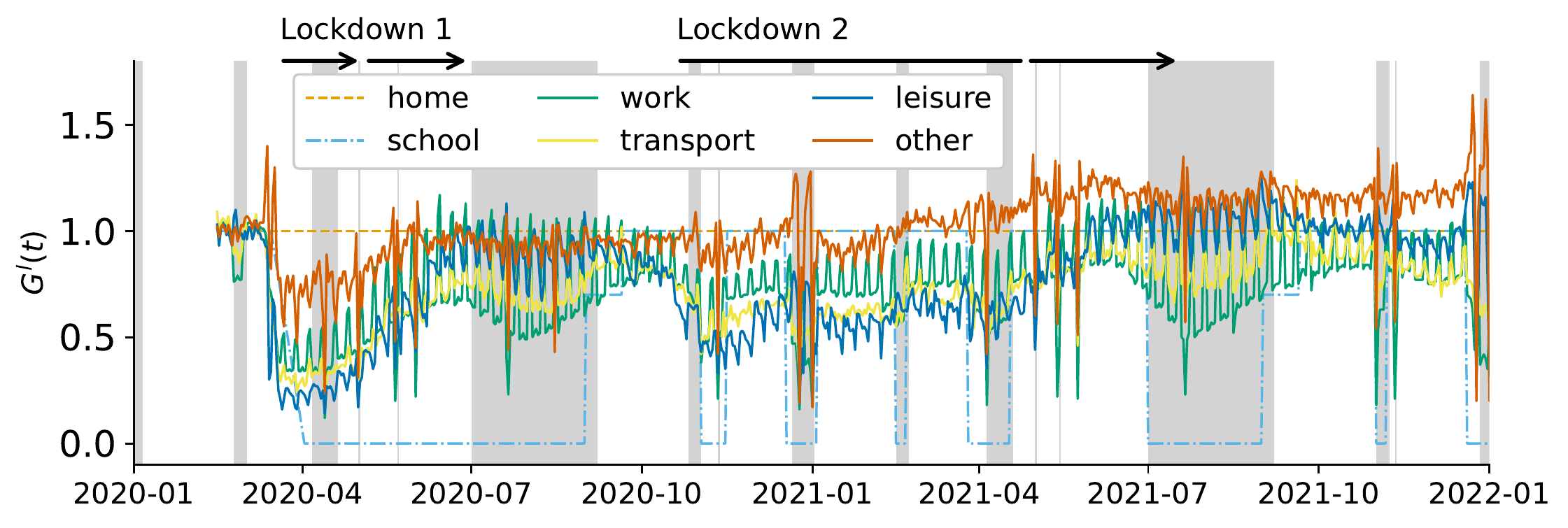}
    \caption{Nationally averaged values of the GCMR indicators $\bm{G}^l(t)$, used for rescaling of the social contact matrices (Eq. \eqref{matrices}). A grey background is used to indicate a holiday period.}
    \label{fig:GCM_resulting_timeseries}
\end{figure}

\noindent The GCMR includes an indicator for residential mobility as well. During lockdowns, residential mobility increases and this is indicative of decreased community mobility. Although the mobility figures indicate people spend more time at home during lockdown (see Fig. 2 in Alleman et al. \citep{Alleman2021}), this does not mean people have more contacts at home. Increasing the fraction of household contacts under lockdown measures would increase the intergenerational mixing of the population and this is not realistic or desired when modeling social restrictions. Hence, we assume home mobility remains the same throughout the entire pandemic, and thus $G^{\text{home}}(t) = 1$. $\bm{H}(t)$ with elements $H^g(t)$ equals 1 when schools are open in province $g$, and 0 otherwise. All $\bm{G}^l(t)$ are assumed equal to 1 before the start of the dataset on February 15th 2020. The resulting time series $\bm{G}^l(t)$ are shown in Fig. \ref{fig:GCM_resulting_timeseries}.\\

\noindent\textbf{Effectivity parameters} Intuitively, the effectivity of a social contact to spread \sars{} in a given location $l$ may not scale linearly with the observed mobility reductions. The net effectivity of the contacts under lockdown depends on a combination of the prepandemic physical proximity and duration of the contact, the effectivity of preventive measures and on behavioural changes during lockdown. As an example, the effects of alcohol gel and face masks might be significant in workplaces and in grocery stores, but not at home or during leisure activities. To account for different effectivities of contacts in different places, we could introduce one additional parameter per location $l$, denoted $\Omega_l$, but this would require inferring six additional  parameters based on the hospitalisation data, which is not possible because of both practical and structural unidentifiability. Simplifications were thus made, reducing the number of effectivity parameters from six to two identifiable parameters. However, as the posterior distributions of these parameters had similar means, significant overlap and little correlation, we could further reduce the number of effectivity parameters to only one.\\

\noindent First, we found that the effectivity parameters of public transport and other places could not be identified. Likely because too few contacts are made in these places \citep{Mossong2008}. Consequently, the effectivity parameters of public transport, other places and leisure contacts were aggregated, as such reducing the number of effectivity parameters from six to four. Second, as previously mentioned, the home contacts are not scaled with the residential GCMR indicator but rather it is assumed that $G^{\text{home}}=1$. The analytical expression of the basic reproduction number $R_0$ of the (equivalent) national model is \citep{Alleman2021},
$$R_0 = \beta (a d_a + \omega) N\,$$
where $\beta$ is the per contact chance of \sars{} transmission, $a$ is the fraction of asymptomatic individuals, $d_a$ the length of the asymptomatic infectious period, $\omega$ the length of the presymptomatic infectious period and $N$ the total number of social contacts. A constant contribution of home contacts during the pandemic (constituent of $N$) strongly correlates with the infectivity parameters in the model ($\beta$), which makes $\Omega^{\text{home}}$ (structurally) unidentifiable. We thus assume $\Omega^{\text{home}}=1$, and the remaining effectivities are thus expressed relatively to the effectivity of home contacts. Third, when calibrating the remaining three effectivity parameters $\Omega^{\text{schools}}$, $\Omega^{\text{work}}$ and $\Omega^{\text{leisure}}$ from March 15th, 2020 until October 14th, 2021, effectivities in schools close to zero are inferred. Changes in school and work contacts often coincide, during holidays schools are closed and workplace mobility is lower. This results in practical unidentifiability between the effectivities of contacts in schools and workplaces. So, we had to assume that $\Omega^{\text{schools}} = \Omega^{\text{work}} = \Omega^{\text{\{work, schools\}}}$. In this way, only two effectivity parameters remained, $\Omega^{\text{\{work, schools\}}}$ and $\Omega^{\text{leisure}}$. When inferring the distributions of these remaining two parameters, we found highly similar average values ($\Omega^{\text{\{work, schools\}}} = \Omega^{\text{leisure}} \approx 0.5$) and overlapping distributions. Thus, we could assume that only one effectivity parameter $\Omega$ is needed to describe the relevant trends in the data. Its physical meaning is the relative effectivity of social contacts in workplaces, schools and during leisure activities for the transmission of \sars{} as compared to contacts at home. It's value should be smaller than one to be consistent with literature \citep{Thompson2021}, which suggests secondary attack rates are higher for household contacts.\\

\begin{figure}[h!]
    \centering
    \includegraphics[width=0.96\linewidth]{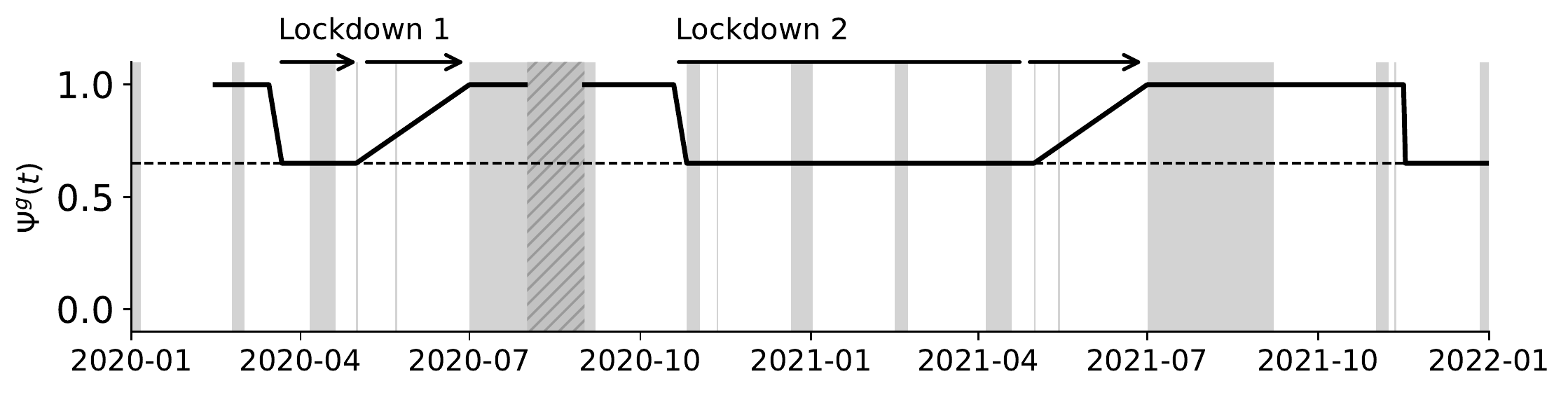}\\
    \includegraphics[width=0.96\linewidth]{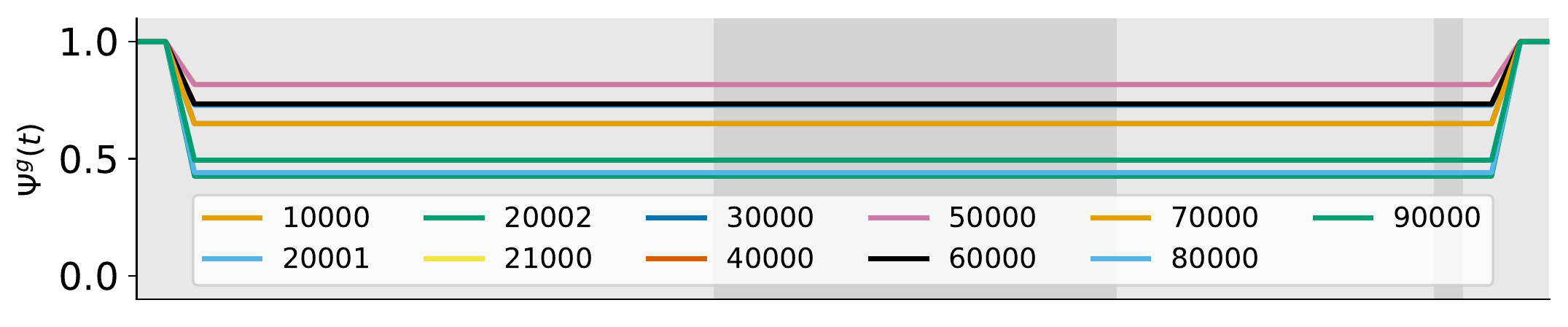}
    \caption{\textit{Top}: time-dependent intervention parameter varying between the values of one when no lockdown is imposed, and $\Psi^g(t) = 0.65\ (95~\% \text{CI}: 0.60-0.70)\ \text{for all provinces}\ g$ under lockdown. The hatched area represents the period in August 2020 where the magnitude of the intervention parameters had to be derived from the peak \covid{} hospitalisation incidence during the second \covid{} wave. A grey background is used to indicate a holiday period.}
    \label{fig:mentality_timeseries}
\end{figure}

\noindent\textbf{Intervention parameter} During model development, we observed that the number of effective social contacts becomes smaller than the number of contacts obtained after rescaling with the GCMR indicators and the effectivity parameters when strict social measures are taken. Thus, one additional parameter was introduced to additionally downscale the number of social contacts when lockdown measures are taken. The so-called \textit{intervention} parameter $\Psi(t)$, with entries $\Psi^g(t)$ for province $g$, is gradually introduced during a two-week period using a ramp function when lockdown measures are taken (2020-03-15 and 2020-10-19) and kept in place throughout lockdowns. Once the lockdown measures are released, it is gradually released from the social contact model over a two-month period using a ramp function. The entries $\Psi^g(t)$ for province $g$ of $\bm{\Psi}(t)$ are always identical, except during a brief period four week period in August 2020, when ad-hoc values were used in order to prevent mistakes during the summer of 2020 from propagating into the second 2020 \covid{} wave (see Fig.~\ref{fig:mentality_timeseries}). First, the \covid{} hospitalisation incidence per 100 000 inhabitants at the peak of the second \covid{} wave were extracted from the hospitalisation data and expressed relative to Antwerp (Table \ref{tab:peak_second_wave}). These values were then rescaled with three parameters: one for Flemish provinces ($0.65,\ 95~\% \text{CI}: 0.60-0.70$), one for Walloon provinces ($0.38,\ 95~\% \text{CI}: 0.29-0.49$) and one for Brussels ($0.73,\ 95~\% \text{CI}: 0.14-0.1.24$). Because no correlation with other model parameters was observed, the values of these parameters were kept constant during the calibrations detailed in this work. Aside from August 2020, the value of the intervention parameter was found to be $\Psi^g(t) = 0.65\ (95~\% \text{CI}: 0.60-0.70)\ \text{for all provinces}\ g$. The introduction of the intervention parameter adds a degree of freedom to the model that can be re-estimated when social context changes in the future or when different measures are taken in different spatial patches of the model.\\

\noindent\textbf{Pandemic contact model} Combining all of the above, the linear combination of prepandemic interaction matrices used to model pandemic social contact is,
\begin{equation}
    \begin{split}
        \widetilde{\bm{N}}^g(t) &= \bm{N^\text{home}} + \Psi^g(t) \Omega \Big \{  G^{g,\ \text{schools}}(t) \bm{N^\text{schools}} + G^{g,\ \text{work}}(t) \bm{N^\text{work}} \\
        & G^{g,\ \text{transport}}(t) \bm{N^\text{transport}} + G^{g,\ \text{leisure}}(t) \bm{N^\text{leisure}} + G^{g,\ \text{other}}(t) \bm{N^\text{other}} \Big \}\,.
    \end{split}
\end{equation}
where $\widetilde{\bm{N}}^g(t)$ is the pandemic social contact matrix in province $g$ at time $t$. $\bm{N}^{\text{home}}$, $\bm{N}^{\text{schools}}$, $\bm{N}^{\text{work}}$, $\bm{N}^{\text{transport}}$, $\bm{N}^{\text{leisure}}$ and $\bm{N}^{\text{other}}$ are the prepandemic social contact matrices. $\Psi^g(t)$ is the intervention parameter in province $g$ at time $t$. $\Omega$ is the relative effectivity of social contacts in workplaces, schools, and during leisure activities to the spread of \sars{} as compared to social contacts at home ($\Omega^{\text{home}} = 1$). $G^{g,\ \text{schools}}$, $G^{g,\ \text{work}}$, $G^{g,\ \text{transport}}$, $G^{g,\ \text{leisure}}$, and $G^{g,\ \text{other}}$ represent the mobility reductions retrieved from the GCMRs.
\pagebreak
Alternatively,
\begin{multline}
    \widetilde{\bm{N}}^g(t) = \alpha^g(t) \bm{N}^\text{home} + \beta^g(t) \bm{N}^\text{schools} + \gamma^g(t)\bm{N}^\text{work}  \\
    + \delta^g(t)\bm{N}^\text{transport} + \epsilon^g(t)\bm{N}^\text{leisure} + \zeta(t)^g\bm{N}^\text{other},
\end{multline}
where,
\begin{equation}
    \left\{
        \begin{array}{rl}
            \alpha^g(t) &= \Psi^g(t),\\
            \beta^g(t) &= \Psi^g(t)\ \Omega\ G^{g,\ \text{schools}}(t),\\
            \gamma^g(t) &= \Psi^g(t)\ \Omega\ G^{g,\ \text{work}}(t),\\
            \delta^g(t) &= \Psi^g(t)\ \Omega\ G^{g,\ \text{transport}}(t),\\
            \epsilon^g(t) &= \Psi^g(t)\ \Omega\ G^{g,\ \text{leisure}}(t),\\
            \zeta(t)^g &= \Psi^g(t)\ \Omega\ G^{g,\ \text{other}}(t).
        \end{array}
    \right.
\end{equation}

\begin{table}[h!]
    \centering
    \caption{Incidence per 100 000 inhabitants at peak of second \covid{} wave, expressed relative to Antwerpen.}
    \begin{tabular}{p{1.2cm}p{2.3cm}p{1.2cm}p{1.5cm}}
        \toprule
        \textbf{NIS} & \textbf{Name} & \textbf{Region} & \textbf{Incidence} \\ \midrule
            10000 & Antwerpen & F & 1.00 \\ 
            20001 & Vlaams-Brabant & F & 1.12 \\
            20002 & Brabant Wallon & W & 1.12 \\
            21000 & Brussels & B & 1.12 \\ 
            30000 & West-Vlaanderen & F & 1.45 \\         
            40000 & Oost-Vlaanderen & F & 1.13 \\
            50000 & Hainaut & W & 2.15 \\
            60000 & Li\`ege & W & 1.93 \\
            70000 & Limburg & F & 1.43 \\
            80000 & Luxembourg & W & 1.16 \\
            90000 & Namur & W & 1.30 \\ \bottomrule 
    \end{tabular}
    \label{tab:peak_second_wave}
\end{table}

\clearpage

\section{Variants of concern, vaccination, and seasonality}
\label{app:VOC_vacc}


\subsection{Variants of concern}

\begin{figure}[t!]
    \centering
    \includegraphics[width=\linewidth]{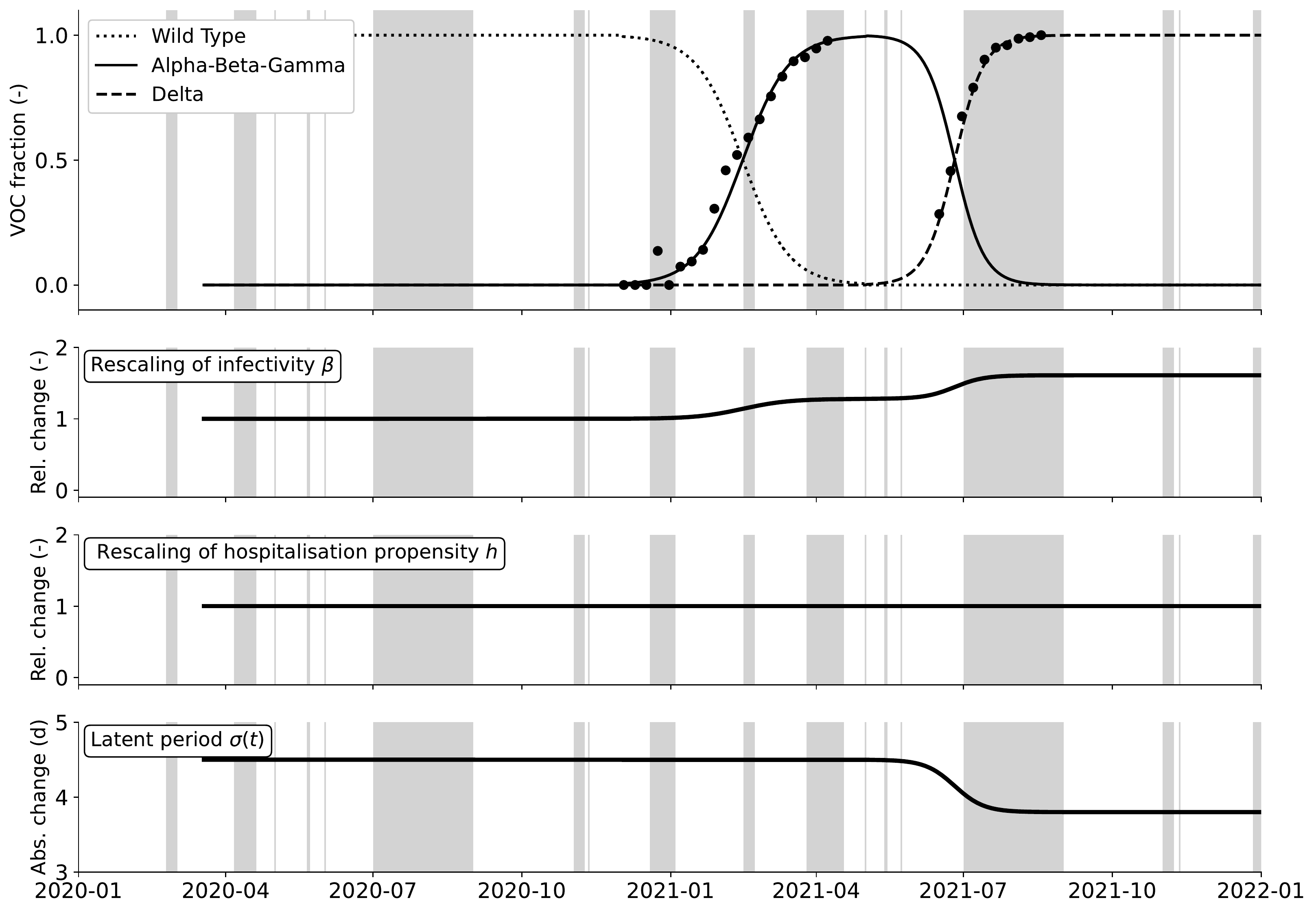}
    \caption{\textit{Top}: Prevalence of the wild-type variant, the Alpha-Beta-Gamma variants (aggregated), and the Delta variant in Belgium. Solid lines show a logistic fit in addition to raw data for the Alpha-Beta-Gamma variant (triangles) and Delta variant (circles) \citep{Wenseleers2021, Sciensano2022}. \textit{Bottom}: Effect of the VOCs on resp. rescaling of the transmission coefficient $\bm{\beta}(t)$,  rescaling of the hospitalisation propensity $\bm{h}(t)$, and time-dependent values of the latent disease period $\sigma(t)$ according to Eq. \eqref{eq:beta-from-VOC}) and values in Table \ref{tab:VOC-dependent-variables}. A grey background is used to indicate a holiday period.}
    \label{fig:VOC_prevalence}
\end{figure}

VOCs are assumed to influence the model dynamics in three ways: 1) VOCs are associated with an increase of the transmission coefficient $\beta$ compared to the wild-type variant, denoted $K_{\text{inf}}$, 2) VOCs can alter the hospital admission propensity of infected individuals compared to the wild-type variant, this is denoted as $K_{\text{hosp}}$,  3) VOCs are associated with different durations of the latent \covid{} period $\sigma$. To account for 1) and 2) respectively, the transmission coefficient $\beta$ is rescaled with the prevalence-weighted average infectivity increase at time $t$, and the hospital admission propensities ($\bm{h}$) are rescaled with the prevalence-weighted average hospital admission propensity gain at every time $t$. The relevant parameter values are listed in Table \ref{tab:VOC-dependent-variables} and graphically illustrated in Fig. \ref{fig:VOC_prevalence}. \\

\begin{table}[!h]
\centering
\caption{VOC prevalence and VOC-dependent variables: infectivity increase of VOC type $n$ compared to the wild type ($K_{\text{inf},n}$), hospitalisation propensity increase, and duration of the latent period. The values of $K_{\text{inf},n}$ were found during model calibration. Values of $K_{\text{hosp},n}$ and  $\sigma_n$ were extracted from \citep{Grint2021, Bager2021, VENETI2022, Hart2022}.}
    \begin{tabular}{ p{2.0cm} p{2.0cm} p{2.0cm} p{2.0cm}} 
    \toprule
        & \multicolumn{3}{l}{\textbf{VOC}} \\[0.5em]
    \textbf{Parameter} & wild type &  Alpha-Beta-Gamma &  Delta \\ \midrule
    $K_{\text{inf},n}$ (-) & 1.00 & $1.40 \pm 0.10$ & $2.00 \pm 0.18$ \\
    $K_{\text{hosp},n}$ (-) & 1.00 & 1.00 & 1.00 \\ 
    $\sigma_n$ (days) & 4.5 & 4.5 & 3.8 \\ \bottomrule
    \end{tabular}
\label{tab:VOC-dependent-variables}
\end{table}

\noindent The VOC prevalence data (on the national level) were obtained from Tom Wenseleers \cite{Wenseleers2021}. The increase in infectivity from the Alpha-Beta-Gamma and Delta VOCs compared to the wild-type was found during model calibration. The combination of the Alpha-Beta-Gamma VOCs were estimated to be $40 \pm 10\%$ more infectious than the wild-type, while the Delta variant was estimated to be $100 \pm 18\%$ more infectious than the wild-type. The combination of the Alpha-Beta-Gamma VOCs almost certainly increased the hospital admission propensity. For instance, Grint et al. \citep{Grint2021} reported an average increase of 62\%. However, we found that applying such multipliers to the model's hospitalisation propensity did not yield satisfactory results. Hence, for the sake of simplicity, we assume no increase of the hospitalisation propensity. The Delta variant was shown to increase the hospital admission propensity for unvaccinated individuals with roughly 70\% \citep{Twohig2022, Bager2021}. On the other hand, a Norwegian study found no significant increase in hospital admission propensity \citep{VENETI2022}. We thus model no hospitalisation propensity increase in this work, as evidence appears to be conflicting.

\subsection{Seasonality}
 The effect of seasonality on \sars{} transmissibility is incorporated in a cosine function with a period of one year (Eq. \eqref{eq:seasonality}, based on \citep{Liu2021season}). The introduction of seasonality rescales the transmission coefficient $\beta$. Maximum transmissibility is assumed at January 1st and minimum transmissibility is assumed at July 1st. The amplitude of the cosine was estimated at $A = 0.18 \pm 0.03$ during model calibration. The seasonality influences viral transmission in ways considered out of this work's scope for this work, hence the simplicity of the seasonal relationship.

\subsection{Vaccination}\label{app_section:vaccination}

Our model uses vaccine incidence data to transfer individuals between the considered vaccination metapopulations (see Section \ref{section:governing_equations}). In every vaccine metapopulation, the vaccine offers protection through three mechanisms, each associated with its own efficacy and waning rate. By dynamically rescaling the efficacies of every vaccine metapopulation, the impact of vaccine waning is accounted for in a computationally inexpensive way (see Section \ref{subsec:voc_and_vac}).\\

\noindent \textbf{Vaccine efficacies} Tartof et al. \citep{Tartof2021} demonstrated that, for an individual fully vaccinated with the mRNA-1273 (Pfizer) vaccine, protection against hospitalisation wanes slower than protection against symptoms. Similar findings were reported by Braeye et al. \citep{Braeye2022b}. From an updated version of Braeye et al. \citep{Braeye2022a} (informal communication), for every relevant VOC, the three vaccine efficacies of a partial vaccination (one dose), full vaccination (two doses) and boosted vaccination (three doses) with mRNA-1273 could be extracted. In addition, the vaccine efficacies 200 days after a full vaccination (two doses) with mRNA-1273 could be extracted (Table \ref{tab:vaccine_properties}). All vaccine efficacies are assumed equal to those of mRNA-1273 because 72\% of vaccines administered by the end of the period considered during the model calibration (2021-10-01) were Pfizer's. We assume that partial vaccination offers half the protection a full vaccination offers, both 25 days and 175 days post-vaccination. No data was available on the waning of booster immunity at the time of writing, so we assumed the immunity of only partially and fully vaccinated individuals to wane. This assumption does not alter any of the results in this work. We assume that the vaccine efficacies for the wild-type variant are the same as the vaccine efficacies of the Alpha-Beta-Gamma variant. This assumption has no impact on the results in this work because an appreciable amount of individuals had only been vaccinated when the Alpha-Beta-Gamma variant became dominant.

\begin{table}[h]
    \centering
    \caption{Efficacies of the vaccines in lowering the susceptiblity to \sars{}, lowering the infectiousness of \sars{}, and the efficacies of the vaccines in lowering the hospitalisation propensity. Partial vaccination is assumed to result in half the efficacy of a full vaccination (both 25 days and 175 days post-vaccination). We assume that the vaccine efficacies for the wild-type variant are the same as the vaccine efficacies of the Alpha-Beta-Gamma variant. Booster shots were not administered under the Alpha-Beta-Gamma VOC. Protection against hospitalisation is retrieved for the Delta VOC from Ref. \citep{Braeye2022b} but assumed to same for the Alpha-Beta-Gamma VOC. All $E_{\text{none,n}}$ are 0.} %
    \begin{tabular}{>{\raggedright\arraybackslash}p{3.0cm} m{1.6cm} m{1.6cm} m{1.6cm} m{1.6cm} m{1.4cm}}
        \toprule
        & $\bm{E_{\text{\textbf{partial}},n}}$ & $\bm{E_{\text{\textbf{full}},n,0}}$ & $\bm{E_{\text{\textbf{full}},n,w}}$ & $\bm{E_{\text{\textbf{booster}},n,0}}$ \\ \midrule
        \multicolumn{5}{l}{\textbf{Susceptibility}} \\
        \quad Alpha-Beta-Gamma & 0.44 & 0.87 & 0.64 & NA  \\
        \quad Delta & 0.40 & 0.79 & 0.54 & 0.80 \\ \midrule
        \multicolumn{5}{l}{\textbf{Infectiousness}} \\
        \quad Alpha-Beta-Gamma & 0.31 & 0.62 & 0.43 & NA \\
        \quad Delta & 0.19 & 0.38 & 0.25 & 0.34 \\ \midrule
        \multicolumn{5}{l}{\textbf{Hospitalisation }} \\
        \quad Alpha-Beta-Gamma & 0.47 & 0.93 & 0.81 & NA\\
        \quad Delta & 0.47 & 0.93 & 0.81 & 0.93 \\
        \bottomrule
    \end{tabular}
    \label{tab:vaccine_properties}
\end{table}

\begin{figure}[h]
    \centering
    \includegraphics[width=\linewidth]{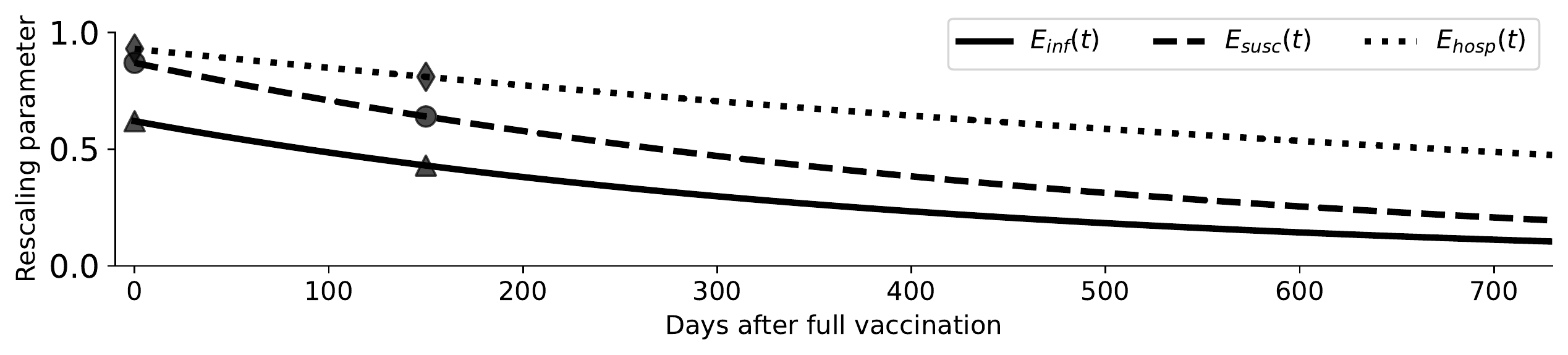}
    \caption{Dynamics of the vaccine efficacy associated with infectivity $\tilde{E}_{\text{full, }\alpha\beta\gamma,\ \text{inf}}(t)$, susceptibility $\tilde{E}_{\text{full, }\alpha\beta\gamma,\ \text{susc}}(t)$, and hospitalisation propensity $\tilde{E}_{\text{full, }\alpha\beta\gamma,\ \text{hosp}}(t)$ under the Alpha-Beta-Gamma VOCs and for a full vaccination. The observations extracted from literature (see Table  \ref{tab:vaccine_properties}) were used to inform the half-life of the fitted exponential decay function.}
    \label{fig:effect_of_waning_delayed}
\end{figure}

\begin{figure}[h]
    \centering
    \includegraphics[width=\linewidth]{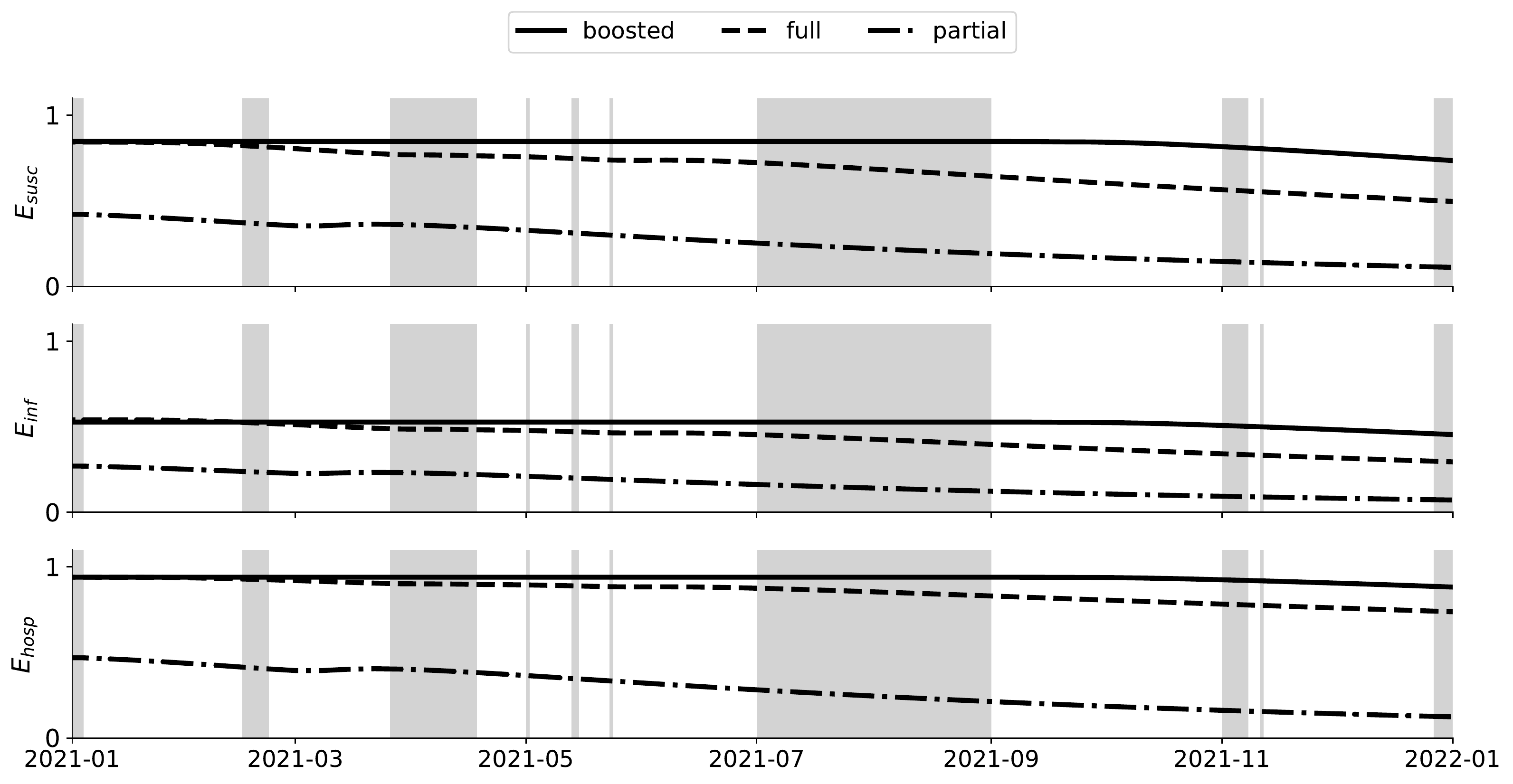}
    \caption{Dynamics of the vaccination efficacy associated with susceptibility $\bm{E}_{v,\text{susc}}$, infectivity $\bm{E}_{v,\text{inf}}$, and hospitalisation propensity $\bm{E}_{v,\text{hosp}}$ for $v \in \{ \text{partial, full, boosted}\}$. VOC-weighted efficacy. Average of all age groups and provinces. Values closer to one denote beter protection. A grey background is used to indicate a holiday period.}
    \label{fig:vaccine_rescaling_effect}
\end{figure}

\begin{figure}[h!]
    \centering
    \includegraphics[width=\linewidth]{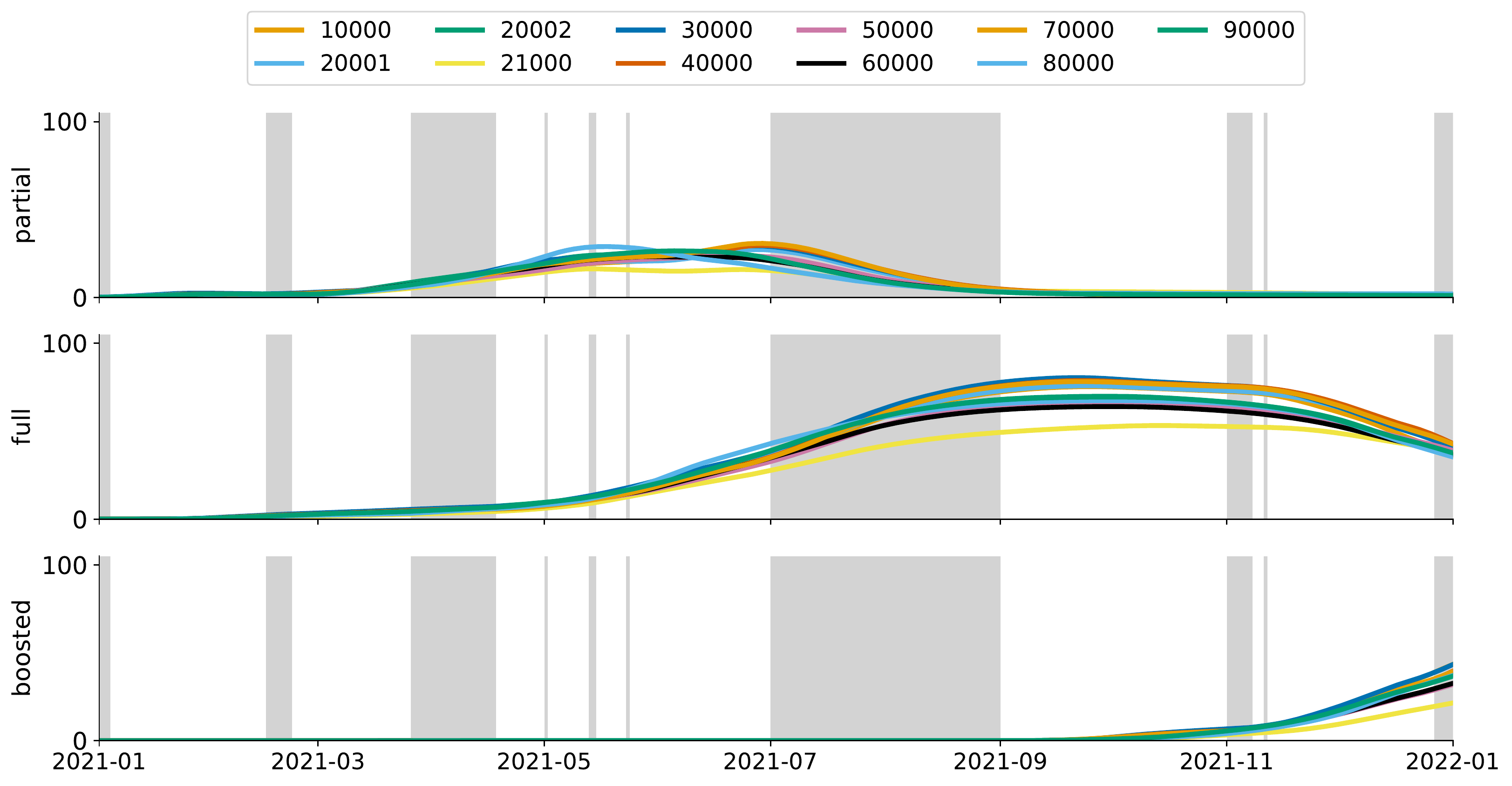}
    \caption{Fraction of individuals in every province in the partially, fully and boosted vaccine metapopulations (indicated by NIS code, see Table \ref{tab:class-NIS-name}). From top to bottom: first dose only (all vaccine types except Janssen), full dose only (second dose and Janssen vaccine), booster shot. A grey background is used to indicate a holiday period.} 
    \label{fig:vaccination_timeseries_NIS}
\end{figure}

\begin{figure}[h!]
    \centering
    \includegraphics[width=\linewidth]{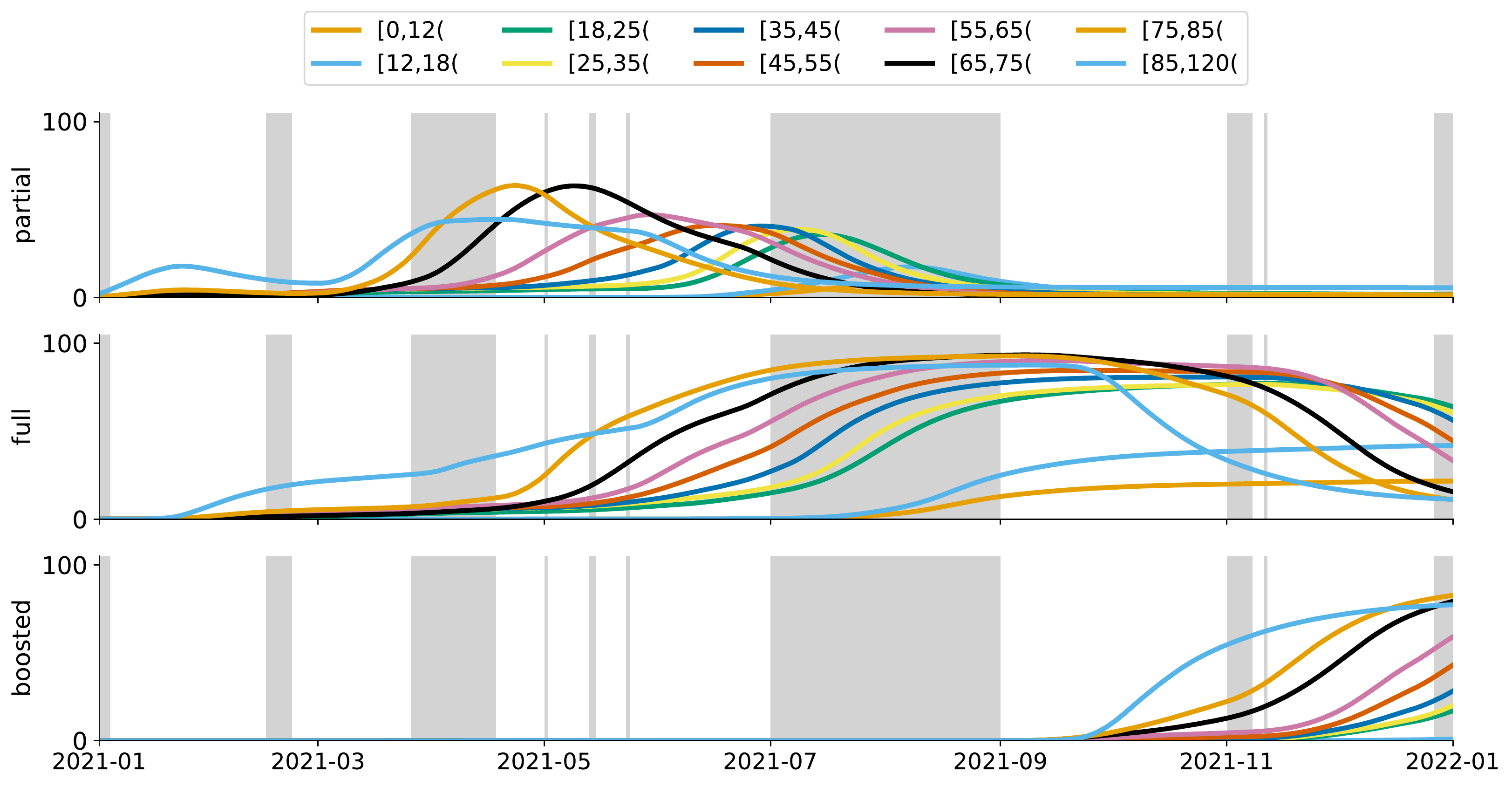}
    \caption{Fraction of individuals in every age group in the partially, fully and boosted vaccine metapopulations. From top to bottom: first dose only (all vaccine types except Janssen), full dose only (second dose and Janssen vaccine), booster shot. A grey background is used to indicate a holiday period.} 
    \label{fig:vaccination_timeseries_age}
\end{figure}

\clearpage
\newpage
\section{Model parameters and assumptions}
\label{app:model-equations-and-model-parameters}

\subsection{Model parameters}

\begin{table}[!h]
\centering
\caption{Fraction of asymptomatic individuals $a_i$ (based on \citep{Poletti2021}), and hospitalisation propensity $h_i$ for symptomatic infections per age class (inferred, see Alleman et al. \citep{Alleman2021}). The hospitalisation propensity $\bm{h}$ is dynamically and spatially rescaled in the model to account for the combined effects of VOCs and vaccination. The baseline values without VOCs or vaccines are shown here.}
\begin{tabular}{ p{3cm} p{1.5cm} p{1.5cm} } 
\toprule
\textbf{Age class $i$ (years)} & $a_i$ (\%) & $h_i$ (\%)\\ \midrule
$[0,12[$ & 81.9 & 1.0 \\
$[12,18[$ & 81.9 & 1.0 \\
$[18,25[$ & 78.8 & 1.5 \\
$[25,35[$ & 77.6 & 2.5 \\
$[35,45[$ & 73.6 & 3.0 \\
$[45,55[$ & 69.5 & 6.0 \\
$[55,65[$ & 67.1 & 12.0 \\
$[65,75[$ & 64.5 & 40.0 \\
$[75,85[$ & 51.1 & 70.0 \\
$[85,\infty[$ & 35.4 & 99.0 \\ \midrule
\textbf{Population average} & 71.4 & 14.7 \\ \bottomrule
\end{tabular}
\label{tab:ageDistributionAsymptomatic}
\end{table}

\begin{table}[!h]
\centering
\caption{Average fraction $c_i$ of hospitalised individuals admitted in a cohort ward (as opposed to an Intensive Care Unit), average mortality in cohort wards ($m_{\text{C},i}$) and average mortality in ICU ($m_{\text{ICU},i}$) per age class. These estimates were obtained by analysing a dataset of \num{22 136} patients in all 133 Belgian hospitals (see Alleman et al. \cite{Alleman2021} for details).}
    \begin{tabular}{ p{3cm} p{1.5cm} p{1.5cm} p{1.7cm} } 
    \toprule
    \textbf{Age class $i$ (years)} & $c_i$ \textbf{(\%)} & $m_{\text{C},i}$ \textbf{(\%)} & $m_{\text{ICU},i}$ \textbf{(\%)} \\ \midrule
    $[0,12[$ & 97.4 & 0.0 & 0.0 \\
    $[12,18[$ & 88.8 & 0.0 & 9.0 \\
    $[18,25[$ & 90.3 & 0.4 & 17.4 \\
    $[25,35[$ & 91.5 & 1.0 & 11.8 \\
    $[35,45[$ & 87.1 & 1.5 & 16.0 \\
    $[45,55[$ & 83.0 & 2.7 & 19.3 \\
    $[55,65[$ & 78.3 & 5.1 & 35.4 \\
    $[65,75[$ & 76.3 & 11.4 & 51.6 \\
    $[75,85[$ & 83.6 & 26.4 & 70.0 \\    
    $[85,\infty[$ & 95.3 & 42.3 & 78.6 \\ \midrule
    \textbf{Population average} & 83.8 & 16.6 & 46.4 \\ \bottomrule
    \end{tabular}
\label{tab:results_hospital_age}
\end{table}

\begin{table}[!h]
\centering
\caption{Hospital length-of-stay in a cohort ward ($C$) or intensive care unit (ICU) in case of recovery or death. NA denotes no deaths were recorded in that particular age class. These estimates were obtained by analysing a dataset of \num{22136} patients in all 133 Belgian hospitals (see Alleman et al. \cite{Alleman2021} for details).}
    \begin{tabular}{ p{1.6cm} p{1.2cm} p{1.2cm} p{1.2cm} p{1.2cm} p{1.2cm}} 
    \toprule
    \textbf{Age class $i$ (years)} & $d_{C,R,i}$ (days) & $d_{C,D,i}$ (days) & $d_{\text{ICU},R,i}$ (days) & $d_{\text{ICU},D,i}$ (days) & $d_{\text{ICU},\text{rec},i}$ (days)\\ \midrule
    $[0,12[$ & 3.5 & NA & 5.9 & NA & 3.0 \\
    $[12,18[$ & 6.8 & NA & 3.2 & 16.0 & 4.0\\
    $[18,25[$ & 5.7 & 2.0 & 5.3 & 3.0 & 4.0 \\
    $[25,35[$ & 4.8 & 8.1 & 9.3 & 12.6 & 4.5 \\
    $[35,45[$ & 5.9 & 6.0 & 10.9 & 16.3 & 5.0\\
    $[45,55[$ & 6.9 & 8.8 & 11.4 & 20.6 & 6.0 \\
    $[55,65[$ & 8.5 & 8.7 & 12.7 & 17.3 & 6.0 \\
    $[65,75[$ & 11.2 & 13.2 & 13.8 & 16.3 & 8.0 \\
    $[75,85[$ & 15.2 & 12.1 & 11.9 & 13.6 & 11.0 \\    
    $[85,\infty[$ & 18.9 & 11.8 & 5.0 & 9.1 & 10.0\\ \midrule
    \textbf{Population average} & 10.8 & 11.8 & 12.0 & 15.2 & 5.6\\ \bottomrule
    \end{tabular}
\label{tab:results_hospital_days}
\end{table}

\pagebreak
\subsection{Model assumptions and simplifications}

Here, we list the assumptions and simplifications underpinning our model our model. While we consider these to not alter the paper's conclusions, we choose to explicitly mention them below as good scientific practice.\\

\begin{enumerate}

    \item The Tau-Leaping method \citep{Gillespie2001} assumes the model's rates of transitions remain constant during one leap. Because some of our model’s parameters and populations change over the course of one leap, due to time dependency, we expect the introduction of a small numerical error. By contrasting the stochastic model with a deterministic (ODE) variant, it was verified that the error was small and did not alter any of the manuscript's in a significant way. We suspect the deviation is not noticeable because the changes to the parameters (f.i. due to seasonality) and populations (due to vaccinations) are small during the Tau-leap of $\tau = 0.5~d$.\\
 
    \item Cross-border mobility is not included in this model, the mobility matrix, $\bm{P}$, is not age-stratified, and the elements $P^{gh}(t)$ were estimated when no data was available at time $t$ (see Appendix \ref{app:proximus-mobility-data}).\\

    \item The GCMR indicators, which are used to inform the degree of social interaction in the model, are not age-stratified. Because the pandemic contact matrices are made by scaling prepandemic contact matrices with the GCMR indicators, our model preserves prepandemic mixing of the population under pandemic circumstances. Our method is thus a more coarse-grained alternative to social-epidemiological contact studies under lockdown measures.\\

    \item The intervention parameter, $\Psi(t)$, is a phenomenological parameter downscaling the number of social contacts when lockdown measures are taken. It is introduced into the model when lockdown measures are taken and gradually eased out of the model when lockdown measures are released. Its value under lockdown measures is determined by fitting to the available hospitalisation data. Alternatively, $\Psi(t)$ could be a function (linear, logistic, etc.) of \sars{} spread and the function's parameters could be determined during the calibration procedure (similar to \citep{Bruno2020}).\\

    \item The average vaccine efficacies and information on vaccine waning used in the model were those of the Pfizer vaccine. The model does not explicitly distinguish between the different vaccines.\\

    \item We aggregate the Alpha, Beta and Gamma VOCs because the effect of their epidemiological properties are comparable in our model, and the aggregation decreases the overall complexity.\\

    \item Our model does not include age-specific increases for transmissibility and disease severity for the VOCs.\\
    
    \item VOC strains are modeled by rescaling the transmission coefficient ($\beta$), hospitalisation propensity ($\bm{h}$) and length of the latent phase ($\sigma$) with the weighted prevalence of VOCs, as obtained from data. The model does not accomodate VOCs explictly by adding compartments and can thus not be used to model competition between strains. We used the former approach because it is computationally cheaper.\\
    
    \item The emergence of the variants was implemented on the national level, thus, the geographic spread of the Alpha-Beta-Gamma and Delta variant was not included in the simulations.\\

    \item We assume that new VOCs and vaccines do not alter the seroreversion rate ($\zeta$).\\

    \item Implementing seasonality using a cosine function is a high-level mathematical abstraction of several factors such as, but not limited to, the effects of humidity and temperature on viral survival in the environment.\\

    \item In order for the negative binomial distribution log-likelihood function to apply to all $G \times n$ data points in the model calibration, the data points should strictly speaking be independent of each other, which they are not.\\

    \item The model does not explicitly account for testing and tracing. These effects are implicitly accounted for in the calibrated parameters, however.\\

    \item Raw vaccination data is only communicated for minors 0-17 years. There is no distinction for 0-12 or 12-17. In our current implementation, all vaccinations are distributed between 0-12 and 12-17 year olds based on demographics. \\
    
    \item Vaccinated people are assumed to have the same number of contacts and the same mobility patterns as non-vaccinated people. Vaccinated people come into contact with the same fraction of vaccinated and non-vaccinated people as the national average, while some degree of segregation between vaccinated and non-vaccinated individuals could be expected.\\
    
    \item The rate of transfer from the recovered to the susceptible pool, which influences the average duration of protection against reinfection does not depend on the vaccine stage.

\end{enumerate}

\newpage

\section{Model calibration}\label{app:calibration}

Eight model parameters are considered to be a priori unknown and must be calibrated using the available data. Here we elaborate on the calibration procedure and the resulting parameter values and uncertainties.

\subsection{Choosing an appropriate observational model}\label{app:appropriate_statistical_model}

\begin{figure}[h!]
    \centering
    \includegraphics[width=\linewidth]{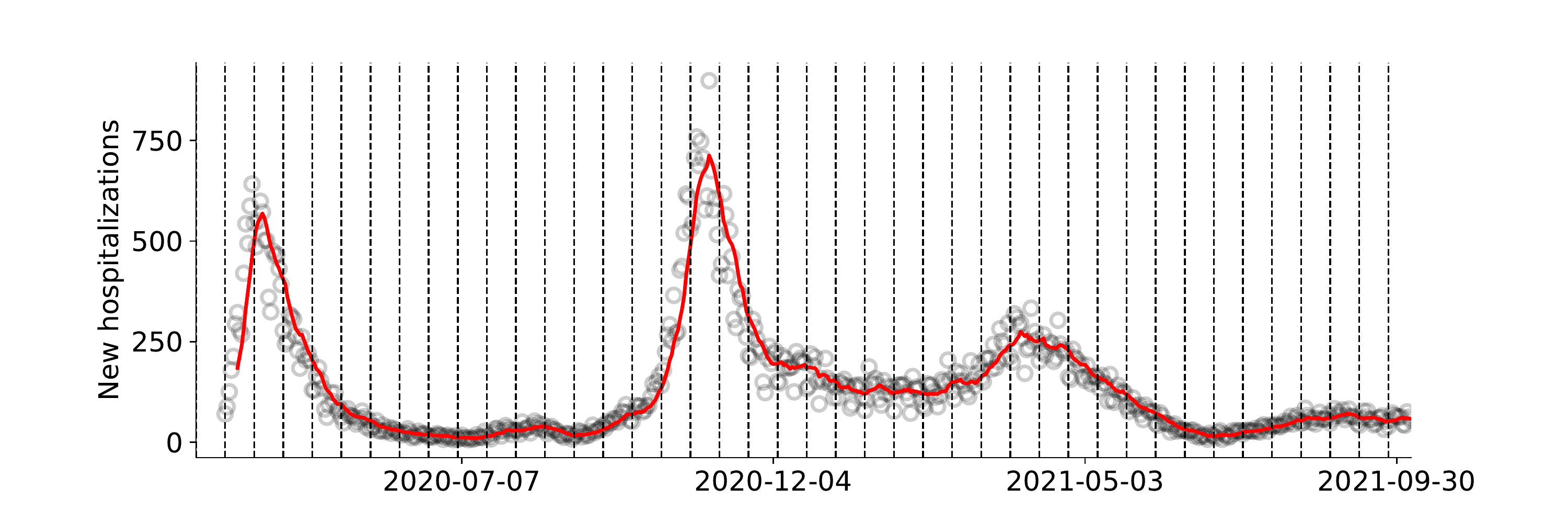}
    \caption{National daily new hospitalisations (markers). Seven-day exponential moving average (solid red line).}
    \label{fig:data_subdivide}
\end{figure}

Given a time series of daily hospitalisations $\bm{x}^g$ for every province $g \in \{1, ..., G\}$ with $n$ observations $x_t^g$ for $t \in \{1, ..., n\}$ corresponding to times $\{t_1, ..., t_n\}$, any choice for model parameters $\bm{\theta}$ combined with an initial condition (IC) will produce a continuous time series $\tilde{x}^g(t)$ for every province $g$ (after summing over all age groups and vaccination stages). This time series may be sampled to produce a set of model-based values $\{\tilde{x}^g(t_1), ..., \tilde{x}^g(t_n)\}$ that we will denote as $\{\tilde{x}^g_1, ..., \tilde{x}^g_n\}$. The aim is to find the model parameters for which it is most likely that the $\bm{x}^g$ are observations of the modelled time series $\bm{\tilde{x}}^g$. An appropriate statistical distribution must be chosen to assess what deviations between $\bm{x}^g$ and $\bm{\tilde{x}}^g$ are tolerable. To find the most appropriate statistical distribution, the relationship between the mean and variance of the time series $\bm{x}^g$ must be studied. The time series for the daily number of hospitalisations consist of one observation per day without information on the variance. Mean-variance couples were \textit{approximated} for all provincial time series using the following procedure,
\begin{enumerate}
    \item Compute the (7-day) exponential moving average of the time series $\bm{x}^g$ (solid red line in Figure \ref{fig:data_subdivide}). Assume it represents the \textit{underlying truth}.
    \item Subdivide the time series $\bm{x}^g$ into discrete windows of length $n$ days. Window lengths $n$ of 7, 14 and 31 days were used with consistent results.
    \item In every window, compute the mean observation and the variance between the exponential moving average and the observations.
\end{enumerate}

\begin{figure}
    \centering
    \includegraphics[width=\linewidth]{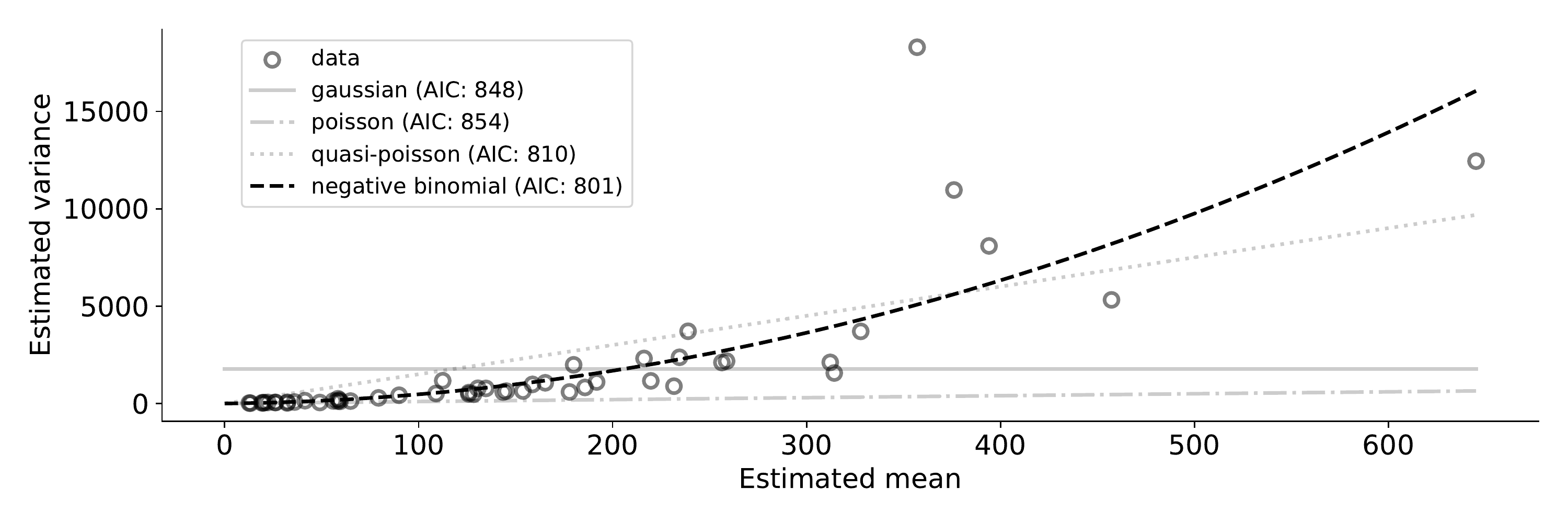}
    \caption{Estimated mean-variance couples of the national time series of daily new hospitalisations, shown together with the fitted mean-variance models and their respective AIC scores.}
    \label{fig:mean_variance_model}
\end{figure}

Next, the most appropriate statistical model was chosen by fitting the mean-variance of several candidate distributions -- the Gaussian model ($\sigma^2 = c $), Poisson model ($\sigma^2 = \mu$), quasi-Poisson model ($\sigma^2 = \alpha \mu$) and negative binomial model ($\sigma^2 = \mu + \alpha \mu^2$) -- and using the Akaike Information Criterion (AIC) to determine what model fits best. As an example, the result of the above analysis is shown for the national time series of daily hospitalisations in Figure \ref{fig:mean_variance_model}. At the provincial level, the negative binomial model best described the variance in the data in all but two provinces, in which the quasi-Poisson model had the lowest AIC. However, for the sake of simplicity, it was assumed that all eleven provincial time series variance are described by the negative binomial model. In this way, we assume that a single observation $x^g_t$ is the result of a counting experiment with an additional unknown error for every province $g$, captured by the estimated overdispersion parameter $\alpha^g$ per province $g$ \cite{cameron1998, Chan2021} (see Table \ref{tab:overdispersions}). The values of which were obtained by fitting the negative binomial mean-variance relationship to our estimated mean-variance couples. In general, the overdispersion in the data becomes larger when the population in a province decreases. The associated negative binomial likelihood for every observation $t$ is
\begin{equation}
    \mathcal{L}( \tilde{x}_t^g \vert x_t^g ) = \frac{\Gamma(x_t^g + 1/\alpha^g)}{\Gamma (x_t^g + 1) \Gamma(1/\alpha^g)} \left( \frac{1/\alpha^g}{1/\alpha^g + \tilde{x}^g_t} \right)^{1/\alpha^g} \left( \frac{\tilde{x}^g_t}{1/\alpha^g + \tilde{x}^g_t} \right)^{x^g_t},
\end{equation}
with $\Gamma$ the gamma function. The negative binomial distribution has mean value $\tilde{x}^g_t$ and variance $\tilde{x}^g_t(1 + \alpha^g\tilde{x}^g_t)$; it is maximised for $\tilde{x}_t = x_t$ and reduces to the Poisson likelihood for $\alpha^g \rightarrow 0$. Adding more observations over time and regions, individual likelihood functions can be multiplied:
\begin{equation*}
    \mathcal{L}( \bm{\tilde{x}} \vert \bm{x} ) = \prod_{g=1}^G\prod_{t=1}^n \mathcal{L}( \tilde{x}_t^g \vert x_t^g ).
\end{equation*}
Again, this value $\mathcal{L}( \bm{\tilde{x}} \vert \bm{x} )$ is maximised if $\forall g,t: \tilde{x}_t^g = x_t^g$, but this is generally not possible: the values $\tilde{x}_t^g$ must be samples of the simulated local time series $\tilde{x}^g(t)$, for particular $\bm{\theta}$ values. Since the logarithmic function is monotonically increasing, the maximum value for $\mathcal{L}(\bm{\tilde{x}} \vert \bm{x})$ occurs at the same location in parameter space as for $\log \mathcal{L}(\bm{\tilde{x}} \vert \bm{x})$, so we may as well consider:
\begin{multline*}
    \log \mathcal{L}(\bm{\tilde{x}} \vert \bm{x}) = -\sum_{g=1}^G\sum_{t=1}^n \left( \log\left[\frac{\Gamma(x^g_t + 1/\alpha^g)}{\Gamma (x_t^g + 1) \Gamma(1/\alpha^g)}\right] + 1/\alpha^g\log\left[ \frac{1/\alpha^g}{1/\alpha^g + \tilde{x}^g_t} \right] \right. \\ \left.+ x^g_t\log\left[ \frac{\tilde{x}^g_t}{1/\alpha^g + \tilde{x}^g_t} \right]\right).
    \label{eq:calibration_loglikelihood_complete}
\end{multline*}
The result is the log-likelihood in Eq. \eqref{eq:calibration_loglikelihood}. The parameter choice $\bm{\theta} = \bm{\hat{\theta}}$ that maximises Eq. \eqref{eq:calibration_loglikelihood} for the obtained values of $\alpha^g$ is considered the \textit{best-fitting} choice. A large collection of such sampled $\bm{\hat{\theta}}$ make up the posterior. The posterior distributions resulting from the calibration MCMC also provide a quantitative measure for the calibrated value's uncertainty interval \citep{emcee2013}, which together with the overdispersion values ($\alpha^g$) determines the uncertainty on the simulated time series. Note that large $\tilde{x}^g_t$ and $x^g_t$ values will contribute more to the total sum in Eq. \eqref{eq:calibration_loglikelihood} than small such values, which means that time series of large provinces will have a larger weight in the overall sum. This effect is further amplified by the fact that less densely populated provinces generally have noisier data and thus larger overdispersion factors $\alpha^g$. In our calibration procedure, we use three sources of data and thus, we optimise the weighted sum of three such log-likelihoods,
\begin{equation*}
    \log \mathcal{L}(\bm{\tilde{x}}_{H_{\text{in}}} \vert \bm{x}_{H_{\text{in}}}) + \log \mathcal{L}(\bm{\tilde{x}}_R \vert \bm{x}_{R,\text{Herzog}}) + \log \mathcal{L}(\bm{\tilde{x}}_R \vert \bm{x}_{R,\text{Sciensano}}),
\end{equation*}
The time series $\bm{\tilde{x}}_{H_{\text{in}}}$ and $\bm{\tilde{x}}_R$ correspond to the simulated daily new hospitalisations per province (summed over age groups and vaccine doses) and the total number of recovered individuals (summed over provinces, age groups, and vaccine doses), respectively. The observed time series are $\bm{x}_{H_{\text{in}}}$, $\bm{x}_{R,H}$ and $\bm{x}_{R,S}$: observed daily new hospitalisations per province \citep{Sciensano2020}, national seroprevalence data from general practitioners by Herzog et al. \citep{Herzog2020}, and national seroprevalence data from Red Cross by Sciensano \citep{Sciensano2020}, respectively (see Appendix \ref{app:sciensano}).\\

\noindent Due to the stochastic nature of the model, the simulated time series $\tilde{x}^g(t)$ will differ slightly for every choice of model parameters and IC. To asses the robustness of the sampled parameter distributions to this stochasticity, the model was calibrated using the mean of 2, 5 and 10 stochastic realisations as $\tilde{x}^g(t)$. No noticeable changes in the distributions were observed, however, the calibration procedure had become respectively 2,5 and 10 times slower. A single stochastic realisation was thus used as $\tilde{x}^g(t)$.

\begin{table}[h]
    \centering
    \caption{Values per province of the estimated overdispersion parameter of the negative binomial distribution associated with the time series of daily \covid{} hospitalisations, used in the log-likelihood function \eqref{eq:calibration_loglikelihood}. The average overdispersion coefficient of 0.034 (population-size weighted) was used for all simulations presented in this work.}
    \begin{tabular}{p{2.2cm}p{1.1cm}p{2.3cm}p{1.1cm}p{1.8cm}p{.5cm}}
        \toprule
        \textbf{Province} & $\bm{\alpha^g}$ & \textbf{Province} & $\bm{\alpha^g}$ & \textbf{Province} & $\alpha^g$ \\ \midrule
        Antwerpen & 0.031        & West-Vlaanderen & 0.041    & Limburg & 0.060 \\
        Vlaams-Brabant & 0.035   & Oost-Vlaanderen & 0.027    & Luxembourg & 0.003 \\
        Brabant Wallon & 0.059   & Hainaut & 0.029            & Namur & 0.007 \\
        Brussels & 0.037         & Li\`ege & 0.039            &  & \\ \bottomrule
    \end{tabular}
    \label{tab:overdispersions}
\end{table}

\pagebreak
\subsection{Results of Model calibration}

Calibrated values of all a priori unknown model parameters, including their interpretation, are listed in Table \ref{tab:calibration_parameters}. The posterior distributions of the estimated parameters and their potential correlations are shown in Fig. \ref{fig:full-calibration-corner-plot}. Simulations of the daily number of new hospitalisations for every province are shown in Figs \ref{fig:provincial-complete-model-fit-0} and \ref{fig:provincial-complete-model-fit-1}. The small difference in goodness-of-fit between the spatially explicit and the national models is demonstrated in Fig. \ref{fig:RMSE-fit-boxplot}.

\begin{figure}
    \centering
    \includegraphics[width=\linewidth]{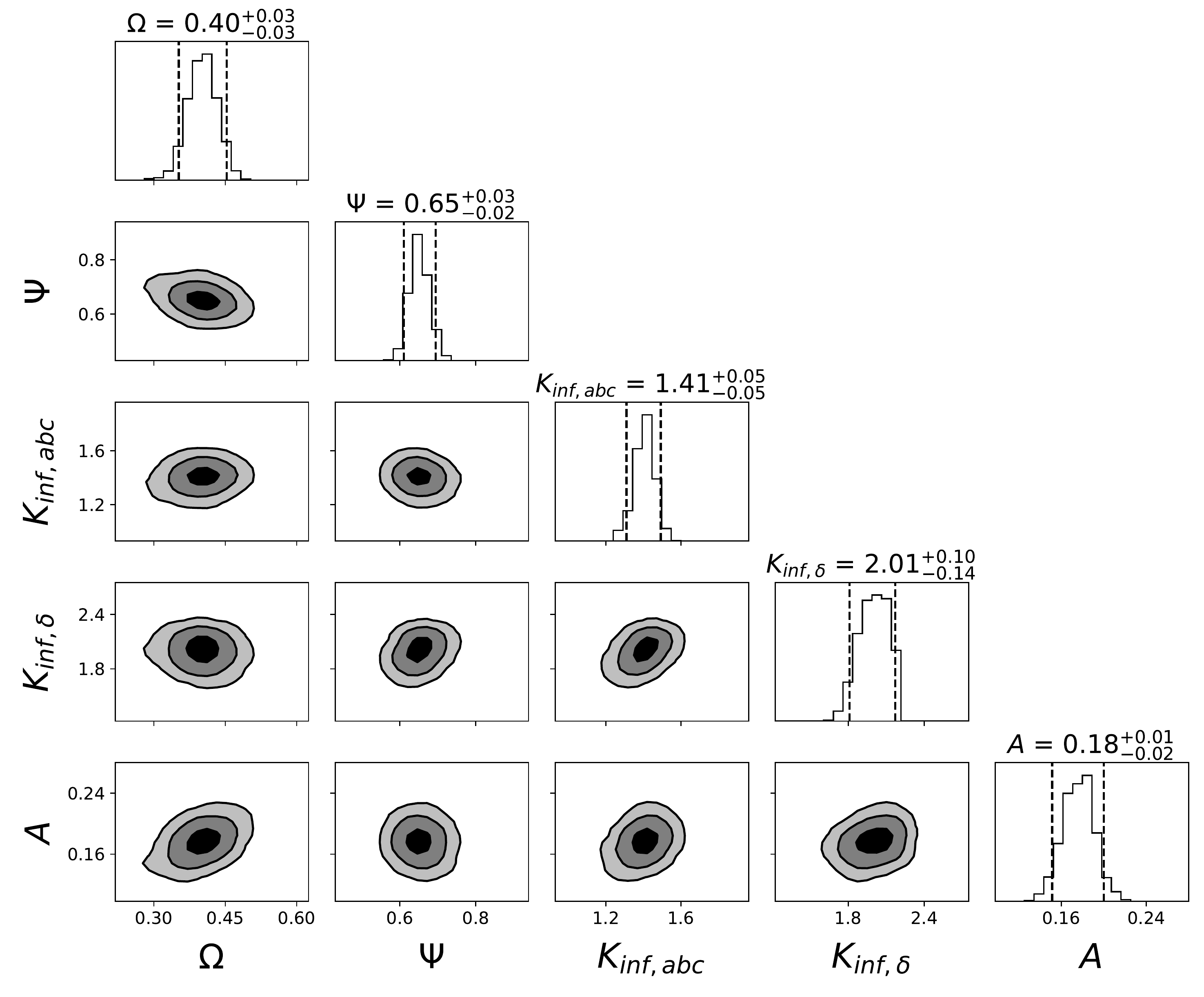} \caption{Corner plot showing the posterior distributions of the five calibrated parameters. Provincial model. Created with the \texttt{corner} package \cite{corner2016}.}
    \label{fig:full-calibration-corner-plot}
\end{figure}

\begin{figure}
    \centering
    \includegraphics[width=\linewidth]{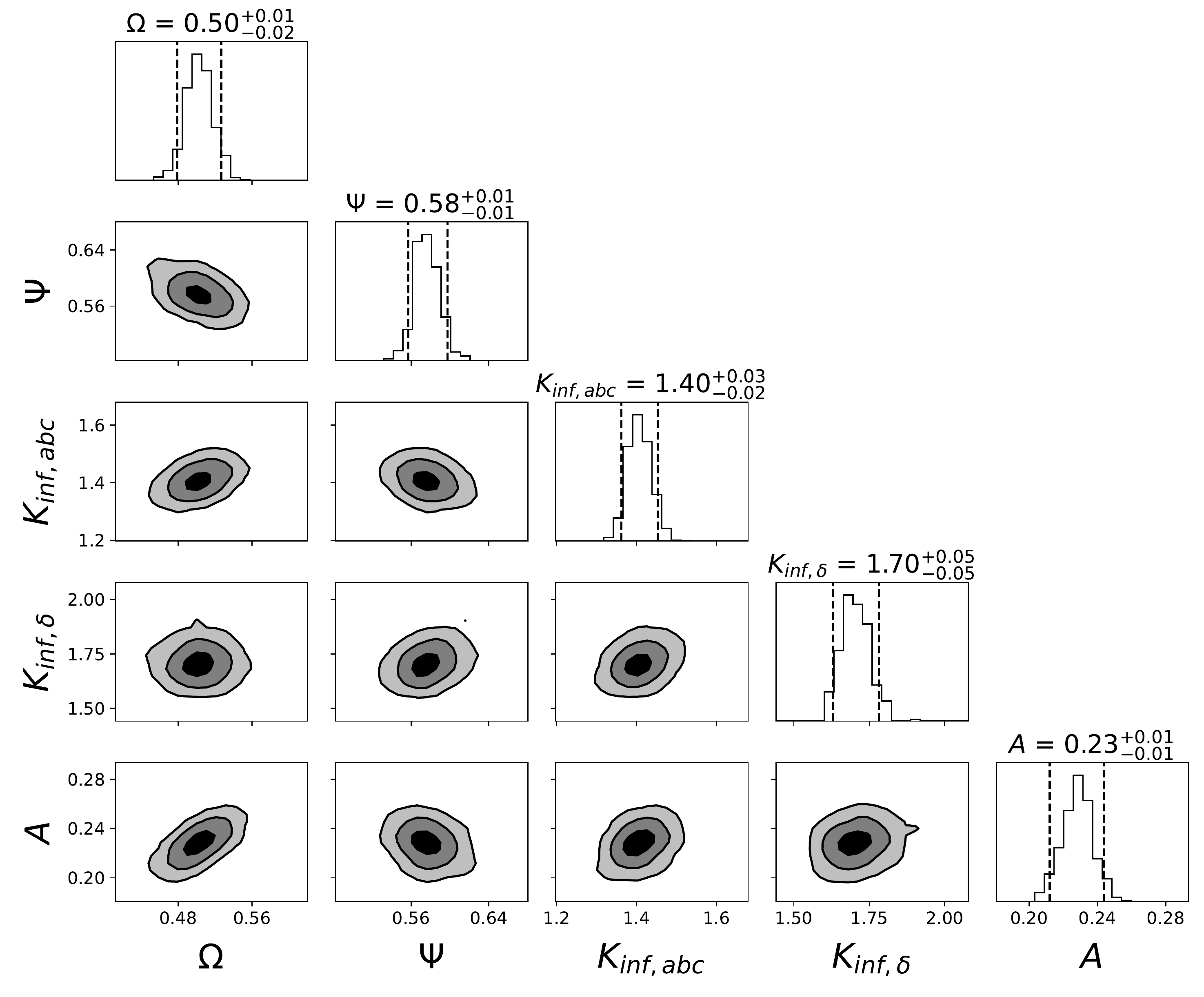} \caption{Corner plot showing the posterior distributions of the five calibrated parameters. Equivalent national model. Created with the \texttt{corner} package \cite{corner2016}.}
    \label{fig:full-calibration-corner-plot}
\end{figure}

\begin{figure}
    \centering
    \includegraphics[width=\linewidth]{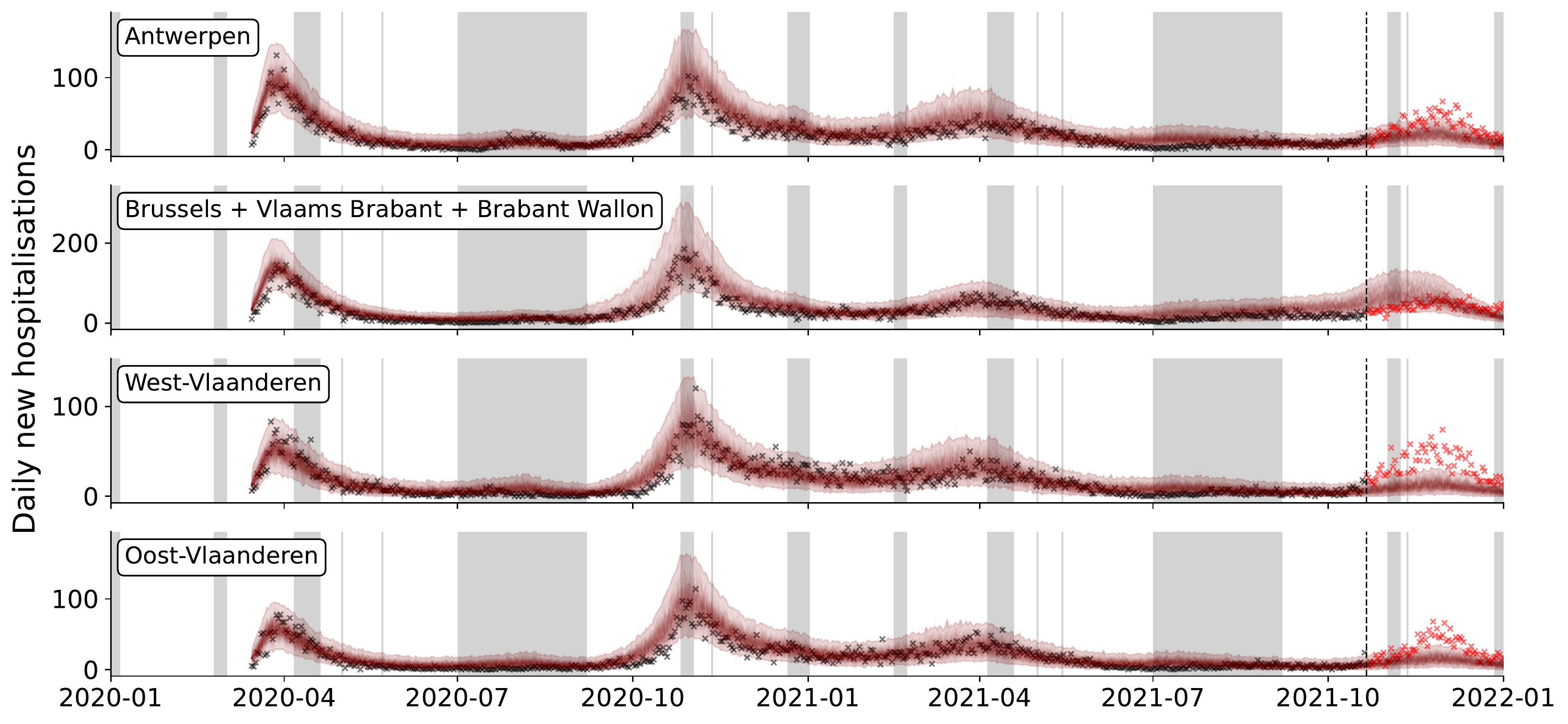}
    \caption{100 model realisations of the daily new hospitalisations between March 15th 2020 and January 1st 2022 (solid lines) with a negative binomial 95\% confidence region (transparent band). Black crosses signify raw data from Sciensano \cite{Sciensano2020} were used in the calibration procedure while red crosses signify data were not used during the calibration procedure. From top to bottom: Antwerpen (10000), Brussels, Brabant Wallon and Vlaams Brabant (20001, 20002, 21000), West-Vlaanderen (30000) and Oost-Vlaanderen (40000). (see Table \ref{tab:class-NIS-name} and Fig. \ref{fig:beta_classes_prov}). A grey background is used to indicate a holiday period.}
    \label{fig:provincial-complete-model-fit-0}
\end{figure}

\begin{figure}
    \centering
    \includegraphics[width=\linewidth]{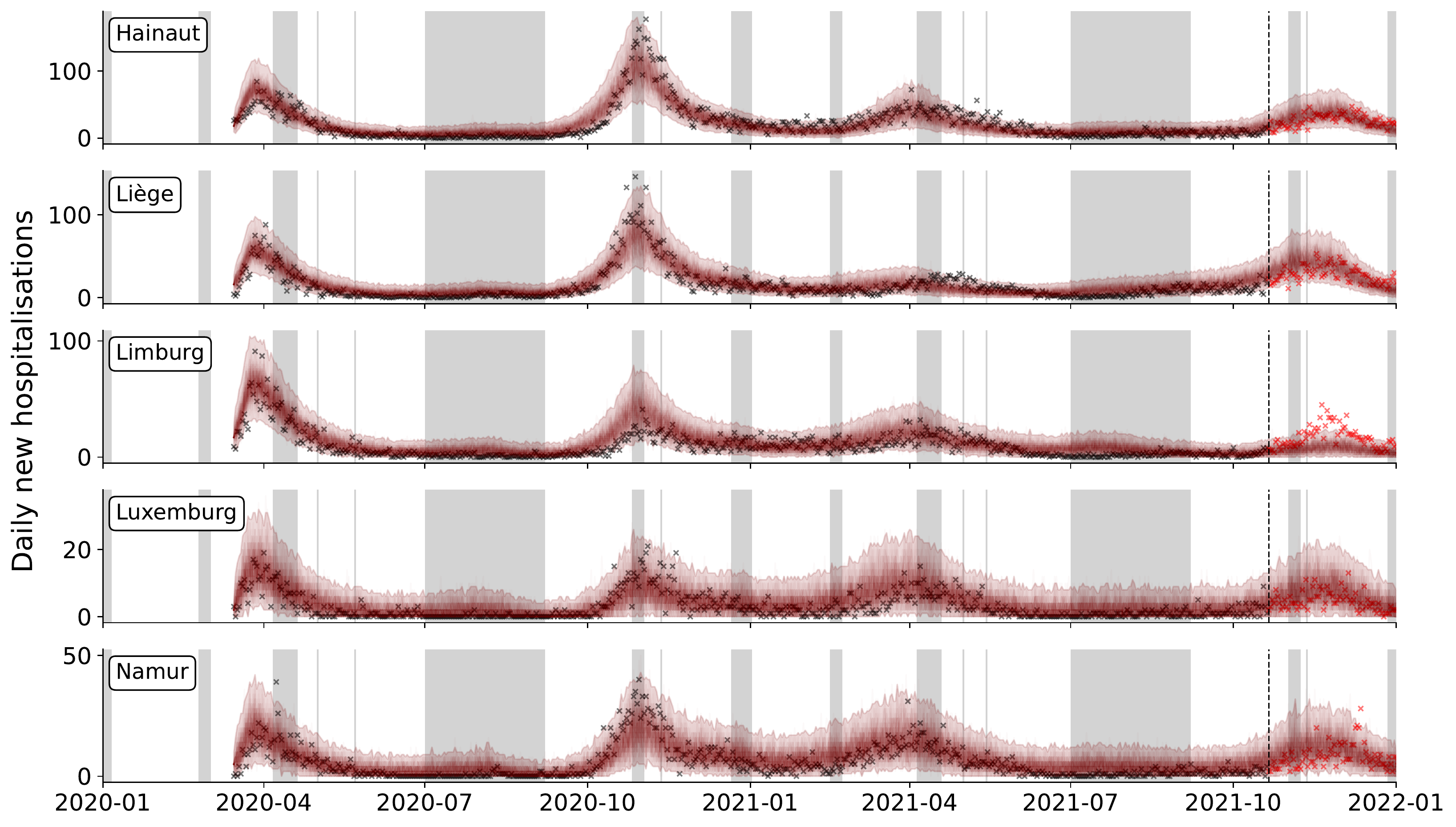}
    \caption{100 model realisations of the daily new hospitalisations between March 15th 2020 and January 1st 2022 (solid lines) with a negative binomial 95\% confidence region (transparent band). Black crosses signify raw data from Sciensano \cite{Sciensano2020} were used in the calibration procedure while red crosses signify data were not used during the calibration procedure. From top to bottom: Hainaut (50000), Li\`ege (60000), Limburg (70000), Luxembourg (80000), Namur province (90000)  (see Table \ref{tab:class-NIS-name} and Fig. \ref{fig:beta_classes_prov}). A grey background is used to indicate a holiday period.}
    \label{fig:provincial-complete-model-fit-1}
\end{figure}

\begin{figure}[h!]
    \centering
    \includegraphics[width=\linewidth]{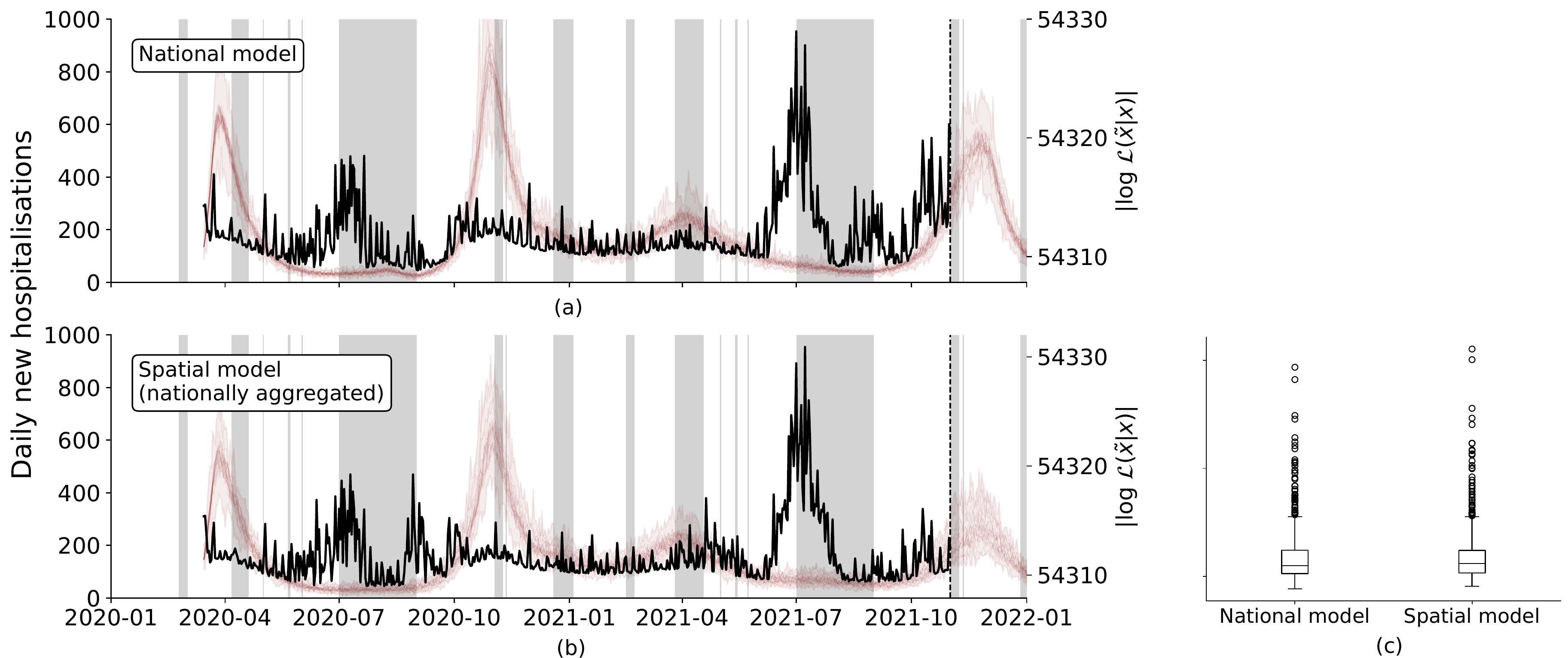}
    \caption{(a) 100 realisations of the equivalent national model and (b) 100 realisations of the spatially explicit model (nationally aggregated) of the daily new hospitalisations between March 15th 2020 and January 1st 2022 (solid lines) with a negative binomial 95\% confidence region (transparent band). The accompanying negative binomial log-likelihood score of the model predictions is given in black on the right hand axis. (c) Boxplot of the log-likelihood values at every time $t$ of the national and spatially explicit model. No difference in log-likelihood was found between the national model and the spatially-explicit model (Mann-Whitney U test; $p=0.81$) Despite morphological differences, the goodness of fit of both models behaves in a similar manner: when \sars{} prevalence is low, both models have difficulties being accurate. A grey background is used to indicate a holiday period.}
    \label{fig:RMSE-fit-boxplot}
\end{figure}

\begin{figure}[h!]
    \centering
    \includegraphics[width=\linewidth]{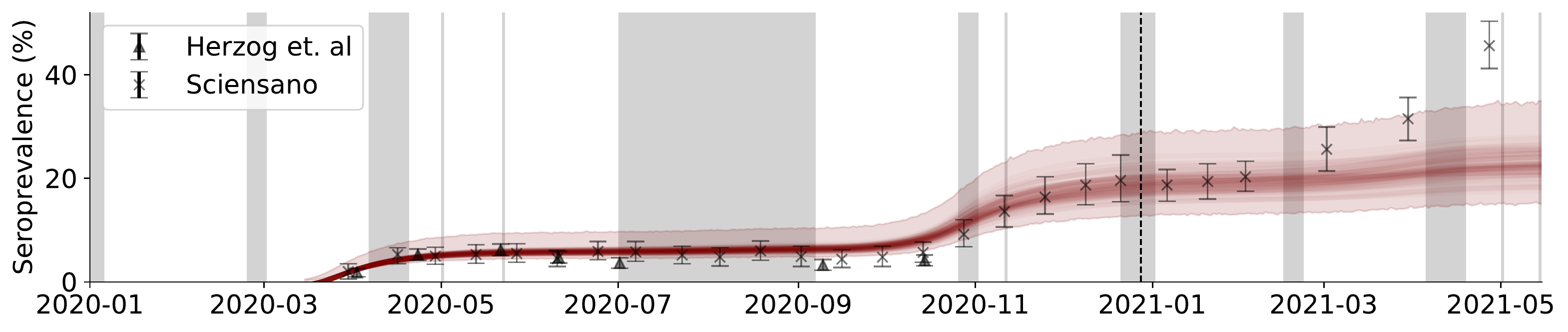}
    \caption{(a) 100 realisations of the estimated fraction of seropositive individuals, as proxied by the Recovered (R) state of the model (solid lines), along with the negative binomial 95\% confidence region (transparent band) versus the fraction of seropositive individuals as measured by Herzog et al. \cite{Herzog2020} and the Belgian Scientific Institute of Public Health \cite{Sciensano2020}. The dashed vertical line indicates the start of the nation-wide vaccination campaign (2020-12-28). After the start of the vaccination campaign, the Recovered (R) state no longer is a valid proxy for seroprevalence. A grey background is used to indicate a holiday period.}
    \label{fig:seroprevalence-spatial-fit}
\end{figure}

\clearpage

\end{appendices}


\bibliography{bibliography}


\end{document}